\documentclass[12pt]{article}

\usepackage{latexsym,amsmath,amssymb,amsthm,amsfonts,graphicx}
\usepackage{natbib}
\usepackage{epsfig,bm,url,blkarray}
\usepackage{setspace}
\usepackage{nicefrac}
\usepackage{comment}
\usepackage{float} 
\usepackage{booktabs} 
\usepackage{graphicx} 
\usepackage[margin=1cm]{caption}
\usepackage{mathtools}
\usepackage{xcolor}
\usepackage{color}
\floatstyle{plain}

\usepackage{algcompatible}
\usepackage{algorithmicx}
\usepackage[noend]{algpseudocode}
\usepackage{algorithm}
\usepackage{diagbox}
\newfloat{Algorithm}{thp}{lop}
\floatname{Algorithm}{Algorithm}
\newtheorem{lemma}{Lemma}

\usepackage{titling}
\newcommand{\eps}{\epsilon}
\newcommand{\wh}{\widehat}

\def\V{{\mathbb V}}
\include{def}
\def\dispmuskip{\thinmuskip= 3mu plus 0mu minus 2mu \medmuskip=  4mu plus 2mu minus 2mu \thickmuskip=5mu plus 5mu minus 2mu}
\def\textmuskip{\thinmuskip= 0mu                    \medmuskip=  1mu plus 1mu minus 1mu \thickmuskip=2mu plus 3mu minus 1mu}
\def\beq{\dispmuskip\begin{equation}}    \def\eeq{\end{equation}\textmuskip}
\def\beqn{\dispmuskip\begin{displaymath}}\def\eeqn{\end{displaymath}\textmuskip}
\def\bea{\dispmuskip\begin{eqnarray}}    \def\eea{\end{eqnarray}\textmuskip}
\def\bean{\dispmuskip\begin{eqnarray*}}  \def\eean{\end{eqnarray*}\textmuskip}

\usepackage{xr}
\makeatletter
\newcommand*{\addFileDependency}[1]{
  \typeout{(#1)}
  \@addtofilelist{#1}
  \IfFileExists{#1}{}{\typeout{No file #1.}}
}
\makeatother

\newcommand*{\myexternaldocument}[1]{
    \externaldocument{#1}
    \addFileDependency{#1.tex}
    \addFileDependency{#1.aux}
}

\myexternaldocument{OnlineSupplement_JCGS}
\def \bvec {\text{\boldmath$b$}}    
\def \cvec {\text{\boldmath$c$}}

\def \yvec {\text{\boldmath$y$}}    
\def \zvec {\text{\boldmath$z$}}    \def \mZ {\text{\boldmath$Z$}}

\def \betavec         {\text{\boldmath$\beta$}}

\def \etavec          {\text{\boldmath$\eta$}}
\def \thetavec        {\text{\boldmath$\theta$}}

\def \lambdavec       {\text{\boldmath$\lambda$}}
\def \muvec           {\text{\boldmath$\mu$}}

\def \pivec           {\text{\boldmath$\pi$}}

\def \ellvec		{\text{\boldmath$\ell$}}

\usepackage{authblk}
\makeatother
\begin{document}
\title{Fast Variational Boosting for Latent Variable Models}
%\date{\empty}
\author[1,3]{David Gunawan}
\author[4,5]{David Nott}
\author[2,3]{Robert Kohn}
\affil[1]{School of Mathematics and Applied Statistics, University of Wollongong}
\affil[2]{School of Economics, University of New South Wales}
\affil[3]{Australian Center of Excellence for Mathematical and Statistical Frontiers}
\affil[4]{Department of Statistics and Data Science, National University of Singapore}
\affil[5]{Institute of Operations Research and Analytics, National University of Singapore}
\renewcommand\Authands{ and }
\maketitle
\vspace{-0.5in}

\begin{abstract}
We consider the problem of estimating complex statistical latent variable models using variational Bayes methods. These methods are used when exact posterior inference is either infeasible or computationally expensive, and they
approximate the posterior density with a family of tractable distributions.  The parameters of the approximating distribution are estimated using optimisation methods.  This article develops a flexible Gaussian mixture variational approximation, where we impose sparsity in the precision matrix of each Gaussian component to reflect the appropriate conditional independence structure in the model. By introducing sparsity in the precision matrix and parameterising it using the Cholesky factor, each Gaussian mixture component becomes parsimonious (with a reduced number of non-zero parameters), while still capturing the dependence in the posterior distribution. Fast estimation methods based on global and local variational boosting moves combined with natural gradients and variance reduction methods are developed. The local boosting moves adjust an existing mixture component, and optimisation is only carried out on a subset of the variational parameters of a new component. The subset is chosen to target improvement of the current approximation in aspects where it is poor.  The local boosting moves are fast because only a small number of variational parameters need to be optimised. The efficacy of the approach is illustrated by using simulated and real datasets to estimate generalised linear mixed models and state space models. 

% \footnote{This sentence reads a little funny. Here is a rewrite or at least a way to clarify if I have it correct. 

%``The local boosting moves adjust an existing mixture component, and optimization is only carried out on the variational parameters of a new component." 
%Do I understand the gist of what you are trying to do? }

\smallskip
\noindent \textbf{Keywords:} Generalised linear mixed models; Global boosting; Local boosting;  Mixture Variational approximation;  State space models; Stochastic gradient ascent algorithm.

\end{abstract}
%XX\footnote{Put the key words in alphabetical order so "Generalised linear ... comes first }
%\citep{} or particle Markov chain Monte Carlo

\section{Introduction}\label{sec:Intro}
Statistical inference for latent variables models is an important and active research area in statistics. One of the main challenges in estimating such models is that the number of latent variables grows with the number of observations. These models are often estimated using exact Bayesian simulation methods, including variants of  Markov chain Monte Carlo \citep{Andrieu:2010, stan}, or importance sampling based methods such as sequential Monte Carlo \citep{gunawan2022flexible}. However, it is usually computationally very expensive to estimate latent variable models using these exact approaches. 

Variational Bayes (VB) is increasingly used as a method for conducting parameter and latent variable inference in a wide range of challenging statistical models with computationally difficult posteriors \citep{blei2017,tan2017,smith+ln20}. VB methods approximate the posterior density by using a family of tractable distributions whose parameters are estimated by optimisation. The variational approximation is used when exact inference for the posterior distribution is either infeasible or computationally expensive. It usually produces posterior inference with much less computational cost compared to the exact methods. 

Much of the current literature on estimating the posterior distribution of the latent variable models using VB methods focuses on Gaussian variational approximation \citep{Titsias2014,tan2017}. 
A major problem with Gaussian approximations is that many posterior distributions of the latent variables and parameters can be
multimodal and heavy-tailed, and Gaussian
variational approximations cannot capture these complex features; see section \ref{sec:example}. 

There are a number of variational approximations that are more flexible than Gaussian variational approximations proposed in the literature. For example, \citet{smith+ln20} 
propose Gaussian copula and skew Gaussian copula-based variational approximations;  \citet{Guo2016}  and \citet{Miller2016} propose a mixture of normals variational approximation; \citet{rezende+m15} propose planar and radial normalising flows. Other flexible normalising flows for variational inference are reviewed by \citet{papamakarios2021normalizing}. 
Much of the work on normalising flows is related to and inspired by transport map approaches to Bayesian computation; see \citet{marzouk2016sampling}
for an introduction.

%Variational inference based on transport maps (see \citet{marzouk2016sampling}, for an introduction) is a related approach that preceded and inspired current work on normalizing flows.

Our article makes a number of contributions. First, we develop versions of the mixture of Gaussian variational approximations for estimating high-dimensional posterior distributions of the latent variable models. The mixture of Gaussians is employed as the variational approximation because, under reasonable assumptions, the mixture of Gaussians is a universal
approximator of multivariate distributions. It is well-known that a mixture of Gaussians is flexible enough to approximate any distribution  \citep{Parzen1962,Epanechnikov1969}.  We consider a multivariate Gaussian distribution with a mean vector $\muvec$ and a precision matrix $\Omega$, $N(\muvec,\Omega)$, where $\Omega=L^{-\top}L^{-1}$ and $L$ is a lower triangular matrix with positive diagonal entries, as the mixture components. The parameterisation provides a way to impose a sparsity structure in the precision matrix of each Gaussian component to reflect the appropriate conditional independence structure in the model. 
Sparsity can significantly reduce computational complexity and enable faster convergence of the optimisation methods, especially for models with a large number of variables. 

%\citet{Miller2016}
%uses stochastic
%gradient ascent (SGA) optimisation with the reparameterization trick to fit a
%mixture of Gaussian densities. They find that the use of the reparameterization trick in the boosting variational method still results in a large variance and it is necessary
% to use many samples to estimate the gradient of the variational lower bound.

Variational optimisation of a Gaussian mixture variational approximation
is challenging in complex latent variable models with a large number of parameters and latent variables because of the large
number of variational parameters that need to be optimised.
The boosting variational inference method used by \citet{Guo2016},  and \citet{Miller2016} provides an efficient approach to fit mixture-type variational approximations by adding a single mixture component at a time. The posterior approximation is refined iteratively over variational boosting iterations. 
\citet{Guo2016}, \citet{locatello+kgr18} and \citet{campbell2019universal} consider similar variational boosting mixture approximations, although they
use different approaches to optimise and specify the mixture components. \citet{jerfel2021variational} consider boosting using the forward Kullback Leibler (KL) divergence and combine variational inference and importance sampling. 

Our second contribution is to develop 
a fast and accurate variational boosting
method by efficiently adding a single mixture component at a time. We propose a global boosting step and two local boosting steps. In the global boosting variational method, only the new mixture weights and Gaussian component parameters are optimised, and the previous components are held fixed. Because only a lower-dimensional subset of the full variational
parameters is being optimised for the new approximation, the optimisation is often easier. Our two local boosting steps start by initialising a new mixture component with the same parameters as an existing component and then optimising over only a subset of the parameters in the new component. The first local variational boosting step only updates the variational parameters that appear in the marginal distributions of the global parameters. The other variational parameters are kept fixed. The second local variational boosting step updates the variational parameters that appear in the conditional distribution of some of the latent variables that are poorly approximated by the current approximation. Novel methods to identify which latent variables are poorly approximated are proposed. Our variational boosting method is combined with the natural gradient \citep{Amira:1998} and variance reduction methods based on the control variates of \citet{Ranganath:2014} and the reparameterisation trick of \citet{Kingma2014}.  
Using natural gradients enables faster convergence than the traditional gradient-based methods because they exploit the
information geometry of the variational approximation. 
Our third contribution is to provide novel methods to initialise the variational parameters of the additional components in the mixtures for global and local boosting steps. 

The rest of the article is organised as follows.
Section~\ref{models}
discusses the latent variable models; section \ref{sec:gaussianmixturevb} discusses the mixture of Gaussians variational approximation and its properties;  section \ref{sec:optimizing} discusses
the variational boosting optimisation algorithm; section \ref{boostingstatespace}
discusses extensions of boosting variational inference for state space models; section~\ref{sec:example}
presents results from both simulated and real datasets; section \ref{sec:conclusion}  concludes
with a discussion of our approach and results. This article has an online supplement containing additional technical details and empirical results.

\section{Latent Variable Models\label{models}}

Suppose we have data $\yvec=(\yvec_1^\top,\dots, \yvec_n^\top)^\top$, where
$\yvec_i\in \mathbb{R}^{n_i}$.  We consider latent variable models for
the data where there are global parameters denoted $\thetavec_G$, and
local parameters or latent variables denoted $\thetavec_L=(\bvec_1^\top,\dots, \bvec_n^\top)^\top$.  
The local parameter $\bvec_i$ appears only in the likelihood for $\yvec_i$, 
$i=1,\dots, n$.  Writing $\thetavec=(\thetavec_G^\top,\thetavec_L^\top)^\top$, the likelihood is 
$p(\yvec|\thetavec) = \prod_{i=1}^n p(\yvec_i |\bvec_i,\thetavec_G),$
and the prior density for $\thetavec$ is $p(\thetavec) = p(\thetavec_G)\prod_{i=1}^n p(\bvec_i|\thetavec_G).$
The prior is generalised later for time series state space models in section \ref{boostingstatespace}.  The posterior density of $\thetavec$ is
\begin{align}
  p(\thetavec|\yvec) & \propto p(\thetavec)p(\yvec|\thetavec)=
  p(\thetavec_G)\prod_{i=1}^n p(\bvec_i|\thetavec_G)p(\yvec_i|\bvec_i,\thetavec_G)\label{posterior} \\ 
  \intertext{so that}
p(\thetavec|\yvec)  & =p(\thetavec_G|\yvec)\prod_{i=1}^n p(\bvec_i|\thetavec_G,\yvec_i).\label{posterior2}
\end{align}
\eqref{posterior2} shows that the local parameters are conditionally
independent in the posterior given $\thetavec_G$.  

%can write the
%posterior density \eqref{posterior} as
%\begin{align}
%  p(\thetavec|\yvec)=p(\thetavec_G|\yvec)\prod_{i=1}^n p(\bvec_i|\thetavec_G,\yvec_i). \label{posterior2}
%\end{align}
%\begin{align}
%  \text{KL}(q_\lambdavec(\thetavec)||p(\thetavec|\yvec)) & = -\int \log \frac{p(\theta)p(\yvec|\thetavec)}{q_\lambda(\thetavec)}q_\lambdavec(\thetavec)\,d\thetavec+\log p(\yvec) \nonumber \\
%  & = -{\cal L}(\lambdavec)+\log p(\yvec),  \label{kld2}
%\end{align} 
%where
%\begin{align*}
%  {\cal L}(\lambdavec) & := \int \log \frac{p(\thetavec)p(\yvec|\thetavec)}{q_\lambdavec(\thetavec)}q_\lambdavec(\thetavec)\,d\thetavec,
%\end{align*}

\section{Gaussian mixture variational approximation\label{sec:gaussianmixturevb}}

Variational inference approximates a Bayesian posterior distribution
by solving an optimisation problem.  Similar to the previous section, 
consider a model for data $\yvec$ having parameter $\thetavec$ with  
density $p(\yvec|\thetavec)$, and   
Bayesian inference with prior density $p(\thetavec)$ 
and posterior density $p(\thetavec|\yvec)\propto p(\thetavec)p(\yvec|\thetavec)$.  
In variational inference, we optimise an approximation 
$q_\lambdavec(\thetavec)$ with
respect to variational parameters $\lambdavec\in \Lambda$ to match
$p(\thetavec|\yvec)$ as closely as possible. The variational inference is discussed in more detail in section \ref{sec:variational} of the online supplement.

This work uses 
 Gaussian mixture densities to flexibly approximate the
 posterior distribution in variational inference
 (e.g. \cite{salimans2013fixed,Lin2019,arenz2023a}, among many others).
 A Gaussian mixture model with a large number of components is very flexible, 
but optimising all the variational parameters simultaneously would lead to a high-dimensional optimisation problem.  Variational boosting algorithms
learn Gaussian mixture variational approximations 
by adding mixture components sequentially to an existing
approximation, resulting in a sequence of lower-dimensional optimisation
problems which are easier to perform \citep{Guo2016,Miller2016,locatello+kgr18,campbell2019universal,jerfel2021variational,gunawan2024flexible}.  
We write a 
$K$ component Gaussian mixture approximating $p(\thetavec|\yvec)$ as
\begin{align}
q_{\lambdavec[K]}(\thetavec) = \sum_{k=1}^K \pi_k[K] q_{\etavec_k[K]}(\thetavec),\label{vapprox}
\end{align}
where $\lambdavec[K]$ is the set of all variational parameters,  
$\pi_k[K]>0$ is a mixing weight for the $k$th component, 
with $\sum_{k=1}^K \pi_k[K]=1$, and $\etavec_k[K]$ are parameters determining
the mean and covariance matrix of the $k$th Gaussian mixture
component density $q_{\etavec_k[K]}(\thetavec)$, $k=1,\dots, K$.  
We write $\pivec[K]=(\pi_1[K],\dots, \pi_{K-1}[K])^\top$, 
$\etavec[K]=(\etavec_1[K]^\top,\dots, \etavec_K[K]^\top)^\top$, and then $\lambdavec[K]=(\pivec[K]^\top,\etavec[K]^\top)^\top$.  

\begin{comment}
\begin{align}
\Omega_k[K] &  = \left[\begin{array}{ccccc}
\Omega_{k1}[K] & 0 & \ldots & 0 & \Omega_{k1G}[K]  \cr
0 & \Omega_{k2}[K] & \ldots & 0 & \Omega_{k2G}[K] \cr
\vdots & \vdots & \ddots & \vdots & \vdots \cr
0 &  0 & \ldots & \Omega_{kn}[K] & \Omega_{knG}[K] \cr
\Omega_{kG1}[K] & \Omega_{kG2}[K] & \ldots & \Omega_{kGn}[K] & \Omega_{kG}[K] 
\end{array}\right].  \label{precision}
\end{align}
\end{comment}

We parameterise the Gaussian mixture components in such a way
that each component has the same conditional independence structure as
$p(\thetavec|\yvec)$.  To achieve this, we consider a parameterisation
of the covariance matrix through the Cholesky factor of the corresponding
precision matrix.  Write
$q_{\etavec_k[K]}(\thetavec)=N(\thetavec;\muvec_k[K],\Omega_k[K]^{-1}),$
where $\muvec_k[K]$ is the mean vector and $\Omega_k[K]=L_k[K]L_k[K]^\top$ 
is the precision matrix with lower-triangular Cholesky factor $L_k[K]$.  
With $\thetavec=(\bvec_1^\top,\dots, \bvec_n^\top,\thetavec_G^\top)^\top$, 
partitioning $\muvec_k$ conformably with this partitioning of $\thetavec$,  
\begin{align}
  \muvec_k[K] & = (\muvec_{k1}[K]^\top,\dots, \muvec_{kn}[K]^\top,\muvec_{kG}[K]^\top)^\top.  \label{muk}
\end{align}
It is useful later to write $\muvec_{kb}[K]=(\muvec_{k1}[K]^\top,\dots, \muvec_{kn}[K]^\top)^\top$.  
If $\thetavec$ is a Gaussian random vector, then to have
$\bvec_i,\bvec_j$, $i\neq j$ conditionally independent given $\thetavec_G$, 
$\Omega_k[K]$ should have the block structure (and partitioning 
$\Omega_k[K]$ conformably with the partitioned definition of $\thetavec$) given in \eqref{precision} in section \ref{sec:addlocalboostingstate} of the online supplement. 
This follows from the well-known result that, for a multivariate normal
random vector $\mZ$, the components 
$\mZ_i$ and $\mZ_j$, $i\neq j$, are conditionally
independent given the remaining elements of $\mZ$ if and only if 
the $(i,j)$th element of the precision matrix is zero.  
For a precision matrix of the form \eqref{precision}, 
the Cholesky factor $L_k[K]$ has the block structured form
(\cite{rothman+lz10}, Proposition 1)
\begin{align}
L_k[K] &  = \left[\begin{array}{ccccc}
L_{k1}[K] & 0 & \ldots & 0 & 0 \cr
0 & L_{k2}[K] & \ldots & 0 & 0 \cr
 \vdots & \vdots & \ddots & \vdots & \vdots \cr
0 &  0 & \ldots & L_{kn}[K] & 0 \cr
L_{kG1}[K] & L_{kG2}[K] & \ldots & L_{kGn}[K] & L_{kG}[K]
\end{array}\right]. \label{Lk}
\end{align}
Here $L_{k1}[K],\dots, L_{kn}[K]$ and $L_{kG}[K]$ are lower-triangular, with
positive diagonal elements.  Write $L'_{kG}[K]$ for the matrix $L_{kG}[K]$
but where the diagonal elements are log-transformed.   It is convenient to work with $L'_{kG}[K]$ rather than $L_{kG}[K]$ as variational parameters because
then the optimisation is unconstrained.  Define $L'_{kj}[K]$
from $L_{kj}[K]$ similarly, for $j=1,\dots, n$.  
For a square matrix $A$, write $\text{vech}(A)$ for the 
half-vectorisation of $A$, 
which stacks the elements on or below the diagonal of $A$ 
column by column into a vector from left to right, and $\text{vec}(A)$ for the vectorisation of 
$A$, which stacks all the entries column by column into a vector.  
Also write $\ellvec_{kb}[K]=(\text{vech}(L'_{k1}[K])^\top,\dots, \text{vech}(L'_{kn}[K])^\top)^\top$, 
$\ellvec_{kGb}=(\text{vec}(L_{kG1}[K])^\top ,\dots, \text{vec}(L_{kGn}[K])^\top)^\top$,
$\ellvec_{kG}=\text{vech}(L'_{kG}[K]))^\top$ and $\ellvec_k=(\ellvec_{kb}[K]^\top,\ellvec_{kGb}[K]^\top,\ellvec_{kG}[K]^\top)^\top$, so that
$\etavec_k[K]=(\muvec_k[K]^\top,\ellvec_k[K]^\top)^\top$.  \\
It is helpful to state the following lemma, which is used later.
\begin{lemma}\label{lemma1}
Suppose that $\thetavec=(\bvec_1^\top,\dots, \bvec_n^\top,\thetavec_G^\top)^\top$ is a Gaussian random vector with mean 
vector $\muvec_k[K]$ given by \eqref{muk} and precision matrix $\Omega_k[K]=L_k[K]L_k[K]^\top$,  where $L_k[K]$ is
given by \eqref{Lk}.  Then, 
\begin{enumerate}
\item[i)] $\mathrm{E}(\thetavec_G)=\muvec_{kG}[K]$ and $\mathrm{Cov}(\thetavec_G)=L_{kG}[K]^{-\top}L_{kG}[K]^{-1}$.  
\item[ii)] $\mathrm{E}(\bvec_i|\thetavec_G)=\muvec_{ki}[K]-L_{ki}[K]^{-\top}L_{kGi}[K]^\top(\thetavec_G-\muvec_{kG}[K])$ and $\mathrm{Cov}(\bvec_i|\thetavec_G)=L_{ki}[K]^{-\top}L_{ki}[K]^{-1}$
\end{enumerate}
\end{lemma} 

The proof is given in section \ref{sec:proofs} of the online supplement.  With a variational approximation having Gaussian mixture components of the
form considered in Lemma \ref{lemma1}, we now suggest localised boosting steps where most parameters
are shared with an existing component when a new mixture component is added, 
lowering the dimension of the optimisation required at each step.  
The localised boosting
iterations are able to focus on only improving certain components in a
decomposition of the posterior following its conditional independence structure
in \eqref{posterior2}.

\section{Optimising the approximation with variational boosting}\label{sec:optimizing}

Variational boosting learns a Gaussian mixture approximation 
to the posterior density iteratively.  
Suppose we have an existing $K$ component approximation and   
we wish to extend it to a more accurate 
one with $K+1$ components, where the variational parameters
for the first $K$ components are held fixed.  Because 
only a low-dimensional subset of the full variational parameters
is being optimised for the new approximation, the optimisation is easier.
The component additions in the algorithm are commonly 
referred to as ``variational boosting" steps. The standard variational boosting approach is discussed in section \ref{sec:standardvariationalboosting} of the online supplement.

To define the boosting steps considered here, 
extra notation is necessary.  
Consider row vectors $\cvec_i$, $i=1,\dots, n$.  For $\mathcal{I}\subseteq \{1,\dots, n\}$ concatenate the row vectors $\cvec_i$, $i\in \mathcal{I}$ into a single row vector by
$[\cvec_i]_{i\in \mathcal{I}}$.  
For $\ellvec_{kb}[K]$ defined in Section \ref{sec:gaussianmixturevb}, we
write
$\ellvec_{kb\mathcal{I}}[K]:=[\mathrm{vech}(L'_{ki}[K])^\top]_{i\in \mathcal{I}}^\top$
and $\ellvec_{kb\overline{\mathcal{I}}}[K]:=[\mathrm{vech}(L'_{ki}[K])^\top]_{i\notin \mathcal{I}}^\top$.  
$\ellvec_{kb\mathcal{I}}[K]$ are the variational parameters in $\ellvec_{kb}[K]$
that appearing in the conditional posterior of $\bvec_i|\theta_G$ for the $k$th
component, $i\in \mathcal{I}$, and $\ellvec_{kb\overline{\mathcal{I}}}[K]$ are
the remaining components of $\ellvec_{kb}[K]$.  
Similarly, write $\ellvec_{kGb\mathcal{I}}[K]:=[\mathrm{vec}(L_{kGi})^\top]_{i\in \mathcal{I}}^\top$ and $\ellvec_{kGb\overline{\mathcal{I}}}:=[\mathrm{vec}(L_{kGi})^\top]_{i\notin \mathcal{I}}^\top$.  Here
$\ellvec_{kGb\mathcal{I}}[K]$ are the variational parameters in $\ellvec_{kGb}[K]$
that appear in the conditional posterior of $\bvec_i|\thetavec_G$ for the $k$th
component, $i\in \mathcal{I}$, and $\ellvec_{kGb\overline{\mathcal{I}}}[K]$ are
the remaining components of $\ellvec_{kGb}[K]$. 

We consider two different types of boosting steps, 
and it is useful to state
the following lemma, which relates to the form of the marginal 
density of $\thetavec_G$ and the conditional distributions of $\bvec_i|\thetavec_G$
for a mixture approximation of the form \eqref{vapprox}.  
This lemma helps us to understand the effect of the local boosting steps
on the approximation of the components of the decomposition
of the posterior distribution given in \eqref{posterior2}.
The proof
is a straightforward application of known results for Gaussian mixture models
and is omitted.
\begin{lemma}
Suppose that $\thetavec=(\bvec_1^\top,\dots, \bvec_n^\top,\thetavec_G^\top)^\top$ has density \eqref{vapprox}.   The $k$th
Gaussian component density can be written as
$$q_{\etavec_k[K]}(\thetavec)=q_{\etavec_k[K]}(\thetavec_G)
\prod_{i=1}^n q_{\eta_k[K]}(\bvec_i|\thetavec_G),$$
where $q_{\eta_k[K]}(\thetavec_G)$ is Gaussian with
mean and covariance matrix given in Lemma \ref{lemma1} i), 
and $q_{\eta_k[K]}(\bvec_i|\thetavec_G)$
is Gaussian with mean and covariance matrix given in Lemma \ref{lemma1} ii).
Then, 
\begin{enumerate}
\item[i.] $q_{\lambdavec[K]}(\thetavec_G)=\sum_{k=1}^K \pi_k[K] q_{\eta_k[K]}(\thetavec_G)$.
\item[ii.] $q_{\lambdavec[K]}(\bvec_i|\thetavec_G)=\sum_{k=1}^K w_k[K](\thetavec_G) q_{\eta_k[K]}(\bvec_i|\thetavec_G)$, where
$$w_k[K](\thetavec_G)=\frac{\pi_k[K] q_{\eta_k[K]}(\thetavec_G)}
{\sum_{k'=1}^K \pi_{k'}[K] q_{\eta_{k'}[K]}(\thetavec_G)}.$$
\end{enumerate}
\end{lemma}

%drawing $c$ randomly according to the mixing 
%weights $\pi_k[K]$, $k=1,\dots, K$.
  
For all the defined boosting steps, we first choose a 
mixture component $c\in \{1,\dots, K\}$ in our existing $K$ component
approximation with the highest mixing weight.
To simplify notation, we will assume that the mixture components
in $q_{\lambdavec[K]}(\thetavec)$ are relabelled so that $c=K$.  Then
we construct a $K+1$ component mixture component approximation
of the form
\begin{align*}
  q_{\lambdavec[K+1]}(\thetavec) & = \sum_{k=1}^{K+1} \pi_k[K+1] q_{\etavec_{k}[K+1]}(\thetavec). 
\end{align*}
where the parameters in $\lambdavec[K+1]$ are related to those
in $\lambdavec[K]$.  We impose the restriction
$$\pi_k[K+1]=\pi_k[K],\, k=1\dots, K-1,\;\;\text{and}\;\;\etavec_k[K+1]=\etavec_k[K],\, k=1\dots, K,$$
so that the mixing weights are kept unchanged for the first $K-1$
components, and
the mixture component parameters are kept unchanged for all
the components of the existing approximation.  
We also assume that 
$$\pi_K[K+1]=\pi\pi_K[K],\;\;\;\pi_{K+1}[K+1]=(1-\pi) \pi_K[K],$$
where $\pi$ is a variational parameter that needs to be optimised.  
Since $\pi_K[K+1]+\pi_{K+1}[K+1]=\pi_K[K]$, we are ``splitting"
the $K$th component into two, keeping the parameters in the first $K$
component densities fixed.
The way that $\etavec_{K+1}[K+1]$ is constructed depends on the particular
type of boosting step considered, and this is discussed next.  

\subsection{Local and global boosting steps}

We define two types of local boosting (LB) steps, which we refer to as type I and 
type II, as well as a global boosting (GB) step.    
For a type I LB step, we set 
$\muvec_{K+1\,b}[K+1]=\muvec_{Kb}[K]$,
$\ellvec_{K+1\,Gb}[K+1]=\ellvec_{KGb}[K]$ and 
$\ellvec_{K+1\,b}[K+1]=\ellvec_{Kb}[K]$.
The remaining variational parameters to be optimised are $\pi$, $\muvec_{K+1\,G}[K+1]$ and $\ellvec_{K+1\,G}[K+1]$.  
Lemma \ref{lemma1} $i)$ shows that we 
are optimising only the variational parameters that 
appear in the marginal distribution of $\thetavec_G$ for the new component, 
$q_{\etavec_{K+1}[K+1]}(\thetavec_G)$.  

For a type II LB step, we choose a subset 
$\mathcal{I}\subseteq \{1,\dots, n\}$ 
and set
$$\muvec_{K+1\,G}[K+1]=\mu_{K\,G}[K],\;\;\ellvec_{K+1\,G}[K+1]=\ellvec_{KG}[K],
\;\;\muvec_{K+1\,b\overline{\mathcal{I}}}[K+1]=\muvec_{Kb\overline{\mathcal{I}}}[K],$$
$$\ellvec_{K+1\,b\overline{\mathcal{I}}}[K+1]=\ellvec_{Kb\overline{\mathcal{I}}}[K],
\;\;\ellvec_{K+1\,Gb\overline{\mathcal{I}}}[K+1]=\ellvec_{KGb\overline{\mathcal{I}}}[K].$$
A new measure to choose a subset $\mathcal{I}$ is discussed in section \ref{subsec:assessqualityposterior}. 
Then, the remaining variational parameters to be optimised consist of
$\pi$, $\muvec_{K+1\,b\mathcal{I}}[K+1]$, $\ellvec_{K+1\,b\mathcal{I}}[K+1]$
and $\ellvec_{K+1\,Gb\mathcal{I}}[K+1]$.  From Lemma \ref{lemma1} $ii)$, we optimise
the variational parameters appearing in the conditional densities
$q_{\etavec_{K+1}[K+1]}(\bvec_i|\thetavec_G)$, $i\in \mathcal{I}$, with
the exception of the ``global" variational parameter $\muvec_{K+1\,G}[K+1]$.  
Finally for a GB step, we optimise $\pi$ and all components of $\boldsymbol{\mu}_{K+1}[K+1]$, $\ellvec_{K+1}[K+1]$.  Alternatively, instead of optimising only $\pi$, we could
optimise all the existing mixing weights.
The following lemma motivates the type II LB by showing that the step only changes the approximating 
conditional posterior densities for $\bvec_i|\thetavec_G$ for $i\in \mathcal{I}$.  Hence, such
a boosting step can be used to target an improvement in
the approximation of $p(\bvec_i|\thetavec_G,\yvec_i)$ by 
$q_{\lambda[K+1]}(\bvec_i|\thetavec_G)$ for a subset 
of the local parameters
$\bvec_i$, $i\in \mathcal{I}$.

\begin{lemma}\label{lemma3} 
For a type II LB step, 
$q_{\lambdavec[K+1]}(\thetavec_G) = q_{\lambdavec[K]}(\thetavec_G),$
and for $i\notin \mathcal{I}$
$q_{\lambdavec[K+1]}(\bvec_i|\thetavec_G)=q_{\lambdavec[K]}(\bvec_i|\thetavec_G).$
\end{lemma}
The proof is in 
section ~\ref{sec:proofs} of the online supplement.  The LB type I step doesn't have such a clear interpretation as the LB type II step
in terms of only influencing certain components in the decomposition of 
the posterior density \eqref{posterior2} because it changes the marginal posterior approximation
for the ``global" parameters $\thetavec_G$, and 
also affects the conditional densities
$q_{\lambda[K+1]}(\bvec_i|\thetavec_G)$, $i=1,\dots, n$. 
Also, even though
the Gaussian mixture components have the conditional independence
structure of the true posterior density, the mixture density does not, although
it will approximately have the correct 
structure if the mixture
components in the marginal density $q_{\lambdavec[K]}(\thetavec_G)$
are well-separated.
We next discuss the algorithm for updating variational parameters for 
performing the boosting optimisation steps.  This is followed by a discussion
of how to choose the subsets $\mathcal{I}$ in an LB type II step.

%\subsection{Sample average approximation}

\subsection{Updating the Variational Parameters\label{subsec:Updating-the-Variational}}

To optimise the free variational parameters in each VB boosting step, we 
use a stochastic gradient ascent algorithm; see section \ref{sec:variational} of the online supplement for an introduction.  Suppose we are at step
$K+1$ of variational boosting, and we wish to add a new component
to an existing $K > 1$ component approximation;  the degenerate cases of learning a single component or
splitting a single component  
($K=1$) are not discussed explicitly, as the discussion below simplifies
in an obvious way in these settings.  
The current variational approximation, 
with $K$ components, is 
$\sum_{k=1}^K \pi_k[K]q_{\etavec_k[K]}(\thetavec),$
for any of the boosting steps, and the new approximation takes the form
$\sum_{k=1}^{K+1} \pi_k[K+1]q_{\etavec_k[K+1]}(\thetavec).$
%and writing
%$$\pi_{<K}[K]:=\sum_{k=1}^{K-1}\pi_k[K],$$
%=\pi_{<K}[K] q_{\lambda[K-1]}(\thetavec)
%+\pi \pi_K[K]q_{\etavec_K[K]}(\thetavec)
%+(1-\pi)\pi_K[K] q_{\etavec_{K+1}[K+1]}(\thetavec)

%The free variational parameters in the boosting 
%step are $\pi_1[K+1],...,\pi_K[K+1]$ and some subvector 
%$\zetavec$ of $\etavec_{K+1}[K+1]$.  

%What $\zetavec$ is depends
%on the type of boosting step considered.

We now outline the updating scheme for the variational parameters
of the full component parameters $\left(\muvec_{K+1}{[K+1]},\ellvec_{K+1}{[K+1]}\right)$
and the mixing weights $\pivec[K+1]=(\pi_1[K+1],\dots, \pi_{K}[K+1], \pi_{K+1}[K+1])^\top$, based on natural-gradient
and variance reduction methods for reducing the variance of the unbiased
estimates of the gradient of the lower bound. Then, the updates for the subset of the variational parameters are discussed. \citet{gunawan2024flexible} and \citet{Lin2019} show 
that natural gradients produce faster convergence
than traditional gradient-based methods because they exploit the information geometry of
the variational approximation $q_{\lambdavec[K]}{(\thetavec)}$ to speed-up the convergence
of the optimisation.

%Many natural-gradient
%methods for variational inference are available
%\citep{Hoffman:2013,Khan2017a}; these
%show that natural-gradients produce faster convergence
%than traditional gradient-based methods.

%\footnote{$\textrm{I}_k(\underline{w})$ denotes the indicator function which is 1 if $\underline{w}=k$, and $0$ otherwise}

\citet{Lin2019} derive natural gradient updates for a mixture of Gaussian distributions without explicitly calculating the Fisher Information Matrix. They apply their natural gradient methods for fitting a mixture of Gaussians variational approximation with a full covariance matrix for each component, which ignores the conditional independence structure in the posterior $p(\thetavec|\yvec)$ and makes it less scalable in the number of latent variables. They also do not implement a boosting approach by adding a mixture component at a time, which makes the optimisation much more challenging. In addition, it is difficult to choose good initial values for all variational parameters. We develop a method to choose good initial values for all variational parameters described in section \ref{sec:Initialising-a-New} of the online supplement.

%All the variational parameters are initialised using the approach
%in section \ref{sec:Initialising-a-New} of the online supplement. 

As a sparse Cholesky of the precision matrix is used in each mixture component, we adopt the natural-gradient updates of \citet{Lin2019} only for updating the weights 
$\pivec[K+1]$ and the mixture means $\muvec_{K+1}[{K+1}]$. 
The natural-gradient update for the mixture weights is 
\begin{equation}
\log\left(\frac{\pi_{k}[K+1]}{\pi_{K+1}[K+1]}\right)^{\left(t+1\right)}=\log\left(\frac{\pi_{k}[K+1]}{\pi_{K+1}[K+1]}\right)^{\left(t\right)}+a^{\pi}_{t}\left(\delta_{k}-\delta_{K+1}\right)\left(\log\left(h\left(\boldsymbol{\thetavec}\right)\right)-\log q_{\boldsymbol{\lambda}[K+1]}\left(\boldsymbol{\thetavec}\right)\right),\label{eq:update pi-1}
\end{equation}
for $k=1,...,K$, where $h\left(\thetavec\right) := p\left(\yvec|\thetavec\right)p\left(\thetavec\right)$ and ${a}^{\pi}_{t}$ is a scalar step size; see section \ref{subsec:Learning-Rate} of the online supplement for further details. The natural-gradient update for the new means
$\muvec_{K+1}[K+1]$ is 
\begin{equation}
\boldsymbol{\mu}_{K+1}[K+1]^{\left(t+1\right)}=\boldsymbol{\mu}_{K+1}[K+1]^{\left(t\right)}+{\boldsymbol{a}}^{\mu}_{t}\circ \left(\delta_{K+1}\Omega^{-1}\left(\nabla_{\boldsymbol{\theta}}\log\left(h\left(\boldsymbol{\theta}\right)\right)-\nabla_{\boldsymbol{\theta}}\log q_{\boldsymbol{\lambdavec}[K+1]}\left(\boldsymbol{\thetavec}\right)\right)\right),\label{eq:update mu-1}
\end{equation}
where\footnote{$\circ$ denotes the Hadamard (element by element) product of
two random vectors} $\delta_{k}=\frac{N\!\left(\boldsymbol{\thetavec}\mid\boldsymbol{\mu}_{k}[K+1],L_{k}[K+1]^{-\top}L_{k}[K+1]^{-1}\right)}{\delta_{\mathrm{tot}}}$,
$\delta_{K+1}=\frac{N\!\left(\boldsymbol{\thetavec}\mid\boldsymbol{\mu}_{K+1}[K+1],L_{K+1}[K+1]^{-\top}L_{K+1}[K+1]^{-1}\right)}{\delta_{\mathrm{tot}}}$,
$\delta_{tot}=\sum_{k=1}^{K+1}\pi_{k}[K+1]N\left(\boldsymbol{\theta}|\boldsymbol{\mu}_{k}[K+1],L_{k}[K+1]^{-\top}L_{k}[K+1]^{-1}\right),
$
and ${\boldsymbol{a}}^{\mu}_{t}$ is a vector of step sizes. We now discuss updating the variational parameters $\ellvec_{K+1}[K+1]$, which are the non-zero elements in the $L_{K+1}[K+1]$. 
We use gradient estimates based on the so-called reparameterisation trick \citep{Rezende2014,Kingma2014}, which is known to reduce variance. 
However, applying the reparameterisation trick for the mixture type variational approximation is difficult because the discrete random variable, which indicates the mixture component, complicates the differentiation procedure. However, following  \citet{Miller2016}, when $q_{\boldsymbol{\lambdavec}[K+1]}\left({\thetavec}\right)$ is
a mixture of $K+1$ components, the lower bound $\mathcal{L}^{(K+1)}\left(\boldsymbol{\lambda}[K+1]\right)$
is 
\begin{align}
\mathcal{L}^{(K+1)}\left(\boldsymbol{\lambda}[K+1]\right) = \sum_{k=1}^{K+1}\pi_{k}\left[K+1\right]E_{q_{\boldsymbol{\eta}_{k}\left[K+1\right]}}(\log h\left(\boldsymbol{\theta}\right)-\log q_{\boldsymbol{\lambda}[K+1]}\left(\boldsymbol{\theta}\right)),
\end{align}

\begin{comment}
\begin{eqnarray*}
\mathcal{L}^{(K+1)}\left(\boldsymbol{\lambda}[K+1]\right) & = & E_{q_{\boldsymbol{\lambdavec}[K+1]}}\left(\log h\left(\boldsymbol{\theta}\right)-\log q_{\boldsymbol{\lambda}[K+1]}\left(\boldsymbol{\theta}\right)\right),\\
 & = & \int\left(\sum_{k=1}^{K+1}\pi_{k}\left[K+1\right]q_{\boldsymbol{\eta}_{k}\left[K+1\right]}\left(\boldsymbol{\theta}\right)\right)\left(\log h\left(\boldsymbol{\theta}\right)-\log q_{\boldsymbol{\lambda}[K+1]}\left(\boldsymbol{\theta}\right)\right)d\boldsymbol{\theta},\\
 & = & \sum_{k=1}^{K+1}\pi_{k}\left[K+1\right]\int q_{\boldsymbol{\eta}_{k}\left[K+1\right]}\left(\boldsymbol{\theta}\right)\left(\log h\left(\boldsymbol{\theta}\right)-\log q_{\boldsymbol{\lambda}[K+1]}\left(\boldsymbol{\theta}\right)\right)d\boldsymbol{\theta},\\
 & = & \sum_{k=1}^{K+1}\pi_{k}\left[K+1\right]E_{q_{\boldsymbol{\eta}_{k}\left[K+1\right]}}(\left(\log h\left(\boldsymbol{\theta}\right)-\log q_{\boldsymbol{\lambda}[K+1]}\left(\boldsymbol{\theta}\right)\right),
\end{eqnarray*}
\end{comment}

\noindent which is a weighted sum of component-specific lower bounds. To apply
the reparameterisation trick to update the variational parameter $\ellvec_{K+1}[K+1]$,
we represent samples from $q_{\boldsymbol{\eta}_{K+1}\left[K+1\right]}\left(\boldsymbol{\theta}\right)$
as $\boldsymbol{\theta}=u\left(\boldsymbol{\epsilon}_{K+1},\boldsymbol{\eta}_{K+1}\left[K+1\right]\right)$,
where $\boldsymbol{\epsilon}_{K+1}$ is a random vector with a density
$p_{\boldsymbol{\epsilon}_{K+1}}\left(\boldsymbol{\epsilon}_{K+1}\right)$.
In the case of the Gaussian component variational distribution parameterised
in terms of a mean vector $\boldsymbol{\mu}_{K+1}\left[K+1\right]$
and the Cholesky factor $L_{K+1}\left[K+1\right]$ of the precision
matrix, we can write $\boldsymbol{\theta}=\boldsymbol{\mu}_{K+1}\left[K+1\right]+L_{K+1}\left[K+1\right]^{-1}\boldsymbol{\epsilon}_{K+1}$,
where $\boldsymbol{\epsilon}_{K+1}\sim N\left(0,I\right)$. Differentiating
under the integral sign, 
\begin{multline}
\nabla_{L_{K+1}[K+1]}\mathcal{L}^{(K+1)}\left(\boldsymbol{\lambda}[K+1]\right)=\pi_{K+1}\left[K+1\right]E_{p_{\boldsymbol{\epsilon}_{K+1}}}\left(\left[\nabla_{{L}_{K+1}}u\left(\boldsymbol{\epsilon}_{K+1},\boldsymbol{\eta}_{K+1}\left[K+1\right]\right)\right.\right.\\
\left.\left.\nabla_{\theta}\log h\left(u\left(\boldsymbol{\epsilon}_{K+1},\boldsymbol{\eta}_{K+1}\left[K+1\right]\right)\right)-\nabla_{\theta}\log q_{\boldsymbol{\lambda}[K+1]}\left(u\left(\boldsymbol{\epsilon}_{K+1},\boldsymbol{\eta}_{K+1}\left[K+1\right]\right)\right)\right]\right);\label{eq:gradLB1}
\end{multline}
this is an expectation with respect to $p_{\boldsymbol{\epsilon}_{K+1}}$
that can be estimated unbiasedly with one or more random draws from
$p_{\boldsymbol{\epsilon}_{K+1}}$. The alternative expression for
the gradient of the lower bound with respect to $L_{K+1}\left[K+1\right]$
is 
\begin{multline}\label{gradientK+1}
\nabla_{L_{K+1}[K+1]}\mathcal{L}^{(K+1)}\left(\boldsymbol{\lambda}[K+1]\right)=\pi_{K+1}\left[K+1\right]E_{p_{\boldsymbol{\epsilon}_{K+1}}}\left( -L_{K+1}[K+1]^{-\top}\right.\\
\left.\boldsymbol{\epsilon}_{K+1}\left[\nabla_{\thetavec}\log h\left(\thetavec\right)-\nabla_{\thetavec}\log q_{\boldsymbol{\lambda}[K+1]}\left(\thetavec\right)\right]^{\top}L_{K+1}[K+1]^{-\top}\right).
\end{multline}

\begin{algorithm}[t]
\caption{Variational Bayes Algorithm \label{alg:Variational-Algorithm}}

%\begin{enumerate}
%\item (a) Initialize $\lambdavec{[K+1]}^{\left(0\right)}=\left(\muvec_{[K+1]}^{\top \left(0\right)},
%    (\ellvec_{[K+1]}^{\left(0\right)})^{\top},
%    \pivec[K+1]^{\left(0\right)}\right)$,
%set $t=0$, and generate $\thetavec_{s}^{\left(t\right)}\sim q_{\lambdavec[{\left(K+1\right)}]}\left(\thetavec\right)$ for $s=1,...,S$. Let $m_{L}$
%be the number of elements in $\ellvec_{K+1}$.

%(b) Evaluate the control variates $\varsigma_{\ellvec{[K+1]}}^{\left(t\right)}=\left(\varsigma_{1,\ellvec{[K+1]}}^{\left(t\right)},...,\varsigma_{m_{L},\ellvec{[K+1]}}^{\left(t\right)}\right)^{'}$,
%with
%\begin{equation}
%\varsigma_{i,L{[K+1]}}^{\left(t\right)}=\frac{\wh {\textrm{Cov}}\left(\left[\log\left(h\left(\thetavec\right)\right)-\log q_{\lambdavec[K+1]}\left(\thetavec\right)\right]\nabla_{\lambda_{i,\ellvec{[K+1]}}}\log q_{\lambdavec[K+1]}\left(\thetavec\right),\nabla_{\lambdavec_{i,\ellvec{[K+1]}}}\log q_{\lambdavec[K+1]}\left(\thetavec\right)\right)}{\wh\V\left(\nabla_{\lambdavec_{i,\ellvec{[K+1]}}}\log q_{\lambdavec[K+1]}\left(\thetavec\right)\right)},\label{eq:cv}
%\end{equation}
%for $i=1,...,m_{L}$, where $\wh {\textrm{Cov}}$ and $\wh\V\left(\cdot\right)$ are the
%sample estimates of covariance and variance based on $S$ samples
%from step (1a).
%\end{enumerate}
Repeat until the stopping rule is satisfied
\begin{itemize}
\item Update $\ellvec_{K+1}{[K+1]}$:
\end{itemize}
\begin{enumerate}
\item Generate $\boldsymbol{\epsilon}^{(t)}_{K+1}$ from $N(0,I)$.
\item Compute the gradients of $\ellvec_{K+1}[K+1]$, the non-zero elements in $L_{K+1}[K+1]$ using \eqref{gradientK+1}. 
\item Compute $\triangle \ellvec_{K+1}{[K+1]}$
using ADAM as described in section \ref{subsec:Learning-Rate} of the online supplement. Then, set
$\ellvec_{K+1}{[K+1]}^{\left(t+1\right)}=\ellvec_{K+1}{[K+1]}^{\left(t\right)}+\triangle \ellvec_{K+1}{[K+1]}^{(t)}$.

%$\widehat{\nabla_{\ellvec{[K+1]}}\mathcal{L}\left(\lambdavec[K+1]^{\left(t\right)}\right)}=\left(g_{1,\ellvec{[K+1]}}^{\left(t\right)},...,g_{m_{L},\ellvec{[K+1]}}^{\left(t\right)}\right)$,
%with
%\begin{equation}
%g_{i,\ellvec{[K+1]}}^{\left(t\right)}=\frac{1}{S}\sum_{s=1}^{S}\left[\log\left(h\left(\thetavec_{s}^{\left(t\right)}\right)\right)-\log q_{\lambdavec[K+1]}\left(\thetavec_{s}^{\left(t\right)}\right)-\varsigma_{i,\ellvec{[K+1]}}^{\left(t-1\right)}\right]\nabla_{\lambdavec_{i,\ellvec_{[K+1]}}}\log q_{\lambdavec[K+1]}\left(\thetavec^{(t)}_{s}\right)\label{eq:gradLB_beta}
%\end{equation}

%\item The gradient of lower bound $\widehat{\nabla_{L_{K+1}}\mathcal{L}\left(\lambdavec^{\left(t\right)}\right)}=\left(g_{1,L_{K+1}}^{\left(t\right)},...,g_{m_{L},L_{K+1}}^{\left(t\right)}\right)$
%can be computed similarly as in Eq. \eqref{eq:gradLB_beta}.

%\item Compute the control variate $\varsigma_{\textrm{vech}\left(L_{K+1}\right)}^{\left(t\right)}$
%as in Eq. \eqref{eq:cv}.
%\item Compute $\triangle \ellvec{[K+1]}$
%using ADAM as described in the online supplement. Then, set
%$\ellvec{[K+1]}^{\left(t+1\right)}=\ellvec{[K+1]}^{\left(t\right)}+\triangle \ellvec{[K+1]}$.
\end{enumerate}
\begin{itemize}
\item Update $\muvec_{K+1}[K+1]$ and $\pivec[K+1]$
\end{itemize}
\begin{enumerate}
\item Generate $\thetavec_{s}^{\left(t\right)}\sim q_{\lambdavec[K+1]}\left(\thetavec\right)$ for $s=1,...,S$.
\item Use \eqref{eq:update pi-1} to update $\pivec[K+1]^{(t+1)}$ and \eqref{eq:update mu-1}
to update $\muvec_{K+1}{[K+1]}^{(t+1)}$, respectively. Set $t=t+1$
\end{enumerate}
\end{algorithm}
\noindent We show in section \ref{subsec:time-varyinglogistic} of the online supplement that updating $\ellvec_{K+1}[K+1]$ with the reparameterisation trick approach is better than updating $\ellvec_{K+1}[K+1]$ with control variate approach of \citet{Ranganath:2014}. Algorithm \ref{alg:Variational-Algorithm} presents the full variational Bayes algorithm. The Cholesky factor $L_{K+1}[K+1]$ is sparse, which means some elements of the matrix $L_{K+1}[K+1]$ are fixed at zero. Only the subset of elements of $L_{K+1}[K+1]$, which correspond to $\ellvec_{K+1}[K+1]$, not fixed at zero, is stored and updated. Algorithm \ref{alg:Variational-Algorithm} can be used for implementing the type I and II LB steps. We multiply the $\triangle \ellvec_{K+1}{[K+1]}$ and $\delta_{K+1}\left(\nabla_{\boldsymbol{\theta}}\log\left(h\left(\boldsymbol{\theta}\right)\right)-\nabla_{\boldsymbol{\theta}}\log q_{\boldsymbol{\lambda}[K+1]}\left(\boldsymbol{\theta}\right)\right)$ with the indicator variables that are equal to 1 for the corresponding elements of $\ellvec_{K+1}{[K+1]}$ and $\muvec_{K+1}{[K+1]}$, which correspond to ($\muvec_{K+1\,G}[K+1]$ and $\ellvec_{K+1\,G}[K+1]$) for type I LB step and
which belong to $\mathcal{I}$ ($\muvec_{K+1\,b\mathcal{I}}[K+1]$, $\ellvec_{K+1\,b\mathcal{I}}[K+1]$
and $\ellvec_{K+1\,Gb\mathcal{I}}[K+1]$) for type II LB step. In our boosting algorithm, we select the type I boosting approach with probability $0.1$, and select the type II boosting approach with probability $0.9$, or we use both the types I and II local boosting approaches in each iteration. 
To draw $S$ samples from the mixture of Gaussians variational approximation, we generate the indicator variables,
$G_{s}$, for $s=1,...,S$; each $G_s$
selects the
component of the mixture from which the sample is to be drawn, with
$G_{s}=k$ with probability $\pi_{k}[K+1]$. Then,
$\boldsymbol{\epsilon}_{G_{s},s}\sim N\left(0,I\right)$ are generated,
where $\boldsymbol{\epsilon}_{G_{s},s}$
is $n+m_G$-dimensional, with $m_G$ is the number of global parameters; $\thetavec_{s}=\muvec_{G_{s}}[K+1]+L_{G_{s}}[K+1]^{-\top}\circ\boldsymbol{\epsilon}_{G_{s},s}$ are then
calculated for $s=1,..,S$.

%Similar approach to Section \ref{subsec:Initialising-a-New} is used.

\subsection{Assessing the quality of current posterior approximation for local parameters\label{subsec:assessqualityposterior}}

This section describes how to find a suitable set $\mathcal{I}\subseteq \{1,\dots, n\}$
to use in an LB type II step.   
First, the mixture component $c$ with the largest weight is selected.
For notational simplicity, we assume components are relabelled 
so that $c=K$.  
Then after relabelling, draw
$\thetavec_G\sim N(\muvec_{KG}[K],\Omega_{KG}[K]^{-1}).$
For this $\thetavec_G$, we assess the quality of the current approximation
for the conditional posterior density of $\bvec_i|\thetavec_G$ as follows.
Define
\begin{align}
 r(\bvec_i) & =\log \left\{p(\bvec_i|\thetavec_G)p(\yvec_i|\bvec_i,\thetavec_G)\right\}- \log \left\{q_{\lambdavec[K]}(\bvec_i|\thetavec_G)\right\},  \label{logratio}
\end{align}
where we suppress the dependence of $r(\bvec_i)$ on the (fixed) $\thetavec_G$.  
We write $s_i[K]$ for an estimate of variance of $r(\bvec_i)$, $\textrm{Var}(r_i(\bvec_i))$,
based on $L$ grid points to evaluate $r_i(\bvec_i)$.  
If the current approximation $q_{\lambdavec[K]}(\bvec_i|\thetavec_G)$ 
were exact for $p(\bvec_i|\thetavec_G,\yvec_i)$, then 
$q_{\lambda[K]}(\bvec_i|\thetavec_G)\propto p(\bvec_i|\thetavec_G)p(\yvec_i|\bvec_i,\thetavec_G)$, which would imply that the variance of $r(\bvec_i)$ is small.  Conversely, 
if $\text{{Var}}(r(\bvec_i))$ is large this indicates the current approximation
to $p(\bvec_i|\thetavec_G,\yvec_i)$ is poor.

Let $R_j[K]$ be the index of the $j$th largest value in $\{s_i[K], i=1,\dots, n\}$.  
When a type II VB boosting step is applied, we recommend setting $\mathcal{I}=\{R_j[K],j=1,\dots, |
\mathcal{I}|\}$, where $|\mathcal{I}|$ is chosen small enough to make the boosting optimisation
fast.  In other words, we concentrate on improving the conditional posterior
approximation for $\bvec_i|\thetavec_G$ for the $|\mathcal{I}|$ local parameters which are
furthest from the target conditional posterior, where the quality of
the approximation is measured by $s_i[K]$, $i=1,\dots, n$. Another way is to select the index of the latent variables with the $s_i[K]$ values bigger than a threshold value.
We also define $\widetilde{s}[K]=\frac{1}{n}\sum_{i=1}^{n}{s}_i[K]$ as the overall measure of the quality of the variational approximations. The smaller $\widetilde{s}[K]$, the better the overall variational approximations.

\section{Local boosting for state space models\label{boostingstatespace}}

We now consider an extension of our methods to state space models where
the data $\yvec$ is a time series, and the latent variables $\bvec$ are 
states, with $\bvec_i$ the state at the $i$th time.  
The likelihood is given by $p(\yvec|\thetavec)  = \prod_{i=1}^n p(\yvec_i |\bvec_i,\thetavec_G),$ 
and the states
have a Markovian prior of order 1 given by 
$ p(\thetavec)  = p(\thetavec_G)p(\bvec_1|\thetavec_G)\prod_{i=2}^n p(\bvec_i|\bvec_{i-1},\thetavec_G).$ %\label{priorss}
In the true posterior distribution for this model, the states are conditionally
independent of each other given their temporal neighbouring states and
$\thetavec_G$, and to match this conditional independence structure
in the $k$th Gaussian component of our Gaussian
mixture boosting approximation with $K$ components, 
the precision matrix takes
the block structured form (partitioning conformably with 
$\thetavec$ partitioned as $\thetavec=(\bvec_1^\top,\dots, \bvec_n^\top,\thetavec_G^\top)^\top$) given in \eqref{precisionss} in section \ref{sec:addlocalboostingstate} of the online supplement. The lower-triangular Cholesky factor has the block sparse form
\begin{align}
L_k[K] & = \left[\begin{array}{ccccccc}
L_{k1}[K] & 0  & 0 & \ldots & 0 & 0 & 0 \cr
\widetilde{L}_{k1}[K] & L_{k2}[K] & 0  & \ldots & 0 & 0 & 0 \cr
0 & \widetilde{L}_{k2}[K] & L_{k3}[K]  & \ldots & 0 & 0 &  0 \cr
\vdots & \vdots & \vdots & \ddots & \vdots & \vdots & \vdots \cr
0 & 0 & 0  &  \ldots  & L_{k\,n-1}[K] & 0 & 0 \cr
0 &  0 & 0  & \ldots & \widetilde{L}_{k\,n-1}[K] & L_{kn}[K] & 0 \cr
L_{kG1}[K] & L_{kG2}[K] & L_{kG3}[K] & \ldots & L_{kG\,n-1}[K] & L_{kGn}[K] & L_{kG}[K] 
\end{array}\right]. \label{cholss}
\end{align}
As previously, we define $L_{kj}'[K]$ as $L_{kj}[K]$ but with diagonal
elements log-transformed.  $L_{kG}'[K]$ is defined similarly from
$L_{kG}[K]$.  Abusing notation slightly, we now write 
$\ellvec_{kbi}[K]=(\text{vech}(L_{ki}'[K])^\top,\text{vec}(\widetilde{L}_{ki}[K])^\top)^\top, \;\;\;i=1,\dots, n-1,$
and
$\ellvec_{kbn}[K]=\text{vech}(L_{kn}'[K])^\top$.  
With this abuse of notation we 
write 
$\ellvec_{kb}[K]=(\ellvec_{kb1}[K]^\top,\dots, \ellvec_{kbn}[K]^\top)^\top,$
and define $\ellvec_{kGb}[K]$,
$\ellvec_{kG}[K]$, $\ellvec_k[K]$, $\muvec_k[K]$ and 
$\etavec_k[K]$ as before.  

%\widetilde{L}_{k\,n-1}[K],L_{knG}[K],\L_{k\,n-1\,G}[K],\widetilde{L}_{k\,n-1}[K]^\top

Now consider boosting variational inference for state space
models with local boosting moves.  We consider mixture
posterior approximations where normal mixture components
have precision matrices of the form \eqref{precisionss} in section \ref{sec:addlocalboostingstate} of the online supplement.  We start with a result
similar to Lemma \ref{lemma1}, which is important for motivating the 
local boosting moves considered here.  

\begin{lemma}
\label{lemma4}
Suppose that $\thetavec=(\bvec_1^\top,\dots, \bvec_n^\top,\thetavec_G^\top)^\top$ is a Gaussian random vector with mean 
vector $\muvec_k[K]$ given by \eqref{muk} and precision matrix $\Omega_k[K]=L_k[K]L_k[K]^\top$ where $L_k[K]$ is
given by \eqref{cholss}.  Then,
\begin{enumerate}
\item[i)] $\mathrm{E}(\thetavec_G)=\muvec_{kG}[K]$ and $\mathrm{Cov}(\thetavec_G)=L_{kG}[K]^{-\top}L_{kG}[K]^{-1}$.  
\item[ii)] $\mathrm{E}(\bvec_n|\thetavec_G)=\muvec_{kn}[K]-L_{kn}[K]^{-\top}L_{kGn}[K]^\top(\thetavec_G-\muvec_{kG}[K])$ and $\mathrm{Cov}(\bvec_n|\thetavec_G)=L_{kn}[K]^{-\top}L_{kn}[K]^{-1}$
\item[iii)] $\mathrm{E}(\bvec_i|\bvec_{i+1},\thetavec_G)=\muvec_{ki}[K]-L_{ki}[K]^{-\top}\widetilde{L}_{ki}[K]^\top (\bvec_{i+1}-\muvec_{k\,i+1}[K])-L_{ki}[K]^{-\top}L_{kGi}[K]^\top(\thetavec_G-\muvec_{kG}[K])$ and $\mathrm{Cov}(\bvec_i|\bvec_{i+1},\thetavec_G)=L_{ki}[K]^{-\top}L_{ki}[K]^{-1}$, $i=1,\dots, n-1$.

\end{enumerate}
\end{lemma} 
The proof is given in section \ref{sec:proofs} of the online supplement. 
Similar to our previous notation, we write $\ellvec_{kb\mathcal{I}}[K]=[\ellvec_{kbi}[K]^\top]_{i\in \mathcal{I}}^\top$, $\ellvec_{kb\overline{\mathcal{I}}}[K]=[\ellvec_{kbi}[K]^\top]_{i\in \overline{\mathcal{I}}}^\top$. 
Also similar to before, we write 
$\ellvec_{kGb\mathcal{I}}[K]:=[\mathrm{vec}(L_{kGi})^\top]_{i\in \mathcal{I}}^\top$ and $\ellvec_{kGb\overline{\mathcal{I}}}:=[\mathrm{vec}(L_{kGi})^\top]_{i\notin \mathcal{I}}^\top$.  In our extended notation, the global, LB type I and LB type II steps are defined exactly as before.  However, for the 
LB type II step, we need to choose the set $\mathcal{I}\subseteq\{1,\dots, n\}$
of the latent variables to use.  

Below, to simplify notation, $q_{\lambdavec[K]}(\bvec_i|\bvec_{i+1},\thetavec_G)$  means $q_{\lambdavec[K]}(\bvec_n|\thetavec_G)$ for $i=n$.  
Similarly, for the target posterior, the induced conditional density 
$p(\bvec_i|\bvec_{i+1},\thetavec_G,\yvec)$ is defined for the case $i=n$
to be $p(\bvec_n|\thetavec_G,\yvec)$.  For the conditional distribution
$p(\bvec_i|\bvec_{i-1},\bvec_{i+1},\thetavec_G,\yvec)$ we also make
the obvious modifications to the definition when $i=1$ and $i=n$.  
Examining Lemma \ref{lemma4} above, we see that $\ellvec_{kb\mathcal{I}}[K]$, 
$\ellvec_{kGb\mathcal{I}}[K]$ and $\muvec_{kb\mathcal{I}}[K]$
are exactly the set of parameters appearing in the conditional
densities $\bvec_i|\bvec_{i+1},\thetavec_G$, $i\in \mathcal{I}$
induced by $q_{\etavec_k[K]}(\thetavec)$ (except for
the mean parameters $\muvec_{k\,i+1}[K]$ and $\muvec_{kG}[K]$, $i\in \mathcal{I}$).  
So in an LB type II boosting
move, to choose $\mathcal{I}$, 
we need to identify states $\bvec_i$ where the conditional posterior
distribution $p(\bvec_i|\bvec_{i+1},\thetavec_G,\yvec)$ is not captured
well by $q_{\lambdavec[K]}(\bvec_i|\bvec_{i+1},\thetavec_G,\yvec)$, $i=1,\dots, n$.  

We suggest to choosing $\mathcal{I}$ in the following way.  
First, select the mixture component $c$ with the highest weight,  
then relabel components so that $c=K$.  Next, we draw
values $(\bvec_{i+1},\thetavec_G)\sim q_{\etavec_K[K]}(\bvec_{i+1},\thetavec_G)$
(or just $\thetavec_G$ if $i=n$).  
Observe that 
\begin{align}
  p(\bvec_i|\bvec_{i+1},\thetavec_G,\yvec) & = 
  \int p(\bvec_i|\bvec_{i-1},\bvec_{i+1},\thetavec_G,\yvec)
  p(\bvec_{i-1}|\bvec_{i+1},\thetavec_G,\yvec) d\bvec_{i-1},  \label{conditional}
\end{align}
and that
  $p(\bvec_i|\bvec_{i-1},\bvec_{i+1},\thetavec_G,\yvec) 
  \propto p(\bvec_{i+1}|\bvec_i,\thetavec_G)p(\bvec_i|\bvec_{i-1},\thetavec_G)
  p(\boldsymbol{y}_i|\bvec_i,\thetavec_G),$
so that
$p(\bvec_i|\bvec_{i+1},\thetavec_G,\yvec) \! \propto  \! \int p(\bvec_{i+1}|\bvec_i,\thetavec_G)p(\bvec_i|\bvec_{i-1},\thetavec_G)
  p(\boldsymbol{y}_i|\bvec_i,\thetavec_G)
  p(\bvec_{i-1}|\bvec_{i+1},\thetavec_G,\yvec) d\bvec_{i-1}.$
If $q_{\lambdavec[K]}(\bvec_i|\bvec_{i+1},\thetavec_G)$ is a perfect
approximation to $p(\bvec_i|\bvec_{i+1},\thetavec_G,\yvec)$, then 
\begin{align*}
 r(\bvec_i) & = \log p(\bvec_i|\bvec_{i+1},\thetavec_G,\yvec) -\log q_{\lambdavec[K]}(\bvec_i|\bvec_{i+1},\thetavec_G)
\end{align*}
would be constant in $\bvec_i$.   From \eqref{conditional}, if we
approximate the expectation with respect to $p(\bvec_{i-1}|\bvec_{i+1},\thetavec_G,\yvec)$ using Monte Carlo sampling of $S$ samples
$\bvec_{i-1}^{(s)}$ from $q_{\lambdavec[K]}(\bvec_{i-1}|\bvec_{i+1},\thetavec_G)$, 
we should have that $r(\bvec_i)\approx \widetilde{r}(\bvec_i)$, where 
\begin{align*}
  \widetilde{r}(\bvec_i) & = \log \frac{1}{S}\left\{\sum_{s=1}^S p(\bvec_{i+1}|\bvec_i,\thetavec_G)p(\bvec_i|\bvec_{i-1}^{(s)},\thetavec_G)p(\yvec_i|\bvec_i,\thetavec_G)\right\}-\log q_{\lambdavec[K]}(\bvec_i|\bvec_{i+1},\thetavec_G),
\end{align*}
should be approximately constant in $\bvec_i$.  For some  $L$ grid points, 
the sample variance of the values 
$\widetilde{r}(\bvec_i^{(l)})$, $l=1,\dots, L$ should be nearly constant.  
Write $s_i[K]$ for this sample variance.  Let $R_j[K]$ be the index
of the $j$th largest value in $\{s_i[K], i=1,\dots, n\}$, and choose
$\mathcal{I}=\{R_j[K], j=1,\dots, |\mathcal{I}|\}$.  
Unlike the hierarchical model case, there is no result similar to Lemma~\ref{lemma3}, in the sense that an LB type II move can change
the conditional posterior distributions $q_{\lambdavec[K+1]}(\bvec_{i+1}|\bvec_i,\thetavec_G)$ for $i\notin \mathcal{I}$, although the marginal distribution
for $\thetavec_G$ is unaffected.

% set of $L$ samples
%$\bvec_i^{(l)}$, $l=1,\dots, L$, from some proposal distribution $h(\bvec_i)$, or

\section{Examples}\label{sec:example}
To illustrate the performance of the mixture of normals variational approximation with global and local boosting steps, we employ them to approximate
complex and high-dimensional posterior distributions of latent variable models, where their greater flexibility may
increase the accuracy of inference of the latent variables compared to
simpler Gaussian variational approximations.

The paper has three examples. The first example is the random effects logistic regression model in section \ref{subsec:RegressionModelWithComplexPriorDistributions}. The second example is the stochastic volatility model in section \ref{subsec:svmodel} of the online supplement. The third example is the time-varying logistic regression model in section \ref{subsec:time-varyinglogistic} of the online supplement. All the examples
are implemented in Matlab. 
Unless otherwise stated, we use the estimates of the $\widetilde{s}[K]$ values to evaluate the improvement of the posterior approximations based on $K$-component mixture of normals.
The variational approximation is useful when the exact MCMC method is impossible or computationally expensive.
The boosting approach, where an existing approximation is improved by adding one new component at a time, allows us to tune the accuracy/computational effort trade-off, where we start with a fast approximation and keep improving until the computational budget is exhausted.  
All the variational parameters are initialised using the approach
in section \ref{sec:Initialising-a-New} of the online supplement. 
\subsection{Random Effects Logistic Regression Models \label{subsec:RegressionModelWithComplexPriorDistributions}}

Let $\boldsymbol{y}_{i}=(y_{i,1},\dots,y_{i,T})^{\top}$, $y_{i,t}\in\{0,1\}$, denote a sequence of outcomes for individual $i$ at time $t$, and let $\boldsymbol{x}_{i,t}$ denote strictly exogenous regressors for individual $i$ at time $t$. We are interested in understanding the impact of $\boldsymbol{x}_{i,t}$ on the probability of a positive outcomes $y_{i,t}=1$, and using a model of the form 
\begin{equation}
	\label{logisticmodel}
	y_{i,t}\sim \mathsf{Bernoulli}(p_{i,t}),\quad p_{i,t}=\frac{\exp\left(y_{i,t}\boldsymbol{x}_{i,t}^{\top}\betavec + b_i\right)}{1+\exp\left(\boldsymbol{x}_{i,t}^{\top}\betavec + b_i \right)},
\end{equation}
for $i=1,...,n$ observations and $t=1,..,T$ time periods, where the $b_i$ denote individual specific latent variables.  The first prior distribution for $b_i$ is the two-component mixture of normals,  
\begin{equation}\label{priors_mixnom_logistic}
p\left(b_{i}|w,\sigma_{1}^{2},\sigma_{2}^{2},\mu_1, \mu_2,\betavec\right)=wN\left(b_{i};\mu_1,\sigma_{1}^{2}\right)+\left(1-w\right)N\left(b_{i};\mu_2,\sigma_{2}^{2}\right),\;i=1,...,n;
\end{equation}
where $N\left(b_{i};\mu,\sigma^{2}\right)$ is the probability density function of normal distribution with mean $\mu$ and variance $\sigma^2$,
setting $w=0.5$, $\sigma_{1}^{2}=\sigma_{2}^{2}=0.01$,
$\mu_1=-2$, and $\mu_2=2$. The second prior distribution for $b_i$ is the $t$-distribution with mean $\mu=0$, scale $\sigma^2=0.01$, and degrees of freedom $\nu=3$. The complex prior distributions for the latent variables are chosen to create a complex posterior structure of the latent variables to evaluate the performance of boosting variational approximations when the posteriors are non-Gaussian. The prior for the $\betavec$ is $N(0,I)$, where $I$ is an identity matrix with an appropriate size.

We use the polypharmacy longitudinal data in \citet{hosmer2013applied}, which features data on 500 individuals over 7 years. The logistic regression includes eight fixed effects, plus one random effect per subject; see \citet{ong2017gaussian} for further details. 
We consider two cases: (I) the mixture of normal and t priors are used for all $n=500$ latent variables, (II) the priors are only used for the first $20$ latent variables, the other $n-20$ use $N(0,1)$ as the priors. We consider the global and the local boosting methods for both cases,
using $S=100$ samples to estimate the lower bound values and the gradients
of the lower bound. Algorithm \ref{alg:Variational-Algorithm} is performed for $5000$ iterations to obtain the optimal variational
parameters for each boosting optimisation.  The values $s_i[K]$ are calculated using grid points between $-5$ and $5$, for each $i$. 

Figure~\ref{fig:Varplot_full_global_logistic_mixnom_t} in section \ref{sec:add_fig_randomeffectlogistic} of the online supplement plots the $\widetilde{s}[K]$ values for the mixture of normals variational approximation (VA) with global boosting steps and sparse precision matrix for each component for the logistic regression example with a mixture of normals and t priors for all latent variables. The figure shows that the values of $\widetilde{s}[K]$ decrease over boosting iterations for both priors, indicating a substantial amount of improvement in the approximation of the marginal distributions of the latent variables. Clearly, the mixture of Gaussians variational approximation is much better than the standard Gaussian variational approximation. 
The optimal number of components for the mixture of normals and $t$ priors are $20$ and $14$ components, respectively, although there are no substantial improvements after $4$ components for the $t$ prior and after $10$ components for the mixture of normals prior. 

%We now compare the accuracy of the mixture of norm to the densities estimated using the Hamiltonian Monte Carlo (HMC) method of \citet{hoffman2014no}, called the No U-Turn Sampler (NUTS); this method is a popular MCMC algorithm for sampling high dimensional posterior distributions. For all examples,
%the NUTS tuning parameters, such as the number of leapfrog steps and the step size, are set to the default values as in the STAN reference manual \footnote{$https://mc-stan.org$}. We ran the HMC method  for $1005000$ iterations, discarding the initial $5000$ iterations as warm-up. The remaining $1000000$ MCMC samples are stored for further analysis. The inefficiency of the HMC method is measured using the integrated autocorrelation time (IACT) defined in section \ref{IACT_section} of the online supplement.

We now compare the accuracy of the mixture of normals variational approximation with sparse precision matrices to the densities estimated using the mean field mixture of Gaussians where the diagonal precision matrix is used for each mixture component and the Hamiltonian Monte Carlo (HMC) method of \citet{hoffman2014no}, called the No U-Turn Sampler. The HMC method is a popular MCMC method for sampling high dimensional posterior distributions. The tuning parameters, such as the number of leapfrog steps and the step size, are set to the default values as in the STAN reference manual\footnote{$https://mc-stan.org$}. The HMC method is run for $100000$ iterations, discarding the initial $5000$ iterations as warm-up.
By comparing our
variational approximations with a mean-field mixture of Gaussians variational approximation, we can investigate the importance of taking into account the posterior dependence structures in the model.
Figure \ref{fig:Varplot_full_global_logistic_mixnom_t} in section \ref{sec:add_fig_randomeffectlogistic} of the online supplement plots  the $\widetilde{s}[K]$ values for the mixture of normal variational approximation with global boosting steps and diagonal precision matrix for each component. The $\widetilde{s}[K]$ values for the mixture of normals variational approximation with the diagonal precision matrix are similar to the approximation with sparse precision matrix for each component for the mixture of normal priors, but the $\widetilde{s}[K]$ values are much higher for the $t$ priors example. The mixture of normals variational approximation with sparse precision matrix is better than the mixture of normals with diagonal precision matrix. 

% and \ref{fig:densityestimates_full_Global_logistic_statespace_mixnom_Precs_MF_globalparam} in section \ref{sec:add_fig_randomeffectlogistic} of the online supplement
%However, the posterior variance of the global parameters estimated using the mixture of normals VA with the diagonal precision matrix is smaller than that of the mixture of normals with the sparse precision matrix. This confirms the findings in \citet{tan2017}. 

\begin{figure}[H]
\caption{Kernel density estimates of the first eight marginals of the latent variables estimated using the mixture of normals variational approximation with sparse precision matrix (Precision), the mean field mixture of normals variational approximation with diagonal precision matrix (MF) with the global boosting steps, and the Hamiltonian Monte Carlo (HMC) method for the logistic regression example with a mixture of normal priors for all of the latent variables. The optimal number of components is chosen using $\widetilde{s}[K]$ values. 
\label{fig:densityestimates_full_Global_logistic_statespace_mixnom_Precs_MF_latent}}
\centering{}\includegraphics[width=15cm,height=6.5cm]{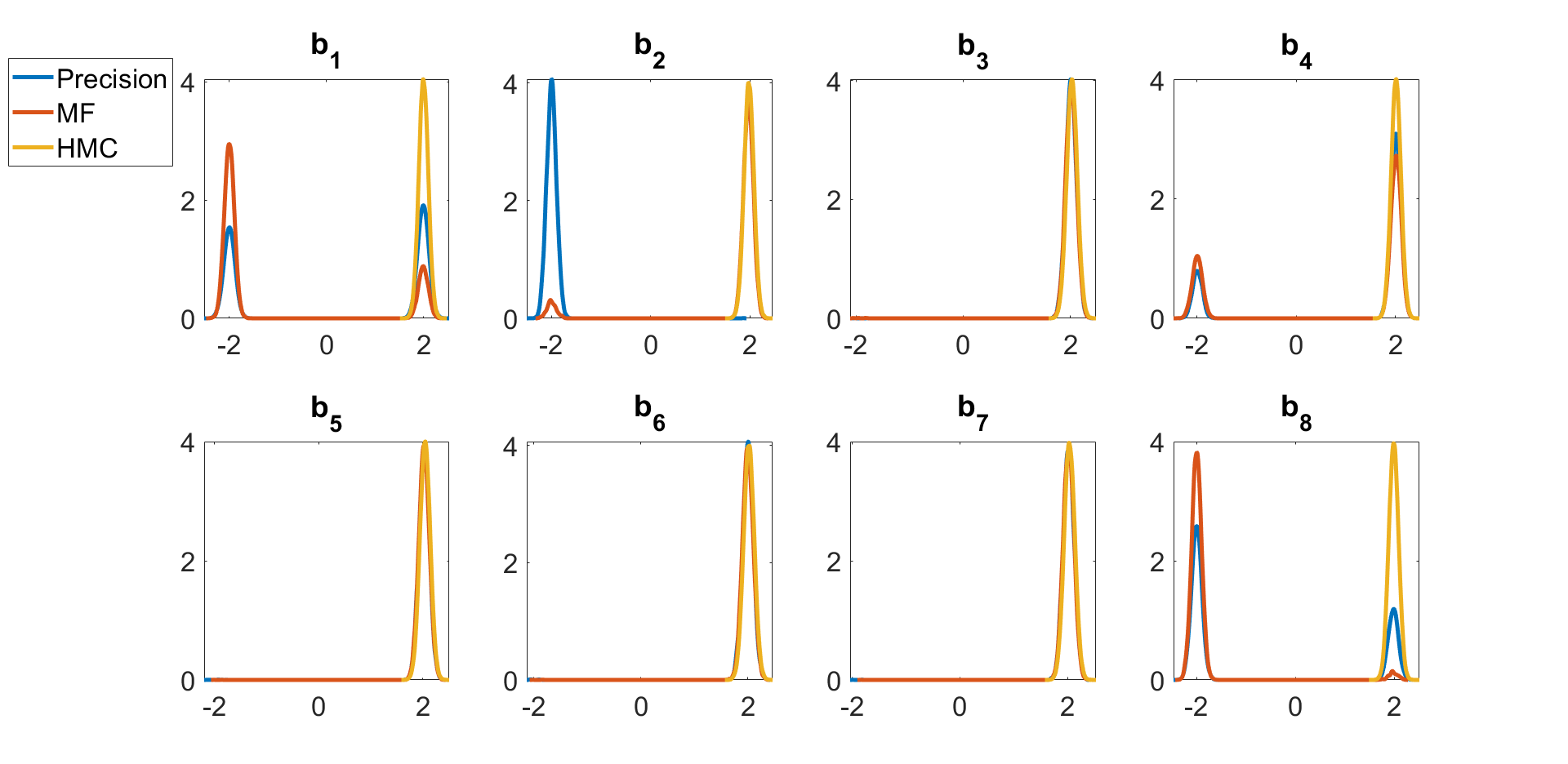}
\end{figure}

Figure \ref{fig:densityestimates_full_Global_logistic_statespace_mixnom_Precs_MF_latent} and Figure \ref{fig:logdensityestimates_full_Global_logistic_statespace_mixnom_Precs_MF_latent} in section \ref{sec:add_fig_randomeffectlogistic} of the online supplement show the kernel density estimates and the log kernel density estimates, respectively, of the first eight latent variables for the logistic regression example with a mixture of normal priors for all observations, respectively, estimated using the mixture of normals with the sparse precision matrix, the mixture of normals with the diagonal precision matrix, and the HMC method.
The figure shows that density estimates and the log density estimates of the latent variables estimated using the mixture of normals VA with the diagonal precision matrix are similar to the density estimates estimated using the mixture of normals VA with the sparse precision matrix. However, the HMC method is unable to produce the multimodal distributions of the latent variables. 

\begin{figure}[H]
\caption{Kernel density estimates of the first eight marginals of the latent variables estimated using the mixture of normals variational approximation with sparse precision matrix (Precision), the mean field mixture of normals variational approximation with diagonal precision matrix (MF) with the global boosting steps, and the Hamiltonian Monte Carlo (HMC) method for the logistic regression example with t priors for all of the latent variables. The optimal number of components is chosen using $\widetilde{s}[K]$ values.\label{fig:densityestimates_full_Global_logistic_statespace_t_Precs_MF_latent}}
\centering{}\includegraphics[width=15cm,height=6.5cm]{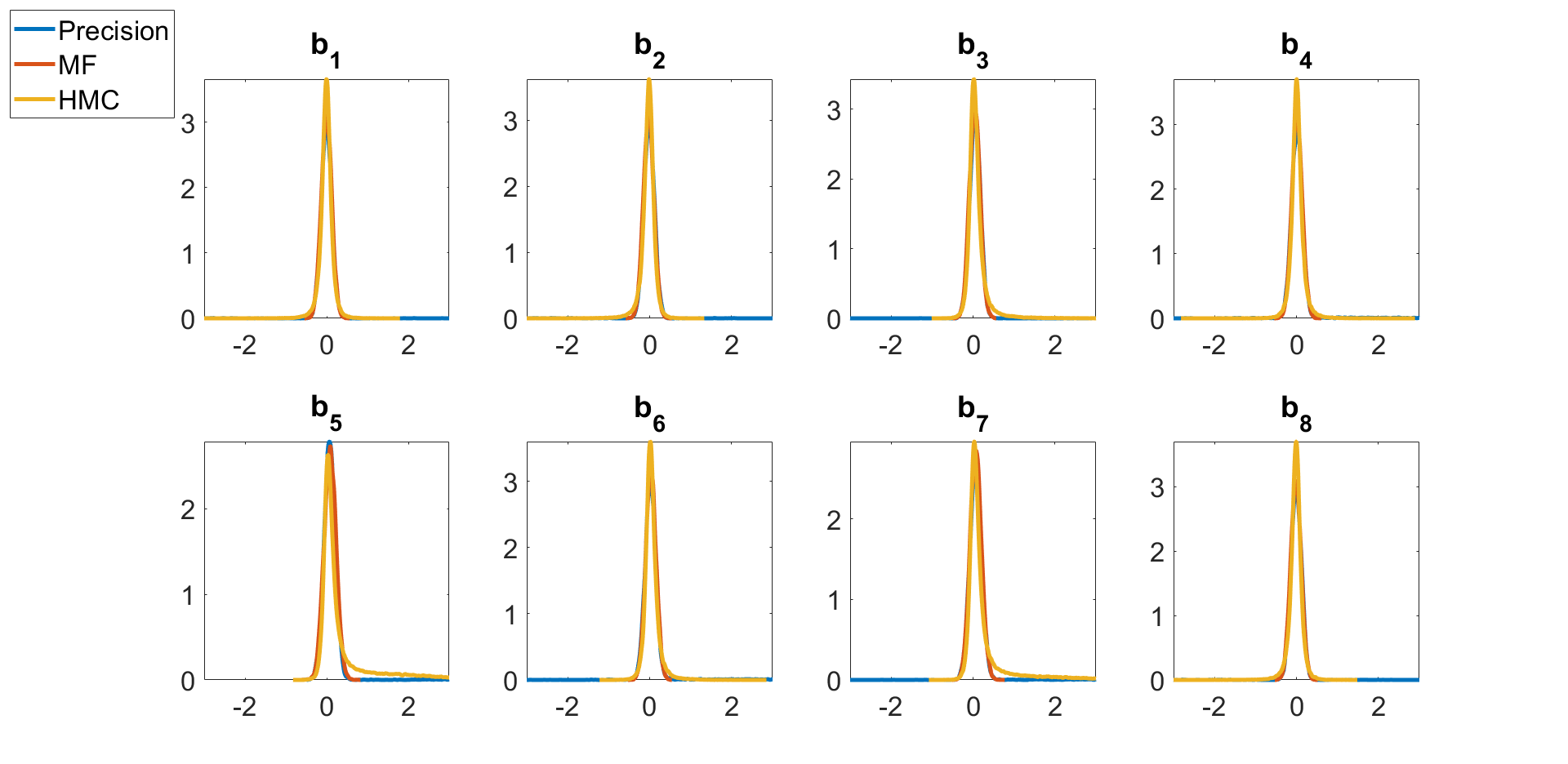}
\end{figure}

Figure \ref{fig:densityestimates_full_Global_logistic_statespace_t_Precs_MF_latent} and Figure \ref{fig:logdensityestimates_full_Global_logistic_statespace_t_Precs_MF_latent} in section \ref{sec:add_fig_randomeffectlogistic} of the online supplement plot
the kernel density estimates and the log kernel density estimates, respectively, of the first eight latent variables for the logistic regression example with $t$ priors for all latent variables, respectively.
The figure shows that the mixture of normals with the diagonal precision matrix does not capture the heavy-tailedness of the posterior distributions of the latent variables. It is, therefore, important to take into account the posterior dependence structure in the model. The posterior densities estimated using the mixture of normals variational approximation have thicker tails than those of the HMC method. 

\begin{figure}[H]
\caption{Plots of the lower bound values for the mixture of normals variational approximations with global and local boosting steps for the logistic regression example with a mixture of normal (left panel) and t (right panel) priors for the first 20 latent variables. \label{fig:LB_some_global_local_mixnom_t}}

\centering{}\includegraphics[width=15cm,height=6.5cm]{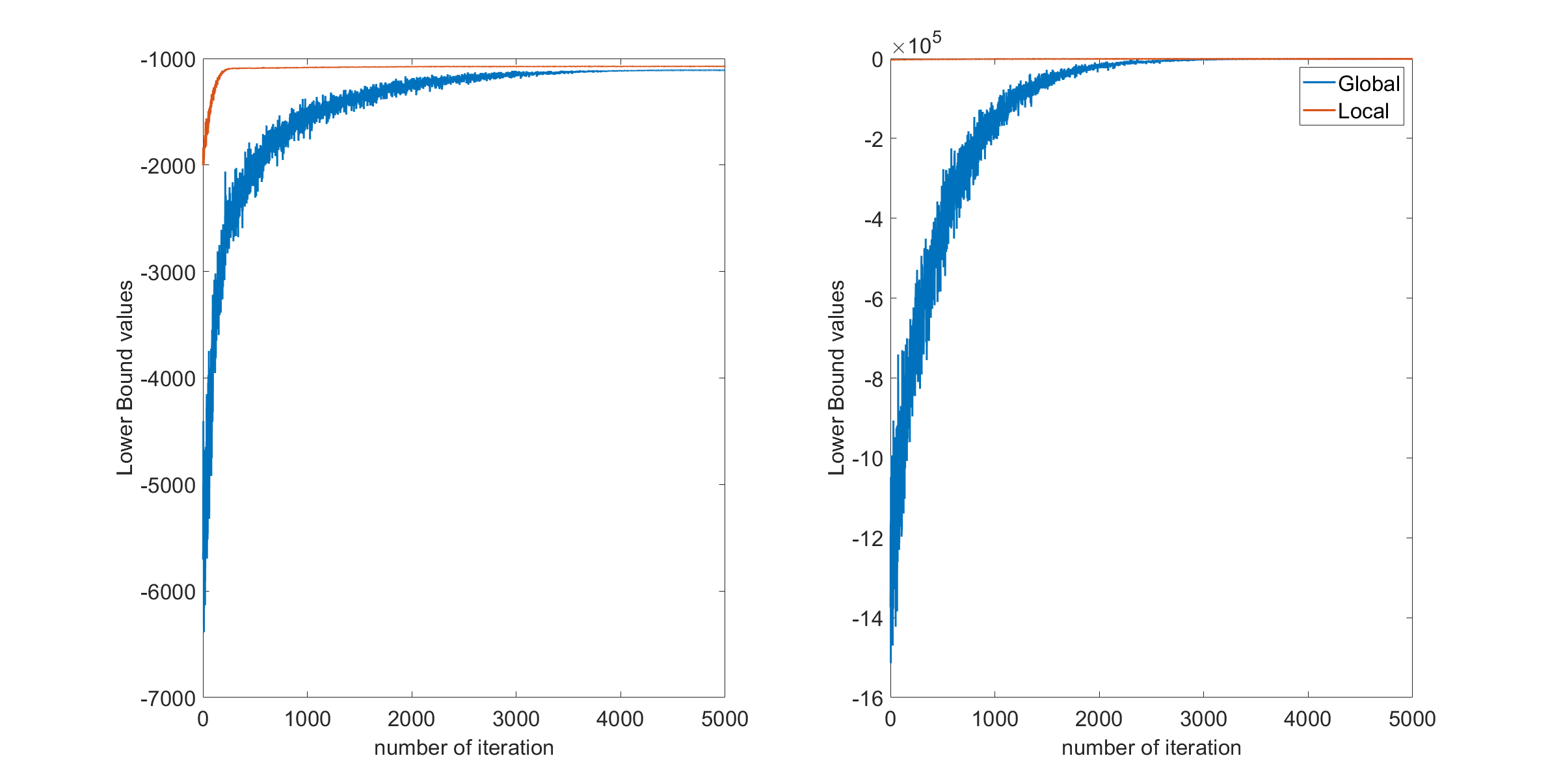}
\end{figure}

Figure \ref{fig:LB_some_global_local_mixnom_t} plots the lower bound values for the mixture of normals variational approximation with global and local boosting steps for the logistic regression example with a mixture of normals and $t$ priors for the first $20$ latent variables. The figure shows that the local boosting steps that only update the variational parameters for the poorly approximated latent variables have better initial lower bound values, are much less noisy, and converge faster than the global boosting steps for both priors. The optimisation for the local boosting steps is easier because it updates fewer parameters. This indicates that if only some of the conditional posterior
distributions of the latent variables are complex, then the local boosting step is more efficient than the
global boosting step.

\begin{figure}[H]
\caption{Plots of the indices of the latent variables selected by the PSIS method (left panel) and the ${s}[K]$ values (right panel) for the mixture of normals variational approximations with local boosting steps for the logistic regression example with a mixture of normal priors for the first $20$ latent variables. \label{fig:index_selected_var_PSIS_mixnom}}

\centering{}\includegraphics[width=15cm,height=6.5cm]{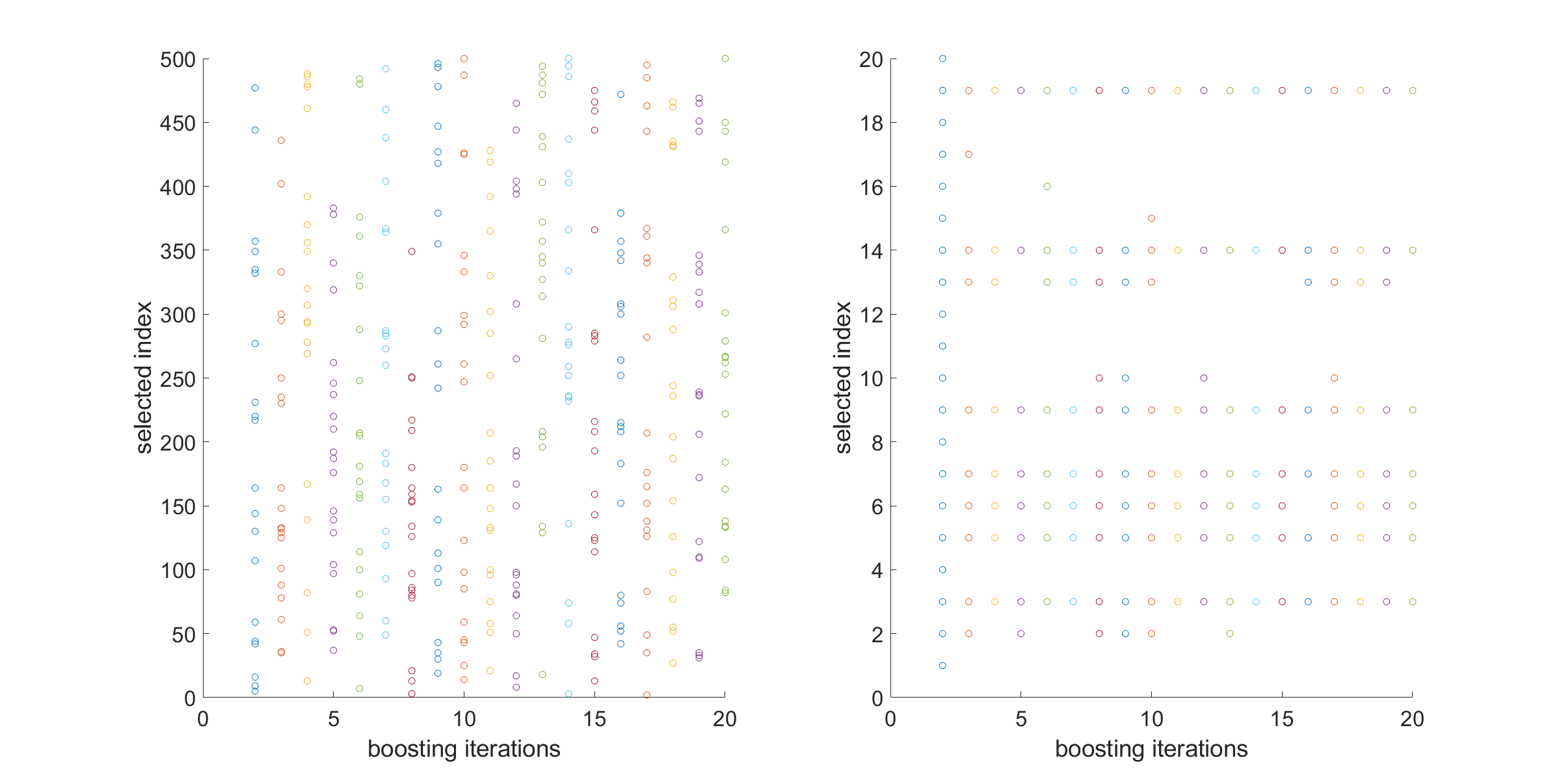}
\end{figure}

We now compare two methods to select the poorly approximated latent variables. The first is based on the Pareto smoothed importance sampling (PSIS) method of \citet{yao2018yes} described in section \ref{sec:PSISdiag} of the online supplement; the second is based on the ${s}[K]$ value. 
Figure \ref{fig:index_selected_var_PSIS_mixnom} plots the indices of latent variables selected by the PSIS method and ${s}[K]$ values for the logistic regression with a mixture of normals prior to the first $20$ observations. The selection method works well if it successfully selects the indices $1 - 20$ for the first few boosting iterations. The figure shows that the selection method based on the PSIS method fails to identify the poorly approximated latent variables. However, the figure clearly shows that the selection method based on the ${s}[K]$ values correctly identifies the poorly approximated latent variables, the first $20$ latent variables. 

\begin{figure}[H]
\caption{Plots of the $\widetilde{s}[K]$ values for the mixture of normals variational approximations with local boosting steps for the logistic regression example with a mixture of normal priors for the first 20 latent variables, where the poorly approximated latent variables are identified using the PSIS method and $s[K]$ values. \label{fig:Varplot_some_local_logistic_mixnom_PSIS}}
\centering{}\includegraphics[width=15cm,height=6.5cm]{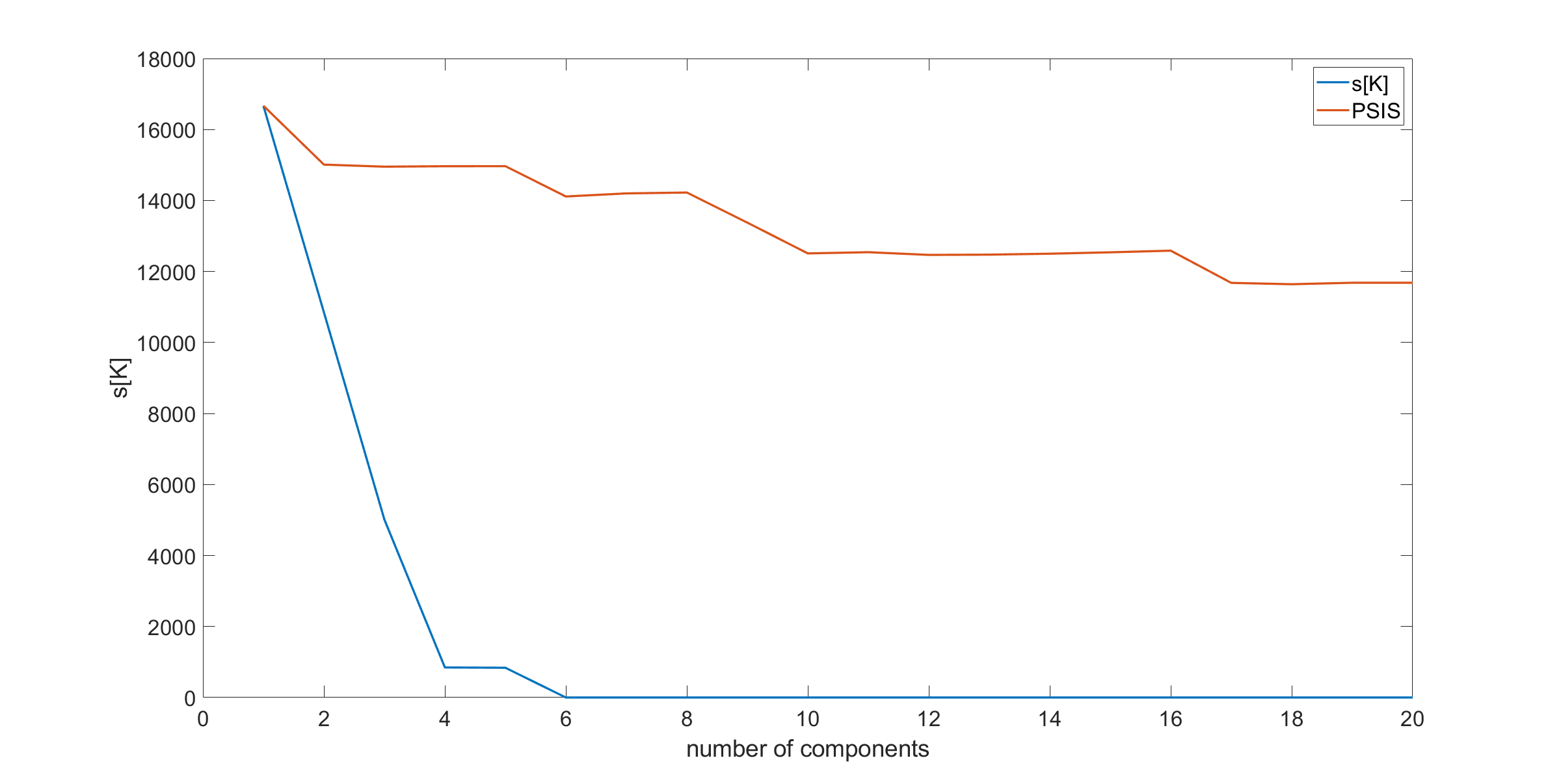}
\end{figure}

Figure \ref{fig:Varplot_some_local_logistic_mixnom_PSIS} shows the $\widetilde{s}[K]$ values obtained from the mixture of Gaussians variational approximation for the logistic regression example with the mixture of normals prior for the first $20$ latent variables using the PSIS and the proposed selection methods. The figure shows that the $\widetilde{s}[K]$ values obtained using the proposed identification method are much lower than the $\widetilde{s}[K]$ values obtained using the PSIS identification method because the PSIS identification method is unable to select the poorly approximated latent variables correctly and therefore the boosting method updates incorrectly.

Figure \ref{fig:Varplot_some_global_local_logistic_mixnom_t} in section \ref{sec:add_fig_randomeffectlogistic} of the online supplement shows the $\widetilde{s}[K]$ values for global and local boosting steps for the logistic regression
example with a mixture of normals and $t$ priors for the first 20 latent variables.
The figure shows that: (1) the optimal number of components for the global boosting step
is $K = 2$ components for the $t$ prior and $K=11$ components for the mixture of normals prior. The optimal number of components for the local boosting step is $K = 3$ components for the $t$ prior and $K=6$ for the mixture of normals prior. The optimal $\widetilde{s}[K]$ values are similar for both global and local boosting steps.  

Figures \ref{fig:densityestimatesmixnom_t_global_local_some} and \ref{fig:logdensityestimatesmixnom_t_global_local_some} in section \ref{sec:add_fig_randomeffectlogistic} of the online supplement show the kernel density estimates and the log kernel density estimates, respectively, of the first marginal latent variable
estimated using the HMC method and the mixture of normals variational approximations with the optimal number
of components chosen using $\widetilde{s}[K]$ values for both global and local boosting steps for the logistic
regression example with a mixture of normals and $t$ priors for the first $20$ latent variables, respectively. 
The figure shows that the mixture of normals variational
approximation can produce bimodal and heavy-tailed posterior distributions of the latent variables. However, HMC produces thinner tailed posterior distributions but is unable to produce multimodal posterior distributions.

The examples in this section suggest that: (1)~The mixture of normals variational approximation with the sparse precision matrix for each component can approximate heavy tails, multimodality and other
complex properties of high dimensional posterior distributions of the latent variables and parameters. It outperforms
the Gaussian variational approximation, the mean-field mixture of normals variational approximation, and the HMC method.
(2)~Adding a few components to the mixture of normals variational approximation generally improves its ability to approximate
complex posterior distributions of the latent variable models. Therefore, the proposed approach
can be considered as a refinement of the Gaussian variational approximation. (3) Adding additional components one at a time provides
a practical method for constructing an increasingly
complicated approximation and applies to a variety
of statistical models with latent variables. (4) The identification method based on the ${s}[K]$ values performs better than the identification method based on the PSIS method. (5) If only some of the conditional posterior
distributions of the latent variables are complex, then the local boosting step is more efficient than
the global boosting step.

\section{Conclusions \label{sec:conclusion}}
The article proposes a mixture of normals variational approximation for latent variable models, where a sparsity structure is imposed in the precision matrix of each Gaussian component to reflect the appropriate conditional independence in the model. A fast optimisation method based on a combination of global and local boosting moves, natural gradients, and the reparameterisation trick is proposed. The type I local boosting updates the variational parameters that appear in the marginal distribution of the global parameters. The type II local boosting   updates the variational parameters that appear in the conditional distributions of some of the latent variables that are poorly approximated by the current approximation. 
The examples show that when only some of the latent variables are poorly approximated, the local boosting steps are much more efficient than the global boosting steps. 
Novel methods to identify which latent variables are poorly approximated are developed. The examples show that the proposed identification method works better than the Pareto smoothed importance sampling method of \citet{yao2018yes}.
The proposed variational approximations are applied to estimate generalised linear mixed models and state space models.

%\section*{Acknowledgements}

\pagebreak
\renewcommand{\thealgorithm}{S\arabic{algorithm}}
\renewcommand{\theequation}{S\arabic{equation}}
\renewcommand{\thesection}{S\arabic{section}}
\renewcommand{\thepage}{S\arabic{page}}
\renewcommand{\thetable}{S\arabic{table}}
\renewcommand{\thefigure}{S\arabic{figure}}
\setcounter{page}{1}
\setcounter{section}{0}
\setcounter{equation}{0}
\setcounter{algorithm}{0}
\setcounter{table}{0}
\setcounter{figure}{0}

\title{\bf Online Supplement: Fast Variational Boosting for Latent Variable Models}
\maketitle
%\author{David Gunawan\textsuperscript{$\star$}, Pratiti Chatterjee\textsuperscript{$\ddagger$} , and Robert Kohn\textsuperscript{$\star\star$}}

We use the following notation in the supplement. (1), algorithm~1,
section~1, etc, refer to the main paper, while (S1),
algorithm~S1, section~S1, etc, refer to the supplement. All the acronyms used without definition in the supplement, are defined in the main paper.

\section{Variational inference}\label{sec:variational}

Variational inference approximates a Bayesian posterior distribution
by solving an optimisation problem.  Consider a model for data $\yvec$ having parameter $\thetavec$ with  
density $p(\yvec|\thetavec)$, and   
Bayesian inference with prior density $p(\thetavec)$ 
and posterior density $p(\thetavec|\yvec)\propto p(\thetavec)p(\yvec|\thetavec)$.  
In variational inference, we optimise an approximation 
$q_\lambdavec(\thetavec)$ with
respect to variational parameters $\lambdavec\in \Lambda$ to match
$p(\thetavec|\yvec)$ as closely as possible.
The most commonly used measure of closeness is the reverse Kullback-Leibler (KL)
divergence, 
\begin{align}
  \text{KL}(q_\lambdavec(\thetavec)||p(\thetavec|\yvec)) & = 
  \int \log \frac{q_\lambdavec(\thetavec)}{p(\thetavec|\yvec)}q_\lambdavec(\thetavec)\,d\thetavec.  \label{kld} 
\end{align}
Writing $p(\thetavec|\yvec)=p(\thetavec)p(\yvec|\thetavec)/p(\yvec)$,
where $p(\yvec):=\int p(\thetavec)p(\yvec|\thetavec)\,d\thetavec$,
and substituting into \eqref{kld} gives
\begin{align}
\text{KL}(q_\lambdavec(\thetavec)||p(\thetavec|\yvec)) & = -{\cal L}(\lambdavec)+\log p(\yvec),  \label{kld2}
\end{align} 
where ${\cal L}(\lambdavec) := \int \log \frac{p(\thetavec)p(\yvec|\thetavec)}{q_\lambdavec(\thetavec)}q_\lambdavec(\thetavec)\,d\thetavec$
is called the evidence lower bound (ELBO).  
It follows from \eqref{kld2} that maximising the ELBO 
with respect to $\lambdavec$ is 
equivalent to minimizing
$\text{KL}(q_\lambdavec(\thetavec)||p(\thetavec|\yvec))$.  
From the non-negativity of the Kullback-Leibler divergence, 
${\cal L}(\lambdavec)\leq \log p(\yvec)$.   Writing
$\lambdavec^*=\arg \max_{\lambdavec\in\Lambda} {\cal L}(\lambdavec)$, 
assuming that this maximum exists, the final posterior approximation
is $q_{\lambdavec^*}(\thetavec)$.  

Although $\mathcal{L}\left(\lambdavec\right)$ is often an intractable integral with no closed
 form solution, we can write it
as an expectation with respect to $q_{\lambdavec}\left(\thetavec\right)$,
\begin{equation}
\mathcal{L}\left(\lambdavec\right)=E_{q_{\lambdavec}}\left(\log h\left(\thetavec\right)-\log q_{\lambdavec}\left(\thetavec\right)\right),\label{eq:LB2}
\end{equation}
where $h\left(\thetavec\right) := p\left(\yvec|\thetavec\right)p\left(\thetavec\right)$.
This interpretation 
%of \Eq{eq:LB2} 
allows the use
of  stochastic gradient ascent (SGA)  methods
to maximise the variational lower bound $\mathcal{L}\left(\lambdavec\right)$.
See, for example, \cite{Nott:2012}, \cite{Paisley:2012}, \cite{Hoffman:2013}, \cite{salimans2013fixed}, \cite{Kingma2014},  
 \cite{Titsias2014}, and \cite{Rezende2014}.
In SGA, an initial value $\lambdavec^{\left(0\right)}$ is updated according
to the iterative scheme
\begin{equation}
\lambdavec^{\left(t+1\right)} := \lambdavec^{\left(t\right)}+\boldsymbol{a}_{t}\circ\widehat{\nabla_{\lambdavec}\mathcal{L}\left(\lambdavec^{\left(t\right)}\right)},
\; \textrm{  for } t=0,1,2,... ; \label{eq:stochastic gradient update}
\end{equation}
$\circ$ denotes the Hadamard (element by element) product of
two random vectors; $\boldsymbol{a}_{t}:=\left(a_{t1},...,a_{tm}\right)^{\top}$
is a vector of step sizes, where
 $m$ is the dimension of  the variational parameters 
$\lambdavec$, and $\widehat{\nabla_{\lambdavec}\mathcal{L}\left(\lambdavec^{\left(t\right)}\right)}$
is an unbiased estimate of the gradient of the lower bound $\mathcal{L}\left(\lambdavec\right)$
at $\lambdavec=\lambdavec^{\left(t\right)}$. The learning rate sequence
satisfies the Robbins-Monro conditions $\sum_{t}a_{t}=\infty$
and $\sum_{t}a_{t}^{2}<\infty$ \citep{Robbins1951}, which ensures
that the iterates $\lambdavec^{\left(t\right)}$ converge to a local
optimum as $t\rightarrow\infty$ under suitable regularity conditions
\citep{bottou2010}. Adaptive step sizes are often used in practice,
and we employ the ADAM method of \citet{Kingma2015}, which uses bias-corrected
estimates of the first and second moments of the stochastic gradients
to compute adaptive learning rates. The update in  \eqref{eq:stochastic gradient update}
continues until a stopping criterion is satisfied.

\section{Standard variational boosting approach \label{sec:standardvariationalboosting}}
This section briefly describes the standard variational boosting approach.
The method starts by fitting an approximation to the posterior distribution
$p(\thetavec|\yvec)$ with a single component mixture distribution, $K=1$, using the stochastic gradient ascent algorithm;  the optimal first component variational
parameters are denoted as $\lambdavec{[1]}^{*}:=\left(\muvec_{1}[1]^{\top},\ellvec_{1}[1]^{\top},\pi_{1}[1]\right)^{\top}$, with
the mixture weight $\pi_{1}[1]$  set to $1$. We
do this by maximising the first lower bound objective function
\begin{align*}
\mathcal{L}^{\left(1\right)}\left(\lambdavec[1]\right) & =  E_{q_{\lambdavec[1]}}\left(\log h\left(\thetavec\right)-\log q_{\lambdavec[1]}\left(\thetavec\right)\right),\quad
\lambdavec[1]^{*}  =  \underset{\lambdavec[1]}{\textrm{arg max}}\; \mathcal{L}^{\left(1\right)}\left(\lambdavec[1]\right),
\end{align*}
where $q_{\lambdavec[1]}\left(\thetavec\right)=N\left(\thetavec;\muvec_{1}[1],L_{1}[1]^{-\top}L_{1}[1]^{-1}\right)$.  After the optimisation algorithm converges,  $\lambdavec[1]$ is fixed as
$\lambdavec[1]^{*}$.

After iteration $K$, the current variational approximation to the posterior distribution
$p(\thetavec|\yvec)$ is a mixture distribution with $K$ components,
\begin{equation*}
q_{\lambdavec{[K]}}\left(\thetavec\right)=
\sum_{k=1}^{K}\pi_{k}[K]N\left(\thetavec|\muvec_{k}[K],L_k[K]^{-\top}L_k[K]^{-1}\right).
\end{equation*}
We can introduce a new mixture component with new component parameters, \\ 
$\left(\muvec_{K+1}[K+1],\ellvec_{K+1}[K+1]\right)$, and a new mixing weight
$\pi_{K+1}[K+1]$. The new
approximating distribution is
\begin{equation*}
q_{\lambdavec{[K+1]}}\left(\thetavec\right)
:=\sum_{k=1}^{K-1}\pi_k[K]q_{\etavec_{K}[K]}\left(\thetavec\right)
+\pi \pi_K[K]q_{\etavec_K[K]}(\thetavec)
 + (1-\pi)\pi_K[K] q_{\etavec_{K+1}[K+1]}(\thetavec).
\end{equation*}
The new lower bound objective function is
\begin{eqnarray*}
\mathcal{L}^{\left(K+1\right)}\left(\lambdavec{[K+1]}\right) & := & E_{q_{\lambdavec[K+1]}}\left(\log h\left(\thetavec\right)-\log q_{\lambdavec[K+1]}\left(\thetavec\right)\right),\\
\lambdavec{[K+1]}^{*} & := & \underset{\lambdavec{[K+1]}}{\textrm{arg max}}\; \mathcal{L}^{\left(K+1\right)}\left(\lambdavec{[K+1]}\right).
\end{eqnarray*}

\section{Additional materials for Gaussian mixture components\label{sec:addlocalboostingstate}}
This section gives more details on the Gaussian mixture components. 
If $\thetavec$ is a Gaussian random vector, then to have
$\bvec_i,\bvec_j$, $i\neq j$ conditionally independent given $\thetavec_G$, 
$\Omega_k[K]$ should have the block structure (partitioning conformably with 
$\thetavec$ partitioned as $\thetavec=(\bvec_1^\top,\dots, \bvec_n^\top,\thetavec_G^\top)^\top$)
\begin{align}
\Omega_k[K] &  = \left[\begin{array}{ccccc}
\Omega_{k1}[K] & 0 & \ldots & 0 & \Omega_{k1G}[K]  \cr
0 & \Omega_{k2}[K] & \ldots & 0 & \Omega_{k2G}[K] \cr
\vdots & \vdots & \ddots & \vdots & \vdots \cr
0 &  0 & \ldots & \Omega_{kn}[K] & \Omega_{knG}[K] \cr
\Omega_{kG1}[K] & \Omega_{kG2}[K] & \ldots & \Omega_{kGn}[K] & \Omega_{kG}[K] 
\end{array}\right].  \label{precision}
\end{align}

In the true posterior distribution for the state space model, the states are conditionally
independent of each other given their temporal neighbouring states and
$\thetavec_G$, and to match this conditional independence structure
in the $k$th Gaussian component of our Gaussian
mixture boosting approximation with $K$ components, 
the precision matrix should take
the block structured form (partitioning conformably with 
$\thetavec$ partitioned as $\thetavec=(\bvec_1^\top,\dots, \bvec_n^\top,\thetavec_G^\top)^\top$), 
\begin{align}
\Omega_k[K] & = \left[\begin{array}{ccccccc}
\Omega_{k1}[K] & \widetilde{\Omega}_{k1}[K]^\top  & 0 & \ldots & 0 & 0 & \Omega_{k1G}[K] \cr
\widetilde{\Omega}_{k1}[K] & \Omega_{k2}[K] & \widetilde{\Omega}_{k2}[K]^\top  & \ldots & 0 & 0 & \Omega_{k2G}[K] \cr
0 & \widetilde{\Omega}_{k2}[K] & \Omega_{k3}[K]  & \ldots & 0 & 0 &  \Omega_{k3G}[K] \cr
\vdots & \vdots & \vdots & \ddots & \vdots & \vdots & \vdots \cr
0 & 0 & 0  &  \ldots  & \Omega_{k\,n-1}[K] & \widetilde{\Omega}_{k\,n-1}[K]^\top & \Omega_{k\,n-1\,G}[K] \cr
0 &  0 & 0  & \ldots & \widetilde{\Omega}_{k\,n-1}[K] & \Omega_{kn}[K] & \Omega_{knG}[K] \cr
\Omega_{kG1}[K] & \Omega_{kG2}[K] & \Omega_{kG3}[K] & \ldots & \Omega_{kG\,n-1}[K] & \Omega_{kGn}[K] & \Omega_{kG}[K] 
\end{array}\right].  \label{precisionss}
\end{align}

\section{Learning Rate \label{subsec:Learning-Rate}}

Setting the learning rate in a stochastic gradient algorithm is
very challenging, especially when the  parameter vector
is high dimensional. The choice of learning rate affects both the
rate of convergence and the quality of the optimum attained. Learning
rates that are too high can cause unstable optimisation,
while learning rates that are too low result in slow convergence
and can lead to a situation where the parameters erroneously
appear to have converged.
In all our examples, the learning rates are set adaptively using the
ADAM method \citep{Kingma2015} that gives different step sizes for
each element of the variational parameters $\lambdavec$. At iteration
$t+1$, the $i$th element $\lambda_{i}$ of $\lambdavec$ is updated as
\[
\lambda^{\left(t+1\right)}_{i}:=\lambda^{\left(t\right)}_{i}+{\triangle}^{\left(t\right)}_{i}.
\]
Let ${g}_{i,t}^{\textrm{nat}}$ denote the natural stochastic gradient estimate (discussed in section \ref{subsec:Updating-the-Variational} of the main paper) 
at iteration $t$ for the $i$th element of $\lambdavec$. ADAM computes (biased) first and second moment
estimates of the gradients using exponential moving averages,
\begin{eqnarray*}
{m}_{i,t} & = & \tau_{1}{m}_{i,t-1}+\left(1-\tau_{1}\right){g}_{i,t}^{\textrm{nat}},\\
{v}_{i,t} & = & \tau_{2}{v}_{i,t-1}+\left(1-\tau_{2}\right)\left({g}_{i,t}^{2}\right)^{\textrm{nat}},
\end{eqnarray*}
where $\tau_{1},\tau_{2}\in\left[0,1\right)$ control the decay rates.
The biased first and second moment estimates are corrected
by \citep{Kingma2015}
\begin{eqnarray*}
\widehat{{m}}_{i,t} & = & {m}_{i,t}/\left(1-\tau_{1}^{t}\right), \quad
\widehat{{v}}_{i,t}  =  {v}_{i,t}/\left(1-\tau_{2}^{t}\right);
\end{eqnarray*}
the change ${\triangle}^{\left(t\right)}_{i}$ is then computed as
\begin{equation*}
{\triangle}^{\left(t\right)}_{i}=\frac{\alpha\widehat{{m}}_{i,t}}{\sqrt{\widehat{{v}}_{i,t}}+\eps}.
\end{equation*}
We set $\tau_{1}=0.9$, $\tau_{2}=0.99$, and $\eps=10^{-8}$
\citep{Kingma2015}. It is possible to use different $\alpha$ for different variational parameters. We set $\alpha_{\muvec}=0.01$ for updating the means, and $\alpha_{\ellvec}=\alpha_{\pivec}=0.001$, for updating the elements of the precision matrix and mixture weights, unless stated otherwise.

%$\muvec$, $\ellvec$, and $\pivec$. We set $\alpha_{\muvec}=0.01$, $\alpha_{\ellvec}=\alpha_{\pivec}=0.001$, unless stated otherwise.

%\footnote{\begin{itemize}
%    \item [i] If $\widehat {\boldsymbol{m}}_{t} $ and $\widehat {\boldsymbol{v}}_{t} $ are vectors, what do you mean by the  ratio of
%    $\widehat {\boldsymbol{m}}_{t} $ and the square root of $\widehat {\boldsymbol{v}}_{t} $ ?  
%   \item [ii] why don't you write $\alpha$ as a vector and use the Hadamard elementwise notation. This comment also applies to part [i]
%   \item Can you give a pointer where the natural gradients are computed in this supplement or elsewhere? Otherwise, this introductory section is less useful. 
%\end{itemize}}

\section{Proofs\label{sec:proofs}}

\vspace{0.2in}
\noindent
{\it Proof of Lemma 1:} \;\;If $\thetavec\sim N(\muvec_k[K],\Omega_k[K]^{-1})$, then we can write 
$\thetavec=\muvec_k[K]+L_k[K]^{-\top}\zvec$, $\zvec\sim N(0,I)$ and rearranging 
\begin{align}
  L_k[K]^\top(\thetavec-\muvec_k[K])= & \zvec.  \label{block}
\end{align}   
Writing $\zvec=(\zvec_1^\top,\dots, \zvec_n^\top,\zvec_G)^\top$ so
that $\zvec$ is partitioned conformably with $\thetavec=(\bvec_1^\top,\dots,\bvec_n^\top,\thetavec_G^\top)^\top$, and writing out the block rows of \eqref{block}, 
we obtain
$$L_{kG}[K]^\top(\thetavec_G-\muvec_G)=\zvec_G,$$
$$L_{ki}[K]^\top(\bvec_i-\muvec_{ki}[K])+L_{kGi}^\top(\thetavec_G-\muvec_G[K])=\zvec_i, \;\;i=1,\dots, n.$$
Rearranging these expressions gives 
$$\thetavec_G=\mu_{G}[K]+L_{kG}[K]^{-\top}\zvec_G,$$ 
$$\bvec_{i}=\muvec_{ki}[K]-L_{ki}[K]^{-\top} L_{kGi}[K]^\top (\thetavec_G-\muvec_G[K])+L_{ki}[K]^{-\top}\zvec_i,\;\;i=1,\dots, n,$$ 
from which the result follows. \\  

\noindent
{\it Proof of Lemma 3:}  We have
\begin{align*}
q_{\lambda[K+1]}(\thetavec_G) & = \sum_{k=1}^{K+1}\pi_k[K+1]q_{\etavec_k[K+1]}(\thetavec_G) \\
 & = \sum_{k=1}^{K-1} \pi_k[K]q_{\etavec_k[K]}(\thetavec_G) +
 (\pi_K[K+1]+\pi_{K+1}[K+1])q_{\etavec_K[K]}(\thetavec_G),
\end{align*}
where the last line follows that for the type II LB step,  
\begin{align*}
q_{\etavec_{K+1}[K+1]}(\thetavec_G) & =
q_{\etavec_K[K]}(\thetavec_G), 
\end{align*}  
\begin{align*}
q_{\etavec_{k}[K+1]}(\thetavec_G) & =
q_{\etavec_k[K]}(\thetavec_G), k=1,\dots, K.
\end{align*}
and
\begin{align*}
  \pi_k[K]& =\pi_k[K+1],\;\;k=1,\dots, K-1.
\end{align*}
Noting that $\pi_K[K+1]+\pi_{K+1}[K+1]=\pi_K[K]$ shows
that $q_{\lambdavec[K+1]}(\thetavec_G)=q_{\lambdavec[K]}(\thetavec_G)$.  

Next, for $i\notin \mathcal{I}$, we have from Lemma 2 $(ii).$ that 
\begin{align*}
  q_{\lambda[K+1]}(\bvec_i|\thetavec_G) = &
  \sum_{k=1}^{K+1} w_k[K+1](\thetavec_G)q_{\etavec_k[K+1]}(\bvec_i|\thetavec_G) \\
   = & \sum_{k=1}^{K-1}w_k[K+1](\thetavec_G)q_{\etavec_k[K]}(\bvec_i|\thetavec_G) + \\
   & \;\;\;\; (w_K[K+1](\thetavec_G)+w_{K+1}[K+1](\thetavec_G))q_{\etavec_K[K]}(\bvec_i|\thetavec_G),
\end{align*}
where the last line follows because for a type II LB step
$q_{\etavec_k}[K+1](\bvec_i|\thetavec_G)=q_{\etavec_k[K]}(\bvec_i|\thetavec_G)$ for $k=1,\dots, K$, and the fact that
$q_{\etavec_{K+1}[K+1]}(\bvec_i|\thetavec_G)=q_{\etavec_K[K]}(\bvec_i|\thetavec_G)$, for $i\notin \mathcal{I}$.  The result follows if
we can show that $w_k[K+1](\thetavec_G)=w_k[K](\thetavec_G)$, $k=1,\dots, K-1$, and $w_{K}[K+1](\thetavec_G)+w_{K+1}[K+1](\thetavec_G)=w_K[K](\thetavec_G)$.  This follows because
$$\pi_K[K+1]q_{\etavec_K[K+1]}(\thetavec_G)+\pi_{K+1}[K+1]q_{\etavec_{K+1}[K+1]}(\thetavec_G)=\pi_K[K]q_{\etavec_K[K]}(\thetavec_G),$$ 
and the expression
for the weights $w_k[K+1](\thetavec_G)$ provided by Lemma 2 $(ii).$   \\ 

\noindent
{\it Proof of Lemma 4:} \;\;If $\thetavec\sim N(\muvec_k[K],\Omega_k[K]^{-1})$, then we can write 
$\thetavec=\muvec_k[K]+L_k[K]^{-\top}\zvec$, $\zvec\sim N(0,I)$; rearranging, 
\begin{align}
  L_k[K]^\top(\thetavec-\muvec_k[K])= & \zvec.  \label{block2}
\end{align}   
Writing $\zvec=(\zvec_1^\top,\dots, \zvec_n^\top,\zvec_G)^\top$ so
that it is partitioned conformably with $\thetavec=(\bvec_1^\top,\dots,\bvec_n^\top,\thetavec_G^\top)^\top$, and writing out the block rows of \eqref{block2}, 
we obtain
$$L_{kG}[K]^\top(\thetavec_G-\muvec_G)=\zvec_G,$$
$$L_{kn}[K]^\top(\bvec_n-\muvec_{kn}[K])+L_{kGn}^\top(\thetavec_G-\muvec_G[K])=\zvec_n,$$
$$L_{ki}[K]^\top(\bvec_i-\muvec_{ki}[K])+\widetilde{L}_{ki}[K]^\top(\bvec_{i+1}-\muvec_{k\,i+1}[K])-L_{kGi}^\top(\thetavec_G-\muvec_G[K])=\zvec_i, \;\;i=1,\dots, n-1.$$
Rearranging these expressions gives 
$$\thetavec_G=\mu_{G}[K]+L_{kG}[K]^{-\top}\zvec_G,$$ 
$$\bvec_{n}=\muvec_{kn}[K]-L_{kn}[K]^{-\top} L_{kGn}[K]^\top (\thetavec_G-\muvec_G[K])+L_{kn}[K]^{-\top}\zvec_n,\;\;i=1,\dots, n,$$ 
$$\bvec_{i}=\muvec_{ki}[K]-L_{ki}[K]^{-\top} L_{kGi}[K]^\top (\thetavec_G-\muvec_G[K])-L_{ki}[K]^{-\top} \widetilde{L}_{ki}[K]^\top (\bvec_{i+1}-\muvec_{k\;i+1})+L_{ki}[K]^{-\top}\zvec_i,\;\;i=1,\dots, n,$$ 
from which the result follows. \\  

\section{Updating Variational Parameters using the Control Variates method \label{sec:controlvariates}}
This section discusses an alternative approach to update the variational parameters $\ellvec_{K+1}[K+1]$, which are the non-zero elements in the $L_{K+1}[K+1]$. We employ control variates as in \citet{Ranganath:2014} to reduce
the variance of an unbiased estimate of the gradient of the $\nabla_{\ellvec_{K+1}[K+1]}\mathcal{L}^{(K+1)}\left(\lambdavec[K+1]\right)$. 
Now, we first discuss the update for the variational parameters $\ellvec_{K+1}[K+1]$.
The gradients of the lower bound require the expression of the gradient
$\nabla_{L_{K+1}[K+1]}\log q_{\boldsymbol{\lambdavec}[K+1]}\left(\boldsymbol{\theta}\right)$.
It is easy to show that the gradient of $\log q_{\boldsymbol{\lambda}[K+1]}\left(\boldsymbol{\theta}\right)$
with respect to $L_{K+1}[K+1]$ is
\begin{multline*}
\nabla_{L_{K+1}[K+1]}\log q_{\boldsymbol{\lambda}[K+1]}\left(\boldsymbol{\theta}\right)=\delta_{tot}^{-1}{\pi_{K+1}[K+1]\left(N\left(\boldsymbol{\theta}|\boldsymbol{\mu}_{K+1}[K+1],L_{K+1}[K+1]^{-\top}L_{K+1}[K+1]^{-1}\right)\right)}\\
\left(\textrm{diag}\left(1/L_{K+1}[K+1]\right)-\left(\boldsymbol{\theta}-\boldsymbol{\mu}_{K+1}[K+1]\right)\left(\boldsymbol{\theta}-\boldsymbol{\mu}_{K+1}[K+1]\right)^{\top}L_{K+1}[K+1]\right),
\end{multline*}
where $\textrm{diag}\left(1/L_{K+1}[K+1]\right)$ denotes the diagonal
matrix with the same dimensions as $L_{K+1}[K+1]$ with $i$th diagonal
entry $1/L_{K+1,ii}[K+1]$. We then take the entries of $ \nabla_{L_{K+1}[K+1]}\log q_{\boldsymbol{\lambdavec}[K+1]}\left(\boldsymbol{\theta}\right)$ corresponding to the elements of $\ellvec_{K+1}[K+1]$.

\begin{algorithm}[H]
\caption{Variational Bayes Algorithm with Control Variates Approach of \citet{Ranganath:2014} \label{alg:Variational-Algorithm-ControlVariate}}

\begin{enumerate}
\item (a) Initialise $\lambdavec{[K+1]}^{\left(0\right)}=\left(\muvec_{K+1}[K+1]^{\top \left(0\right)},
    (\ellvec_{K+1}{[K+1]}^{\left(0\right)})^{\top},
    \pivec[K+1]^{\top\left(0\right)}\right)$,
set $t=0$, and generate $\thetavec_{s}^{\left(t\right)}\sim q_{\lambdavec[{\left(K+1\right)}]}\left(\thetavec\right)$ for $s=1,...,S$. Let $m_{L}$
be the number of elements in $\ellvec_{K+1}[K+1]$.

(b) Evaluate the control variates $\boldsymbol{\varsigma}_{\ellvec_{K+1}{[K+1]}}^{\left(t\right)}=\left(\varsigma_{1,\ellvec_{K+1}{[K+1]}}^{\left(t\right)},...,\varsigma_{m_{L},\ellvec_{K+1}{[K+1]}}^{\left(t\right)}\right)^{\top}$,
with
\begin{equation}\footnotesize
\varsigma_{i,\ellvec_{K+1}{[K+1]}}^{\left(t\right)}=\frac{\wh {\textrm{Cov}}\left(\left[\log\left(h\left(\thetavec\right)\right)-\log q_{\lambdavec[K+1]}\left(\thetavec\right)\right]\nabla_{\lambda_{i,\ellvec_{K+1}{[K+1]}}}\log q_{\lambdavec[K+1]}\left(\thetavec\right),\nabla_{\lambdavec_{i,\ellvec_{K+1}{[K+1]}}}\log q_{\lambdavec[K+1]}\left(\thetavec\right)\right)}{\wh\V\left(\nabla_{\lambdavec_{i,\ellvec_{K+1}{[K+1]}}}\log q_{\lambdavec[K+1]}\left(\thetavec\right)\right)},\label{eq:cv}
\end{equation}
for $i=1,...,m_{L}$, where $\wh {\textrm{Cov}}$ and $\wh\V\left(\cdot\right)$ are the
sample estimates of covariance and variance based on $S$ samples
from step (1a).
\end{enumerate}
Repeat until the stopping rule is satisfied
\begin{itemize}
\item Update $\ellvec_{K+1}{[K+1]}$:
\end{itemize}
\begin{enumerate}
\item Generate $\thetavec_{s}^{\left(t\right)}\sim q_{\lambdavec[{K+1}]}\left(\thetavec\right)$ for $s=1,...,S$.
\item Compute the gradients of $\ellvec_{K+1}[K+1]$, the non-zero elements in $L_{K+1}[K+1]$, as 

$\widehat{\nabla_{\ellvec_{K+1}{[K+1]}}\mathcal{L}^{(K+1)}\left(\lambdavec[K+1]^{\left(t\right)}\right)}=\left(g_{1,\ellvec_{K+1}{[K+1]}}^{\left(t\right)},...,g_{m_{L},\ellvec_{K+1}{[K+1]}}^{\left(t\right)}\right)$,
with
\begin{equation}\footnotesize
g_{i,\ellvec_{K+1}{[K+1]}}^{\left(t\right)}=\frac{1}{S}\sum_{s=1}^{S}\left[\log\left(h\left(\thetavec_{s}^{\left(t\right)}\right)\right)-\log q_{\lambdavec[K+1]}\left(\thetavec_{s}^{\left(t\right)}\right)-\varsigma_{i,\ellvec_{K+1}{[K+1]}}^{\left(t-1\right)}\right]\nabla_{\lambdavec_{i,\ellvec_{K+1}{[K+1]}}}\log q_{\lambdavec[K+1]}\left(\thetavec^{(t)}_{s}\right)\label{eq:gradLB_beta}
\end{equation}

%\item The gradient of lower bound $\widehat{\nabla_{L_{K+1}}\mathcal{L}\left(\lambdavec^{\left(t\right)}\right)}=\left(g_{1,L_{K+1}}^{\left(t\right)},...,g_{m_{L},L_{K+1}}^{\left(t\right)}\right)$
%can be computed similarly as in Eq. \eqref{eq:gradLB_beta}.

\item Compute the control variate $\boldsymbol{\varsigma}_{\ellvec_{K+1}[K+1]}^{\left(t\right)}$
as in \eqref{eq:cv}.
\item Compute $\triangle \ellvec_{K+1}{[K+1]}$
using ADAM as described in this supplement. Then, set
$\ellvec_{K+1}{[K+1]}^{\left(t+1\right)}=\ellvec_{K+1}{[K+1]}^{\left(t\right)}+\triangle \ellvec_{K+1}{[K+1]}^{(t)}$.
\end{enumerate}
\begin{itemize}
\item Update $\muvec_{K+1}[K+1]$ and $\pivec[K+1]$
\end{itemize}
\begin{enumerate}
\item Generate $\thetavec_{s}^{\left(t\right)}\sim q_{\lambdavec[K+1]}\left(\thetavec\right)$ for $s=1,...,S$.
\item Use \eqref{eq:update pi-1} to update $\pivec[K+1]$ and \eqref{eq:update mu-1}
to update $\muvec_{K+1}{[K+1]}$, respectively. Set $t=t+1$.
\end{enumerate}
\end{algorithm}

\section{Pareto Smoothed Importance Sampling (PSIS) \label{sec:PSISdiag}}

Algorithm \ref{PSISdiagnostics} describes how to find a suitable set $\mathcal{I}\subseteq \{1,\dots, n\}$
to use in an LB type II step using the PSIS method of \citet{yao2018yes}.      

\begin{algorithm}[H]
\caption{Pareto Smoothed Importance Sampling (PSIS) Diagnostic\label{PSISdiagnostics}}

\begin{enumerate}
\item Select the mixture component $c$ with the largest weight.
For notational simplicity, we assume components are relabelled 
so that $c=K$.  
After relabelling, draw
$$\thetavec_G\sim N(\muvec_{KG}[K],\Omega_{KG}[K]^{-1}).$$
For this $\thetavec_G$,  draw $L$ samples $\boldsymbol{b}^{l}_i$ from $q_{\lambdavec[K]}(\bvec_i|\thetavec_G)$ for $l=1,...,L$.
\item Calculate 
\begin{align}
 r(\bvec^{l}_i) & =\log \left\{p(\bvec^{l}_i|\thetavec_G)p(\yvec_i|\bvec^{l}_i,\thetavec_G)\right\}- \log \left\{q_{\lambdavec[K]}(\bvec^{l}_i|\thetavec_G)\right\},  \label{logratio3}
\end{align}
for $l=1,...,L$.
\item Fit a generalised Pareto distribution to the $M$ largest values of $r(\bvec^{l}_i)$. 
\item Report the shape parameter $\widehat{k}_{i}$.
\item if $\widehat{k}_{i}<0.7$ then we conclude that variational approximation
for $\bvec_{i}$ is close enough to the $p\left(\bvec_{i}|\thetavec,\yvec_{i}\right)$.
\end{enumerate}
\end{algorithm}

\section{Stochastic Volatility Model \label{subsec:svmodel}}
This section considers the stochastic volatility model, a popular
model in financial time series \citep{kim1998}, which is an example
of a non-linear state space model. The observation $y_{i}$ is generated
from a Gaussian distribution with mean zero and a time-varying variance.
The unobserved log volatility $b_{i}$ is modeled using an AR(1) process
with a mixture of normal and $t$ disturbances. Let 
\begin{eqnarray}
y_{i} & \sim & \textrm{N\ensuremath{\left(0,\exp\left(\kappa+b_{i}\right)\right)},}\;\textrm{for}\;i=1,...,n,\nonumber \\
b_{1} & = & \eta_{1},\nonumber \\
b_{i} & = & \phi b_{i-1}+\eta_{i},\;\textrm{for}\;i=2,...,n,
\end{eqnarray}
where $\kappa\in R$, and $0<\phi<1$. We transform the constrained
parameter $\phi$ to $\psi\in R$ using
$\psi = {\rm logit} (\phi)$; so that 
$\phi=\exp\left(\psi\right)/ ({1+\exp\left(\psi\right)})$. 
 We use a normal prior for 
$\kappa\sim N(0,1)$ and a beta prior for $(\phi+1)/2 \sim B(20,1.5)$. We use complex distributions for the disturbance. The first is 

\begin{equation}\label{priors_mixnom}
p\left(\eta_i|w,\sigma_{1}^{2},\sigma_{2}^{2},\mu_1, \mu_2\right)=wN\left(\eta_i;\mu_1,\sigma_{1}^{2}\right)+\left(1-w\right)N\left(\eta_i;\mu_2,\sigma_{2}^{2}\right),\;i=1,...,n.
\end{equation}
We set $w=0.5$, $\sigma_{1}^{2}=0.01$, $\sigma_{2}^{2}=1$,
$\mu_1=0$, and $\mu_2=0$.
The second is the $t$-distribution with mean $\mu=0$, scale $\sigma^2=0.01$, and degree of freedom $\nu=3$.
%The distribution of the disturbance is
\begin{figure}[H]
\caption{Plots of the $\widetilde{s}[K] $ values for the mixture of normals variational approximations with local boosting steps for the stochastic volatility example with a mixture of normal (left panel) and t (right panel) priors for the first $20$ latent variables. \label{fig:Varplot_some_Local_sv_mixnom_t_Precs}}

\centering{}\includegraphics[width=15cm,height=8cm]{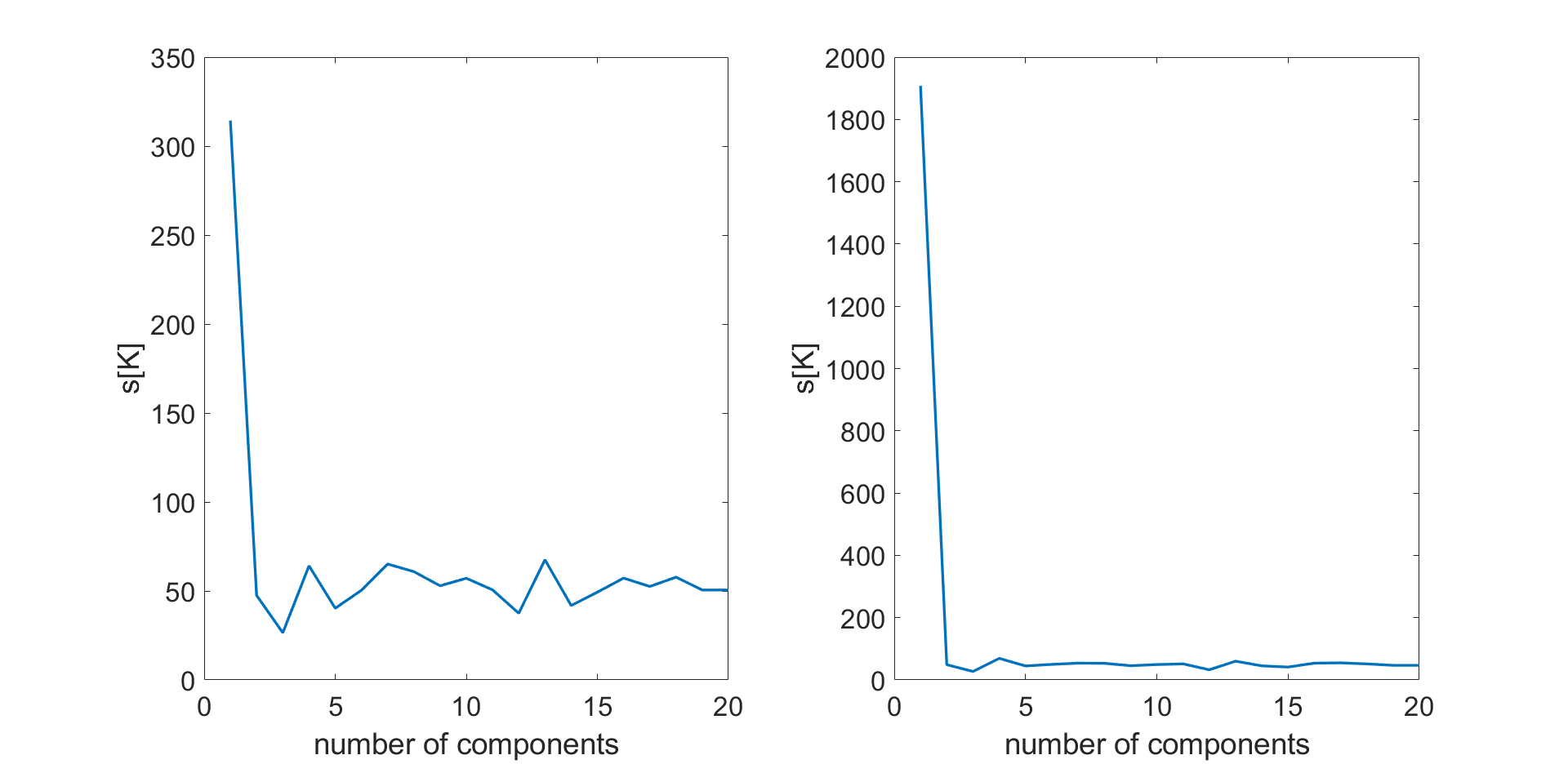}
\end{figure}

We use the demeaned daily return $(r_i)$ data from Apple Inc. stock from 9th October 2009 to 3rd October 2011 (a total of $500$ observations). We consider the case where the complex distributions of the disturbances are applied only for the first $20$ latent variables. The other $n-20$ use $N(0,0.01)$ as the priors. We consider the local boosting methods for both priors. The values $s_i[K]$ is calculated using grid points between $-5$ and $5$, for each $i$. Figure \ref{fig:Varplot_some_Local_sv_mixnom_t_Precs} shows that the $\widetilde{s}[K]$ values decrease over boosting iterations. The mixture of normals variational approximation is clearly better than the commonly used Gaussian variational approximation.

\section{Additional figures for random effects logistic regression models \label{sec:add_fig_randomeffectlogistic}}

This section contains additional figures for the random effects logistic regression model example in section \ref{subsec:RegressionModelWithComplexPriorDistributions}.

\begin{figure}[H]
\caption{Plots of the $\widetilde{s}[K] $ values for the mixture of normals variational approximations with global boosting steps (with sparse precision matrix (Precision) and diagonal precision matrix (MF)) for the logistic regression example with a mixture of normal (left panel) and t (right panel) priors for all the latent variables. \label{fig:Varplot_full_global_logistic_mixnom_t}}

\centering{}\includegraphics[width=15cm,height=8cm]{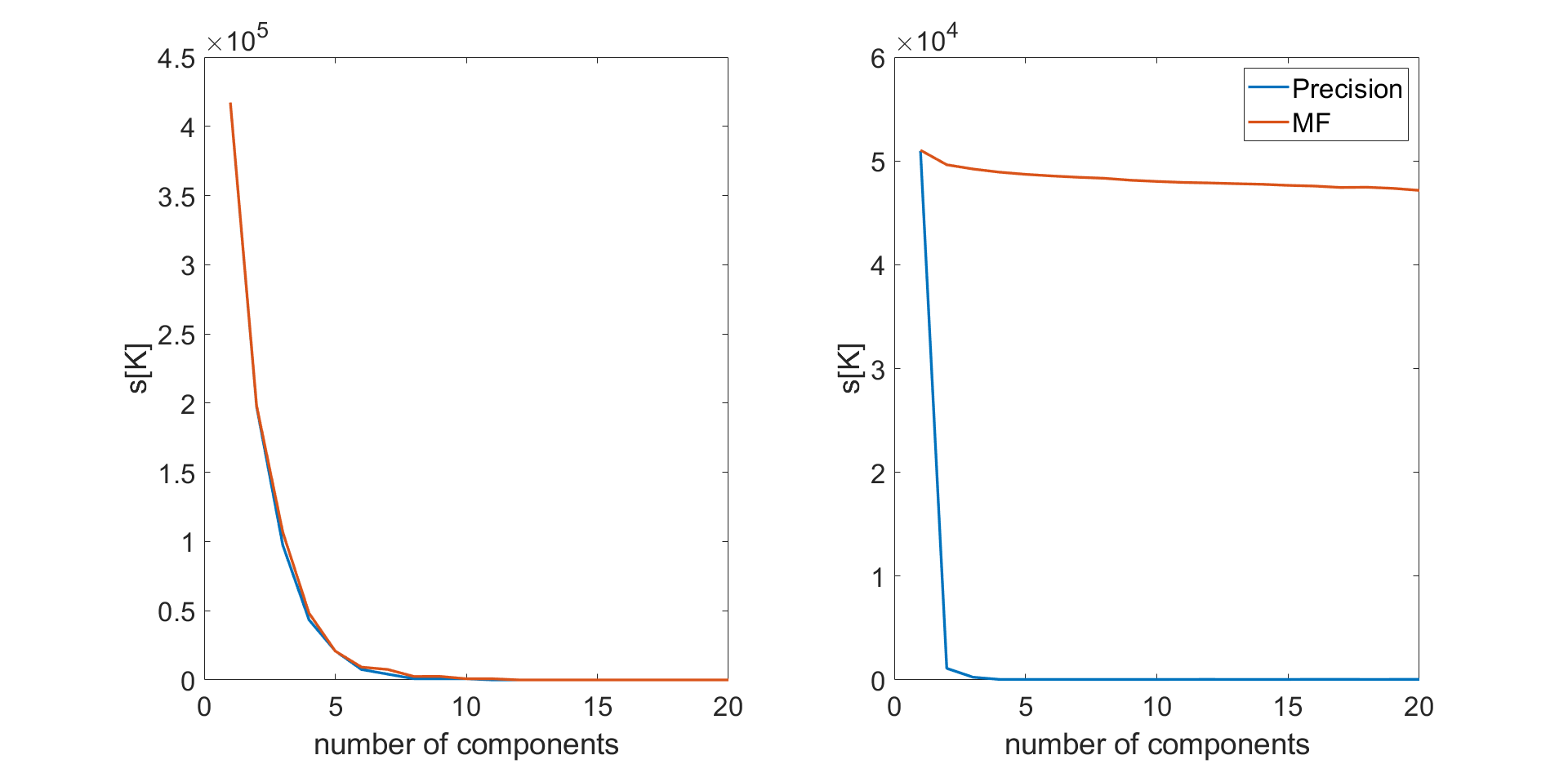}
\end{figure}

Figure \ref{fig:Varplot_full_global_logistic_mixnom_t}  plots the $\widetilde{s}[K]$ values for the mixture of normal variational approximation with global boosting steps and diagonal precision matrix for each component. The $\widetilde{s}[K]$ values for the mixture of normals variational approximation with the diagonal precision matrix are similar to the approximation with a sparse precision matrix for each component for the mixture of normal priors, but the $\widetilde{s}[K]$ values are much higher for the $t$ priors example. The mixture of normals variational approximation with sparse precision matrix is better than the mixture of normals with a diagonal precision matrix. 

\begin{figure}[H]
\caption{Plots of the $\widetilde{s}[K]$ values for the mixture of normals variational approximations with global and local boosting steps for the logistic regression example with a mixture of normal (left panel) and t (right panel) priors for the first $20$ latent variables, where the poorly approximated latent variables are identified using the $s[K]$ values. \label{fig:Varplot_some_global_local_logistic_mixnom_t}}
\centering{}\includegraphics[width=15cm,height=8cm]{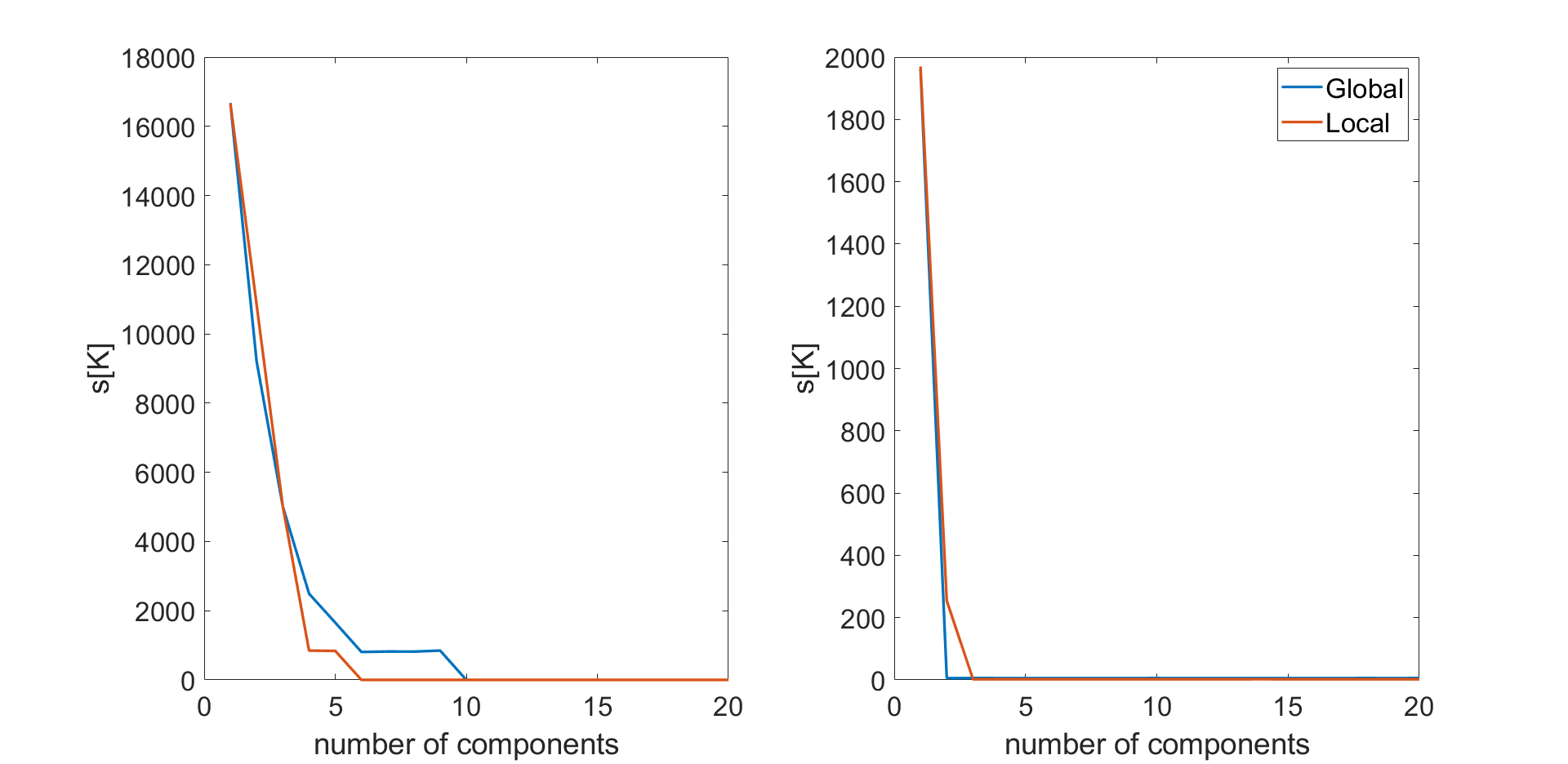}
\end{figure}

Figure \ref{fig:Varplot_some_global_local_logistic_mixnom_t}  plots the $\widetilde{s}[K]$ values for the global and local boosting steps for the logistic regression
example with a mixture of normals and $t$ priors for the first 20 latent variables.
The figure shows that: (1) the optimal number of components for the global boosting step
is $K = 2$ components for the $t$ prior and $K=11$ components for the mixture of normals prior. The optimal number of components for the local boosting step is $K = 3$ components for the $t$ prior and $K=6$ for the mixture of normals prior. The optimal $\widetilde{s}[K]$ values are similar for both global and local boosting steps.  

\begin{figure}[H]
\caption{Kernel density estimates of the first marginal of the latent variables estimated using the mixture of normals variational approximation with the optimal number of components chosen using $\widetilde{s}[K]$ for both global and local boosting steps for the logistic regression example with a mixture of normal (left panel) and t (right panel) priors for the first $20$ latent variables. \label{fig:densityestimatesmixnom_t_global_local_some}}
\centering{}\includegraphics[width=15cm,height=8cm]{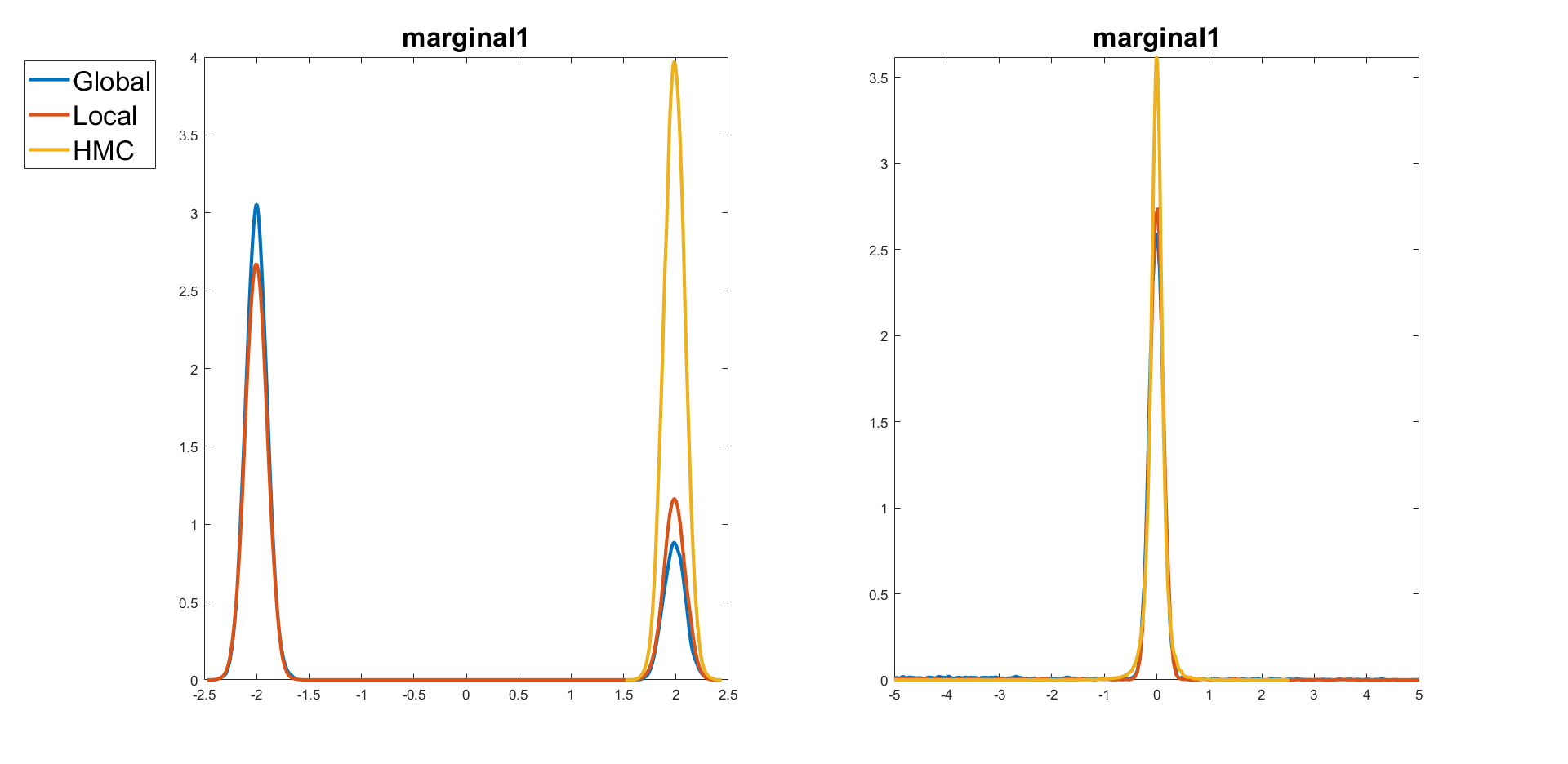}
\end{figure}

\begin{figure}[H]
\caption{Log of density estimates of the first marginal of the latent variables estimated using the mixture of normals variational approximation with the optimal number of components chosen using $\widetilde{s}[K]$ for both global and local boosting steps for the logistic regression example with a mixture of normal (left panel) and t (right panel) priors for the first $20$ latent variables. \label{fig:logdensityestimatesmixnom_t_global_local_some}}
\centering{}\includegraphics[width=15cm,height=8cm]{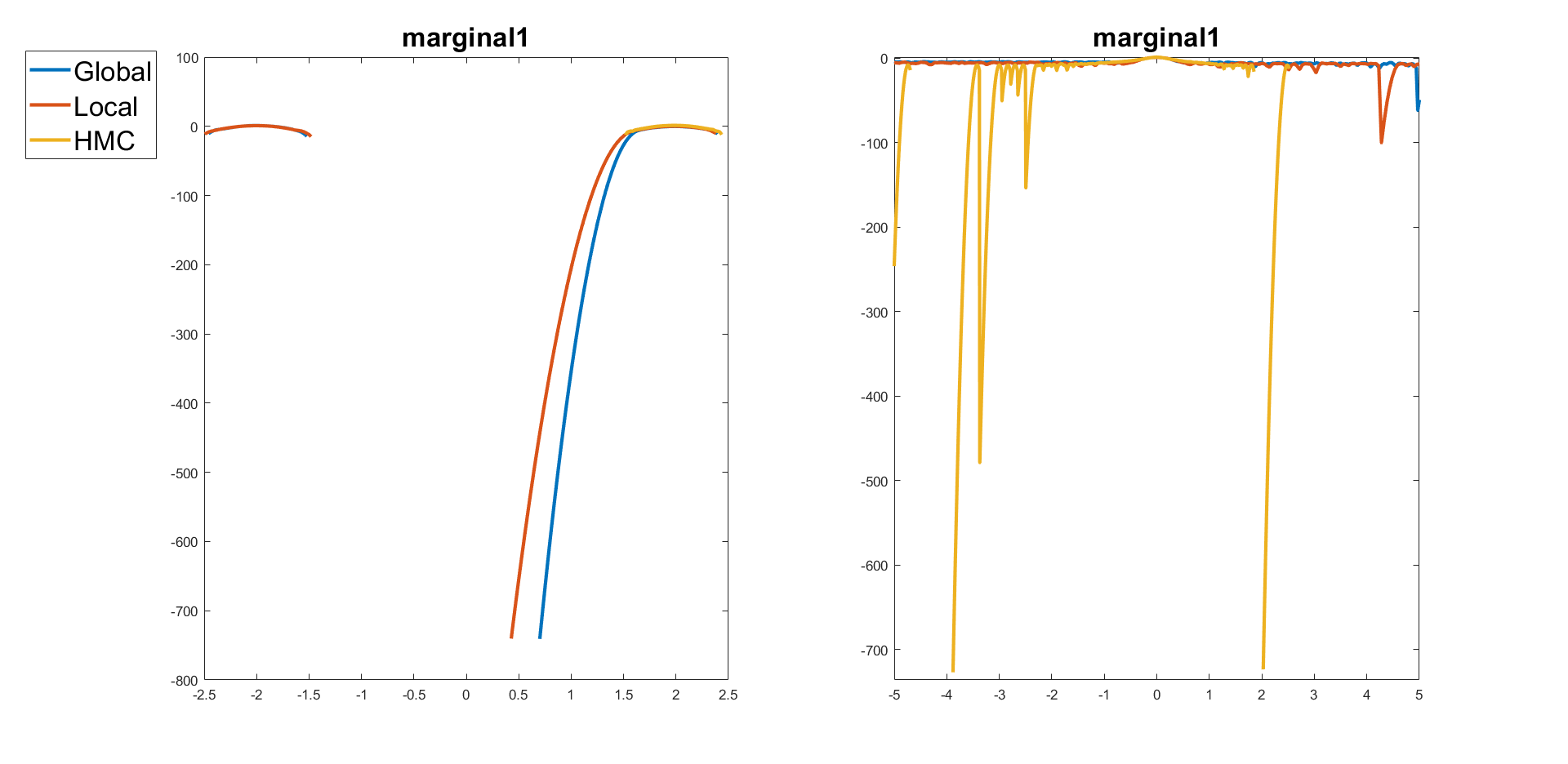}
\end{figure}

Figures \ref{fig:densityestimatesmixnom_t_global_local_some} and \ref{fig:logdensityestimatesmixnom_t_global_local_some}  plot the kernel density estimates and the log kernel density estimates, respectively, of the first marginal latent variable
estimated using the HMC method and the mixture of normals variational approximations with the optimal number
of components chosen using $\widetilde{s}[K]$ values for both the global and local boosting steps for the logistic
regression example with a mixture of normals and $t$ priors for the first $20$ latent variables, respectively. 
The figure demonstrates that the mixture of normals variational approximation can produce bimodal and heavy-tailed posterior distributions for the latent variables. In contrast, Hamiltonian Monte Carlo (HMC) yields posterior distributions with slightly thinner tails but is unable to produce multimodal posterior distributions. 

\begin{figure}[H]
\caption{Log of kernel density estimates of the first eight marginals of the latent variables estimated using the mixture of normals variational approximation with sparse precision matrix (Precision), the mean field mixture of normals variational approximation with diagonal precision matrix (MF) with the global boosting steps, and the Hamiltonian Monte Carlo (HMC) method for the logistic regression example with a mixture of normal priors for all of the latent variables. The optimal number of components is chosen using $\widetilde{s}[K]$ values. 
\label{fig:logdensityestimates_full_Global_logistic_statespace_mixnom_Precs_MF_latent}}
\centering{}\includegraphics[width=15cm,height=6.5cm]{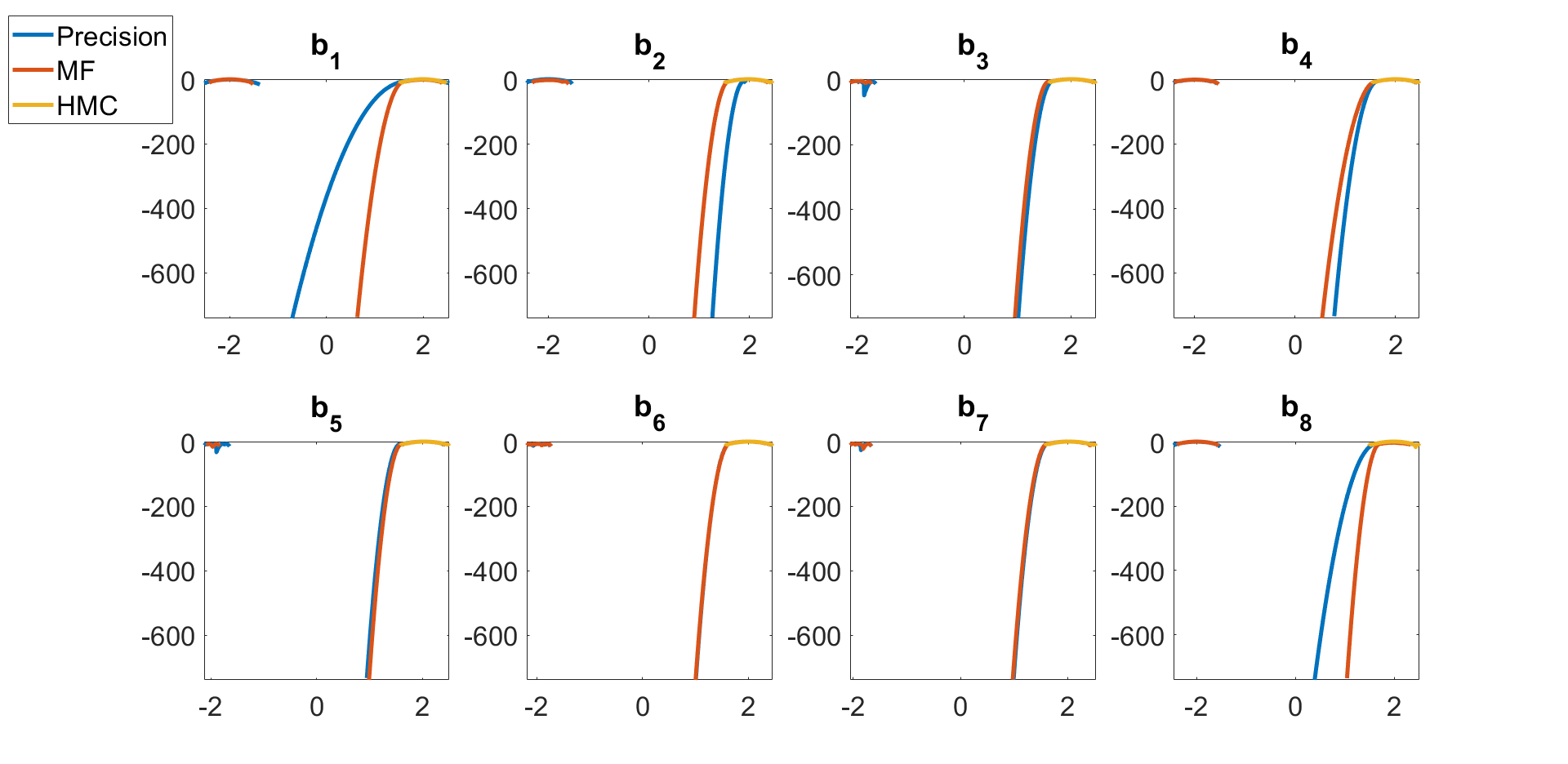}
\end{figure}

\begin{figure}[H]
\caption{Log of kernel density estimates of the first eight marginals of the latent variables estimated using the mixture of normals variational approximation with sparse precision matrix (Precision), the mean field mixture of normals variational approximation with diagonal precision matrix (MF) with the global boosting steps, and the Hamiltonian Monte Carlo (HMC) method for the logistic regression example with t priors for all of the latent variables. The optimal number of components is chosen using $\widetilde{s}[K]$ values.\label{fig:logdensityestimates_full_Global_logistic_statespace_t_Precs_MF_latent}}
\centering{}\includegraphics[width=15cm,height=6.5cm]{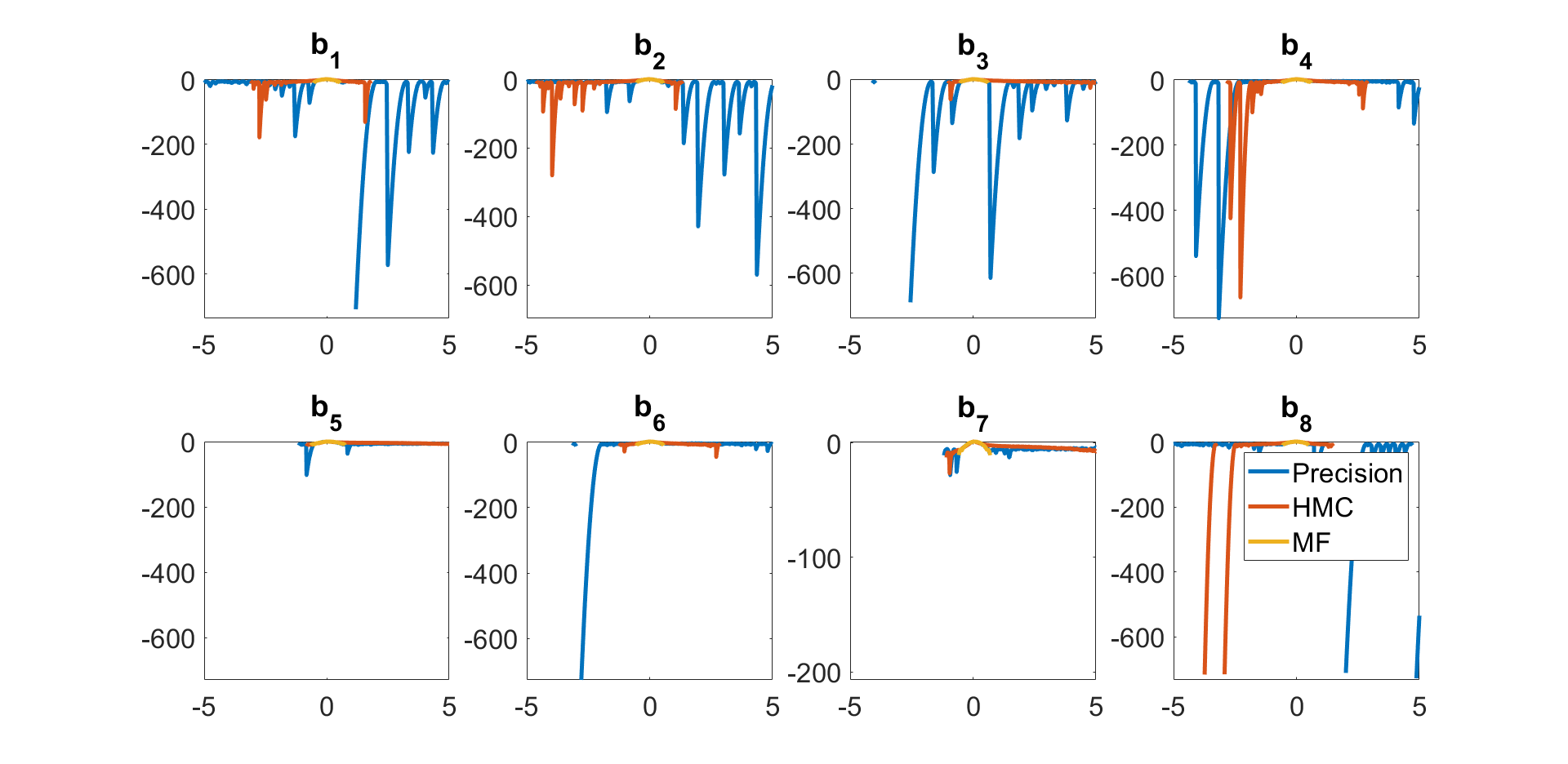}
\end{figure}

%\begin{figure}[H]
%\caption{Kernel Density Estimates of the global parameters estimated using the mixture of normals variational approximation with sparse precision matrix and the mean field mixture of normals variational approximation with diagonal precision matrix (Mean-Field) with the global boosting steps for the logistic regression example with a mixture of normal priors for all of the latent variables. The optimal number of components is chosen using $\widetilde{s}[K]$ values. \label{fig:densityestimates_full_Global_logistic_statespace_mixnom_Precs_MF_globalparam}}
%\centering{}\includegraphics[width=15cm,height=8cm]{densityestimates_full_Global_logistic_statespace_mixnom_Precs_MF_MCMC_globalparam}
%\end{figure}

%\begin{figure}[H]
%\caption{Kernel Density Estimates of the global parameters estimated using the mixture of normals variational approximation with sparse precision matrix and the mean field mixture of normals variational approximation with diagonal precision matrix (Mean-Field) with the global boosting steps for the logistic regression example with $t$ priors for all of the latent variables. The optimal number of components is chosen using $\widetilde{s}[K]$ values.  \label{fig:densityestimates_full_Global_logistic_statespace_t_Precs_MF_globalparam}}
%\centering{}\includegraphics[width=15cm,height=8cm]{densityestimates_full_Global_logistic_statespace_t_Precs_MF_MCMC_globalparam}
%\end{figure}

\section{Time-Varying Logistic Regression Model\label{subsec:time-varyinglogistic}}
This section considers the time series logistic regression model.
Let $\boldsymbol{y}=\left(y_{1},...,y_{n}\right)^{\top},$ $y_{i}\in\left\{ 0,1\right\} $,
denote an outcome at time $i$. We are interested in modeling the
probability of a positive outcome $y_{i}=1$ using the model
of the form

\begin{align}
y_{i} = \textrm{Bernoulli\ensuremath{\left(p_{i}\right)},} \qquad p_{i} = \frac{\exp\left(b_{i}\right)}{1+\exp\left(b_{i}\right)}, \qquad b_{i} = \phi b_{i-1}+\eta_{i},
\end{align}

%\begin{eqnarray}
%y_{i} & \sim & \textrm{Bernoulli\ensuremath{\left(p_{i}\right)},}\nonumber \\
%p_{i} & = & \frac{\exp\left(b_{i}\right)}{1+\exp\left(b_{i}\right)},\nonumber \\
%b_{i} & = & \phi b_{i-1}+\eta_{i},
%\end{eqnarray}
\noindent for $i=1,..,n$ time periods, where $b_{i}$ is the time-varying latent
variable following a first order autoregression  (AR1). We use complex prior distributions for each
error term $\eta_{i}$ for $i=1,...,n$ to produce complex posteriors. 
The first prior distribution for the $\eta_i$ is the two-component mixture of normals,  
\begin{equation}\label{priors_mixnom_timevarying}
p\left(\eta_i|w,\sigma_{1}^{2},\sigma_{2}^{2},\mu_1, \mu_2,\betavec\right)=wN\left(\eta_i;\mu_1,\sigma_{1}^{2}\right)+\left(1-w\right)N\left(\eta_i;\mu_2,\sigma_{2}^{2}\right),\;i=1,...,n.
\end{equation}
We set $w=0.5$, $\sigma_{1}^{2}=0.01$, $\sigma_{2}^{2}=10$,
$\mu_1=0$, and $\mu_2=0$. The second prior distribution for each error term $\eta_i$ is the $t$-distribution with mean $\mu=0$, scale $\sigma^2=0.01$, and degrees of freedom $\nu=3$.
We transform the constrained
parameters to the real space $\phi={\exp\left(\psi\right)}/\left ( {1+\exp\left(\psi\right)}\right)$,
where $\psi\in R$ and use a normal prior for $\psi\sim N(0,1)$.

The daily demeaned return $(r_i)$ data from Apple Inc. stock from 9th October 2009 to 3rd October 2011 (a total of $500$ observations) is  transformed into a binary variable $y_i=1$ if $r_i>0$ and $y_i=0$, otherwise.  
We consider two cases: (I) the mixture of normals and t priors are used for all $n=500$ latent variables, (II) the priors are only used for the first $20$ latent variables, the other $n-20$ use $N(0,1)$ as the priors. The global and local boosting methods are used for both cases.
$S=100$ Samples are used to estimate the lower bound values and the gradients
of the lower bound. Algorithm \ref{alg:Variational-Algorithm} is performed for $5000$ iterations to obtain the optimal variational
parameters for each boosting optimisation. The values $s_i[K]$ are calculated using grid points between $-5$ and $5$, for each $i$. 

\begin{figure}[H]
\caption{Left Panel: Plots of the lower bound values obtained using control variates and the reparameterisation trick methods for updating $\ellvec_{K+1}[K+1]$ for the mixture of normals variational approximations with global boosting steps for the time-varying logistic regression example with a mixture of normals prior for all the latent variables. Right Panel: Plots of the trajectory of the second value of the diagonal matrix of the variational parameters obtained using control variate and the reparameterisation trick methods for updating $\ellvec_{K+1}[K+1]$ for the mixture of normals variational approximations with global boosting steps for the time-varying logistic regression example with a mixture of normals prior for all the latent variables.\label{fig:controlreparamstatespacelogistic}}

\centering{}\includegraphics[width=15cm,height=7cm]{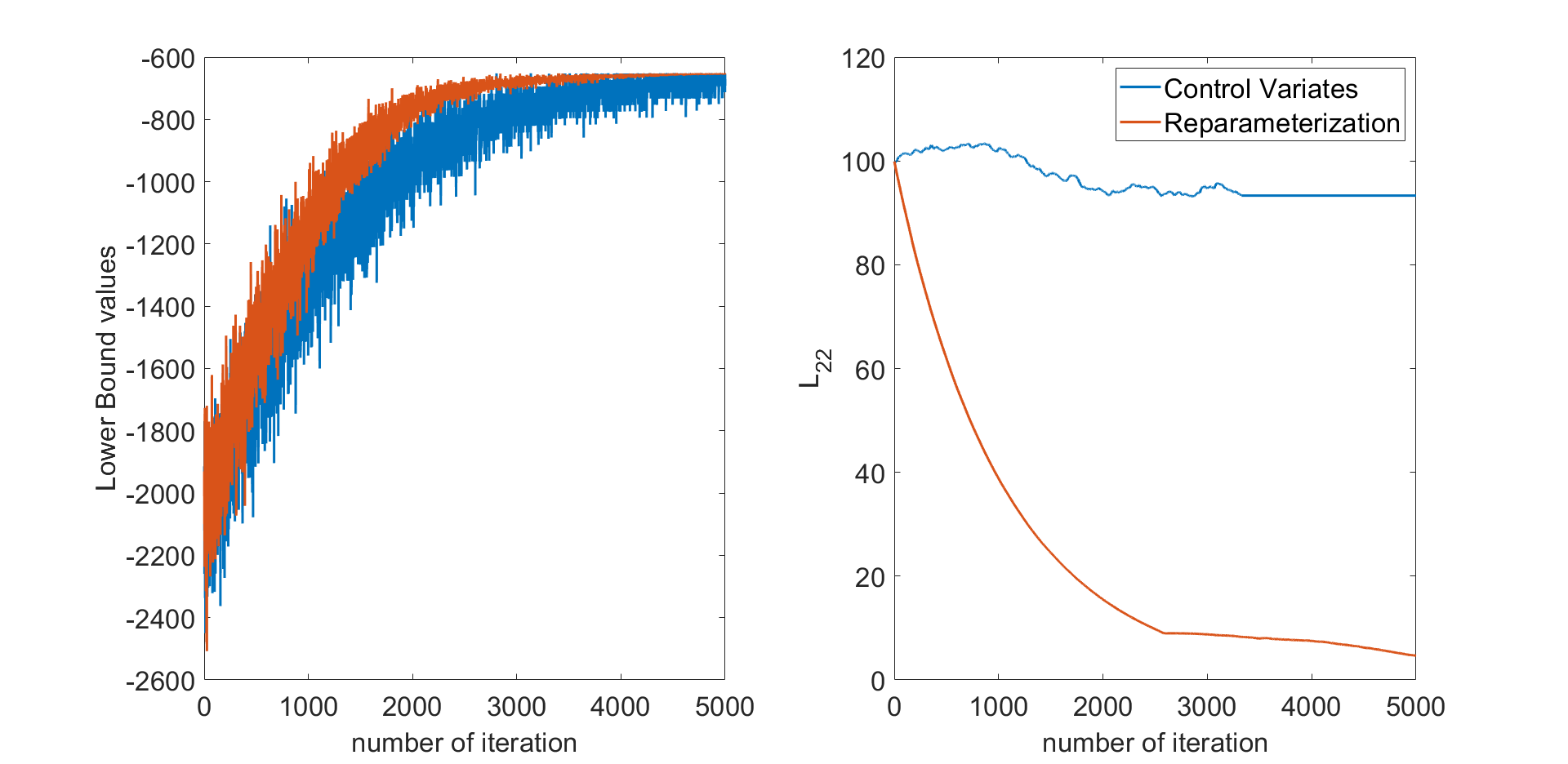}
\end{figure}

We now compare two algorithms to update the variational parameters $\ellvec_{K+1}[K+1]$. The first is based on the reparameterisation trick discussed in section \ref{subsec:Updating-the-Variational}. The second is based on the control variates approach of \citet{Ranganath:2014} described in 
section~\ref{sec:controlvariates}. The left panel of 
Figure~ \ref{fig:controlreparamstatespacelogistic} plots the lower bound values obtained using control variates and the reparameterisation trick. The figure shows that the lower bound 
obtained using control variates
is more noisy and converges much slower than the lower bound obtained using the reparameterisation trick.  The right panel shows the trajectory of the second value of the diagonal of the precision matrix $L_{K+1}[K+1]$ obtained using the control variates method and the reparameterisation trick. The figure shows that the trajectory obtained using the reparameterisation trick converges much faster. 

\begin{figure}[H]
\caption{Plots of the $\widetilde{s}[K] $ values for the mixture of normals variational approximations with global boosting steps for the logistic regression example with a mixture of normal (left panel) and t (right panel) priors for all the latent variables. \label{fig:Varplot_full_global_logistic_statespace_mixnom_t}}

\centering{}\includegraphics[width=15cm,height=8cm]{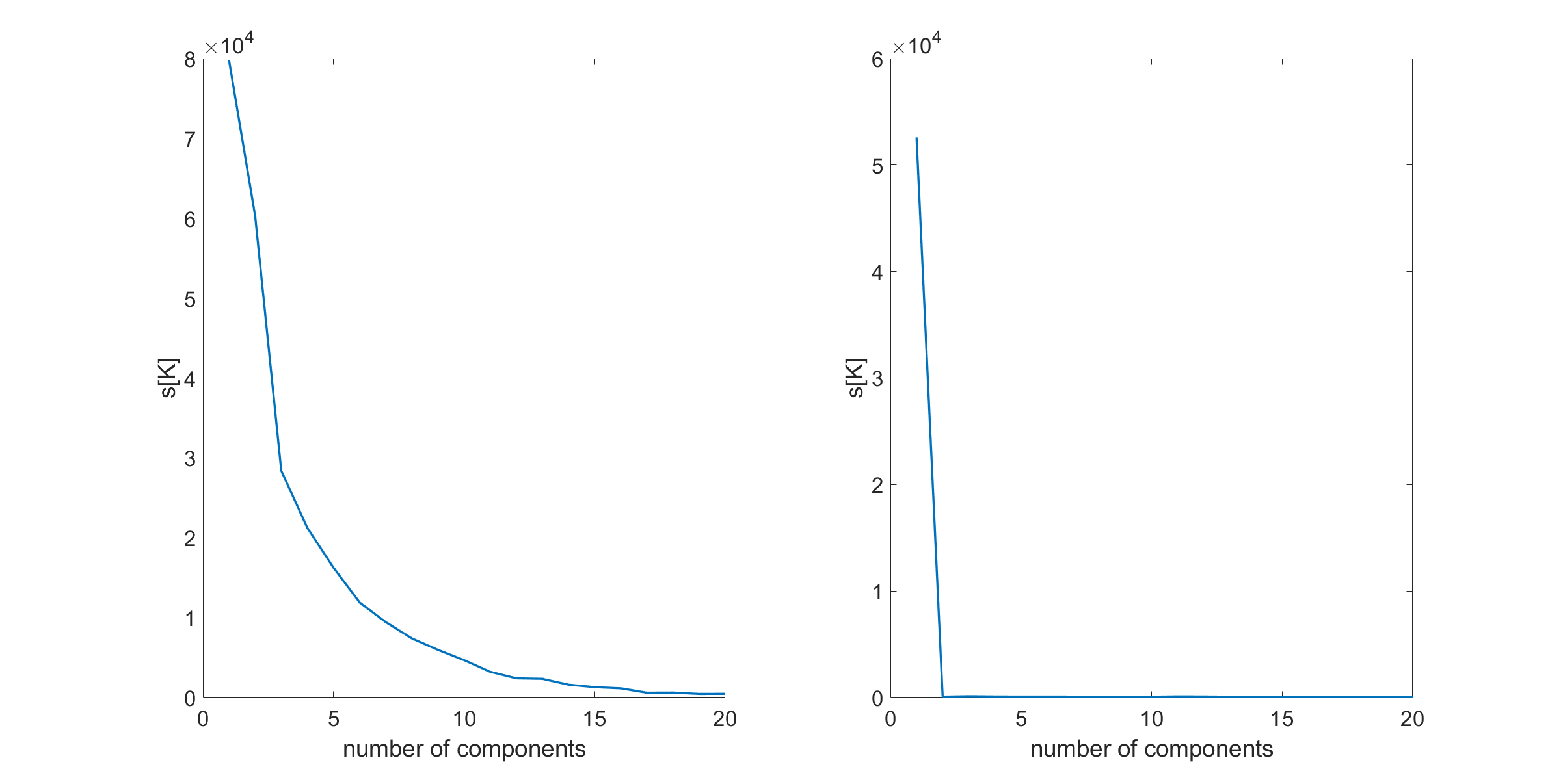}
\end{figure}

Figure~\ref{fig:Varplot_full_global_logistic_statespace_mixnom_t} plots the $\widetilde{s}[K]$ values with global boosting steps for the time-varying logistic regression example with a mixture of normals and $t$ priors for all the latent variables. The figure shows that the values of $\widetilde{s}[K]$ decrease over boosting iterations for both priors, indicating a substantial amount of improvement in the approximation of the marginal distributions of the latent variables. The mixture of Gaussians variational approximation is clearly better than the standard Gaussian variational approximation. 

\begin{figure}[H]
\caption{Plots of the lower bound values for the mixture of normals variational approximations with global and local boosting steps for the logistic regression example with a mixture of normal (left panel) and t (right panel) priors for the first $20$ latent variables. \label{fig:LB_some_global_local_logistic_statespace_mixnom_t}}

\centering{}\includegraphics[width=15cm,height=8cm]{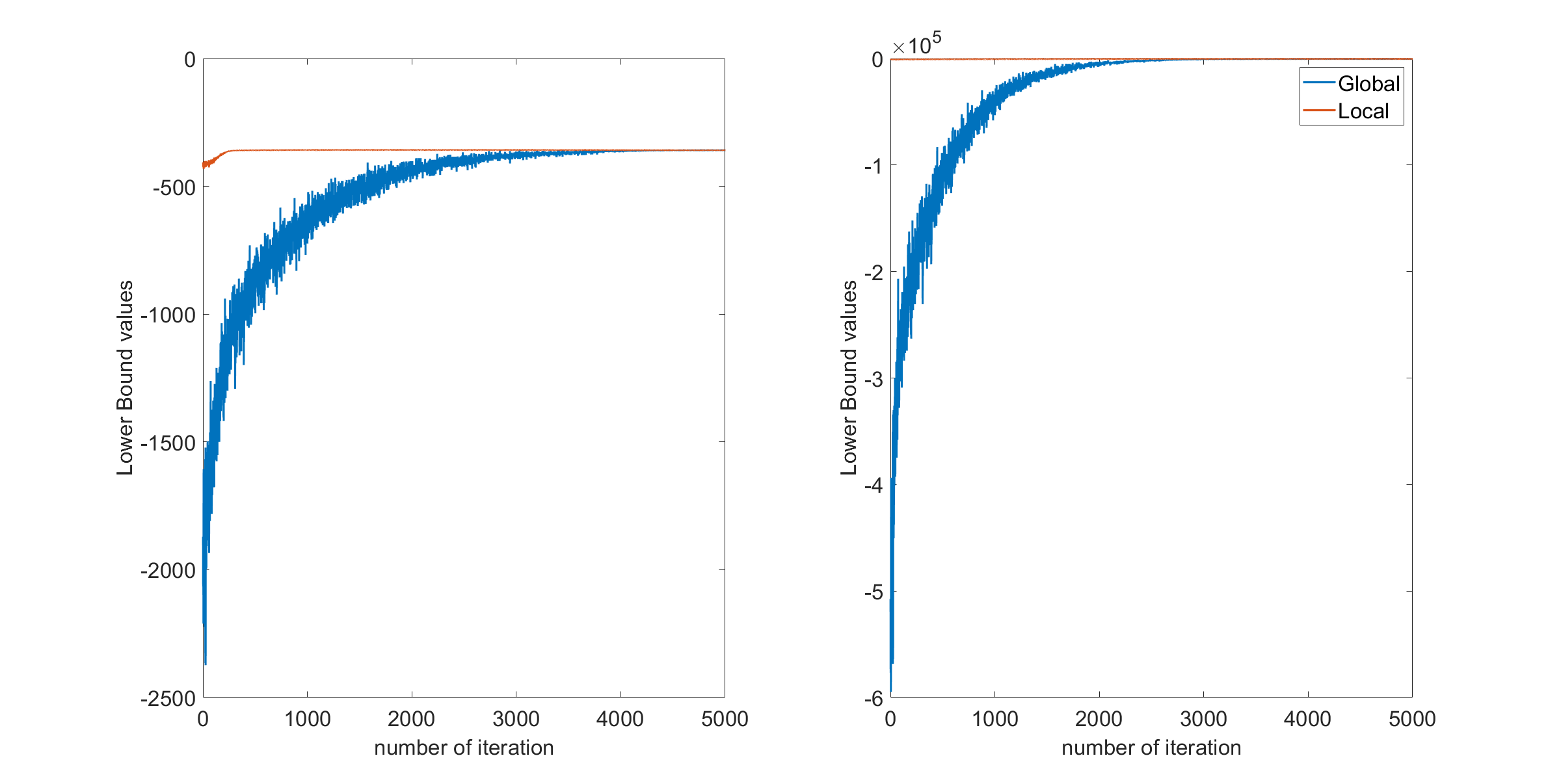}
\end{figure}

Figure \ref{fig:LB_some_global_local_logistic_statespace_mixnom_t}  plots the lower bound values for the mixture of normals variational approximation with global and local boosting steps for the time-varying logistic regression example with a mixture of normals (left panel) and $t$ (right panel) priors for the first $20$ latent variables. The figure shows that the local boosting steps that only update the variational parameters for the poorly approximated latent variables have better initial lower bound values, are much less noisy, and converge faster than the global boosting steps for both priors. The optimisation for the local boosting steps is easier than the global boosting steps because it updates fewer parameters.

\begin{figure}[H]
\caption{Plots of the indices of the latent variables selected by the ${s}[K]$ values for the mixture of normals variational approximations with local boosting steps for the time-varying logistic regression example with a mixture of normals (left) and $t$ (right) priors for the first 20 latent variables. \label{fig:index_selected_var_mixnom_t_statespace_logistic}}

\centering{}\includegraphics[width=15cm,height=8cm]{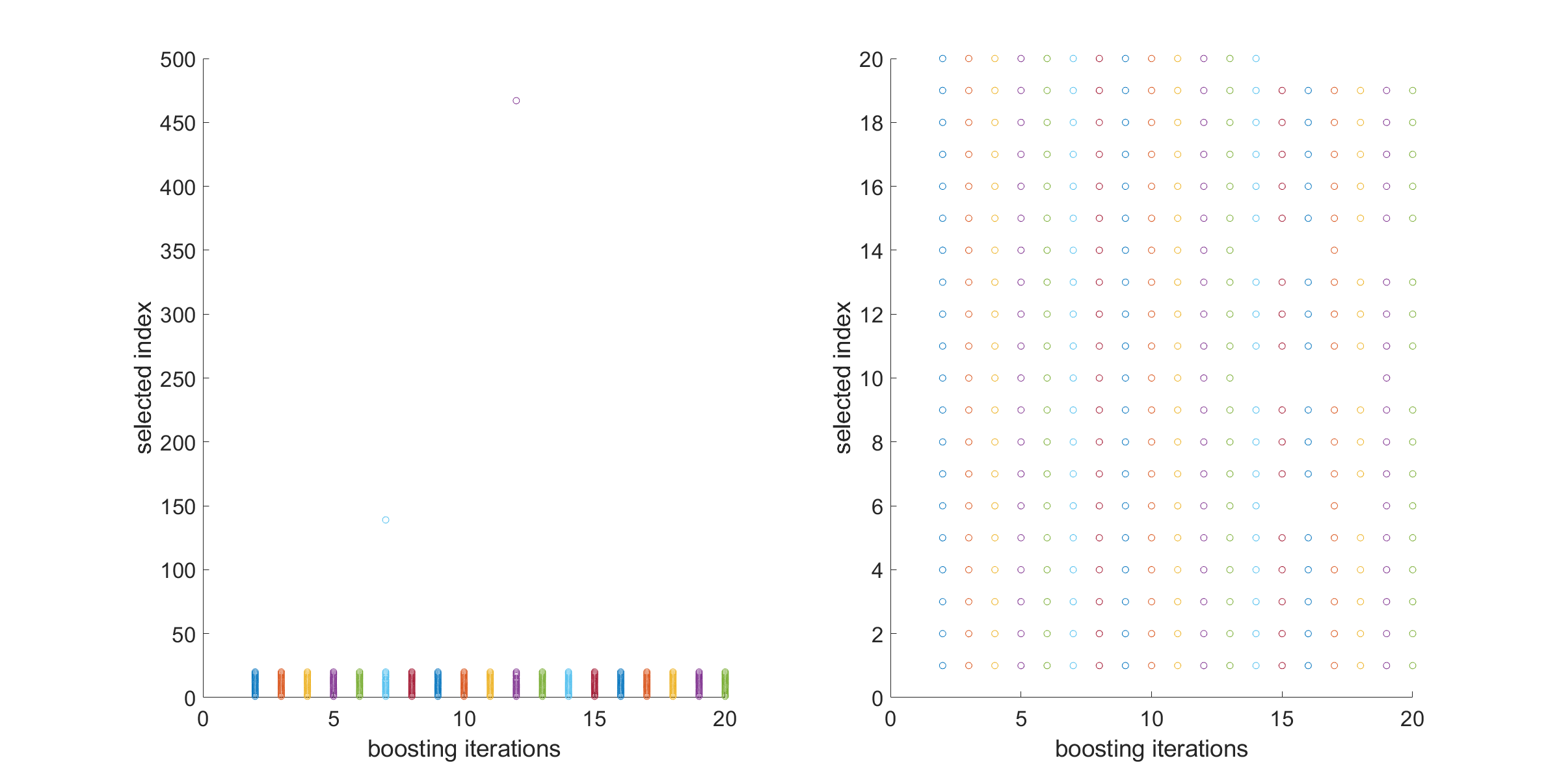}
\end{figure}

We now investigate the performance of the ${s}[K]$ values to select the poorly approximated latent variables in the state space model. 
Figure \ref{fig:index_selected_var_mixnom_t_statespace_logistic}  plots the indices of the latent variables selected by the ${s}[K]$ values for the time-varying logistic regression model with a mixture of normals (left panel) and $t$ (right panel) priors to the first $20$ latent variables. Note that the selection method works well if it successfully selects the indices $1 - 20$ for the first few boosting iterations. The figure shows that the selection method based on ${s}[K]$ correctly identifies the poorly approximated latent variables. 

\begin{figure}[H]
\caption{Plots of the $\widetilde{s}[K]$ values for the mixture of normals variational approximations with global boosting steps for the time-varying logistic regression example with a mixture of normals (left panel) and t (right panel) priors for the first $20$ latent variables. \label{fig:Varplot_some_global_logistic_statespace_mixnom_t}}
\centering{}\includegraphics[width=15cm,height=7cm]{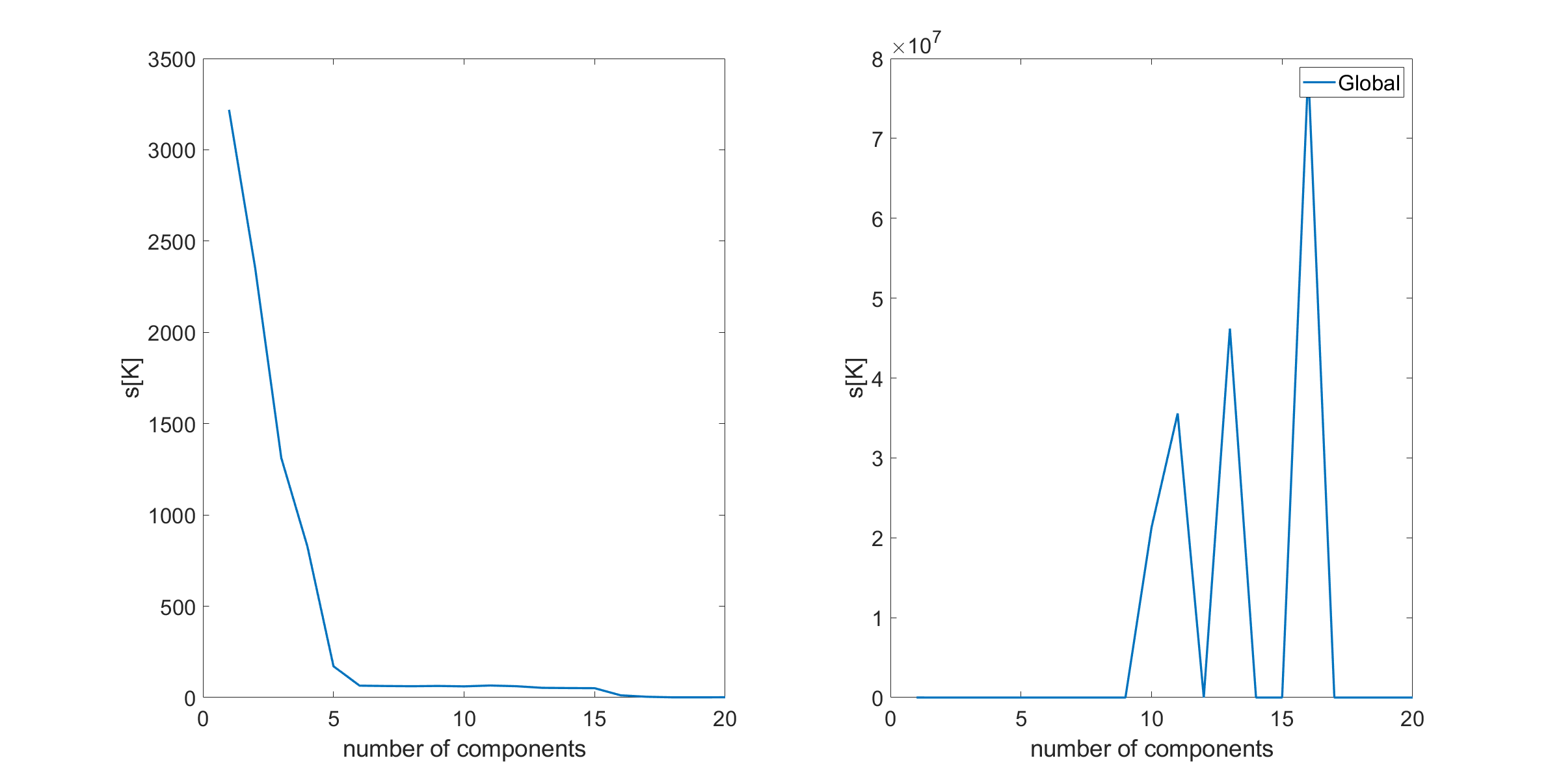}
\end{figure}

\begin{figure}[H]
\caption{Plots of the $\widetilde{s}[K]$ values for the mixture of normals variational approximations with local boosting steps for the time-varying logistic regression example with a mixture of normal (left panel) and $t$ (right panel) priors for the first $20$ latent variables. \label{fig:Varplot_some_local_logistic_statespace_mixnom_t}}
\centering{}\includegraphics[width=15cm,height=8cm]{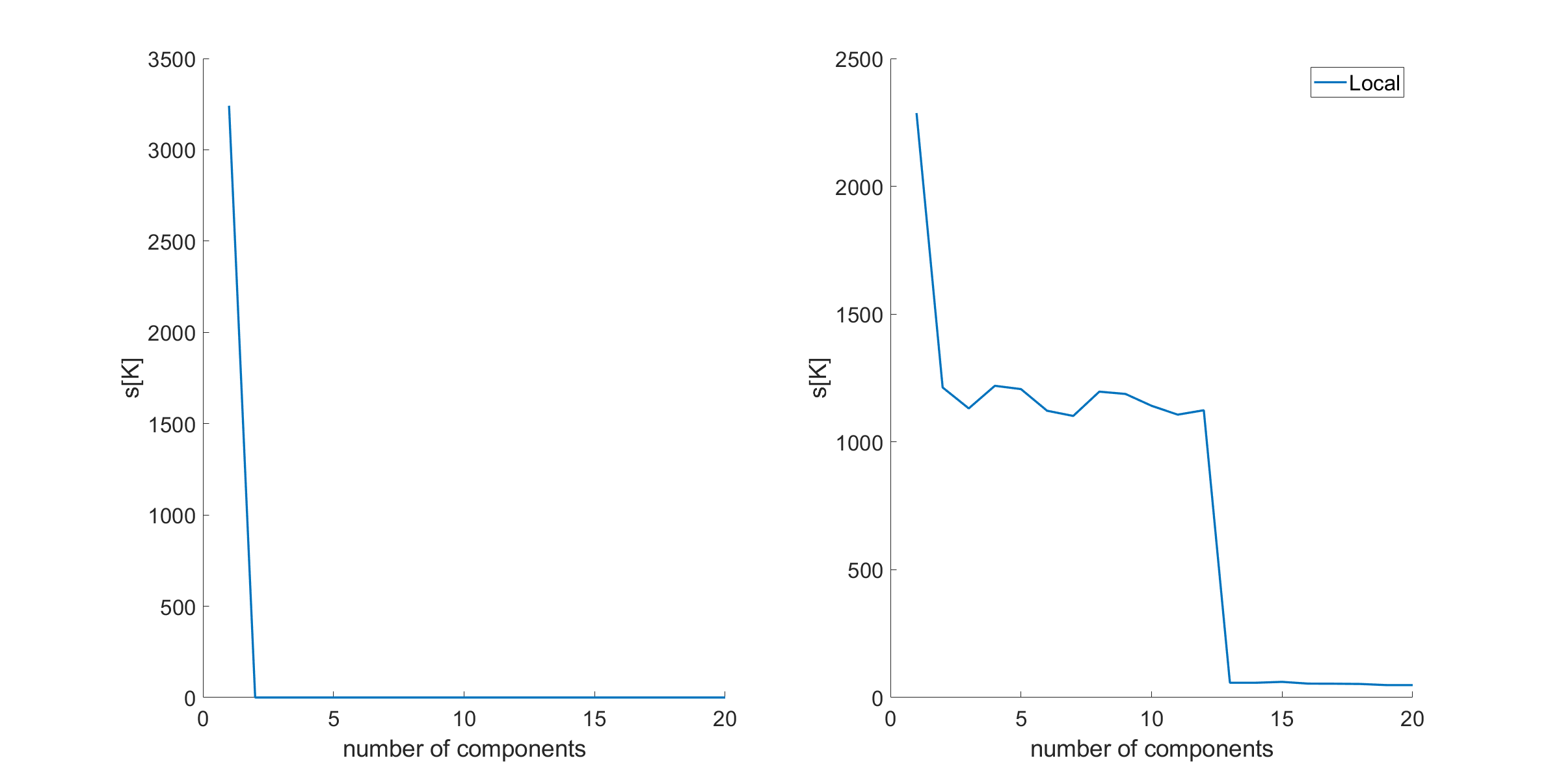}
\end{figure}

Figure \ref{fig:Varplot_some_global_logistic_statespace_mixnom_t} shows the $\widetilde{s}[K]$ values for global boosting steps for the time-varying logistic regression
example with a mixture of normals (left panel) and $t$ (right panel) priors for the first $20$ latent variables.
The figure shows that: (1) the $\widetilde{s}[K]$ values are decreasing over boosting iterations for both priors. (2) Interestingly, the $\widetilde{s}[K]$ values become unstable and very large for some of the boosting iterations for the t prior example. This shows that using global boosting steps, which update all the variational parameters, may worsen some of the posterior distributions that are  already approximated well in the previous boosting iterations. Figure \ref{fig:Varplot_some_local_logistic_statespace_mixnom_t} shows that the local boosting steps, which only update the variational parameters of the posterior distributions of the poorly approximated latent variables, are more stable.

\begin{figure}[H]
\caption{Kernel density estimates of two of the marginals of the latent variables estimated using the mixture of normals variational approximation with the optimal number of components chosen using $\widetilde{s}[K]$ values for both the global and local boosting steps for the logistic regression example with a mixture of normals (left panel) and t (right panel) priors for the first $20$ latent variables. \label{fig:densityestimatesmixnom_t_global_local_some_statespace}}
\centering{}\includegraphics[width=15cm,height=7cm]{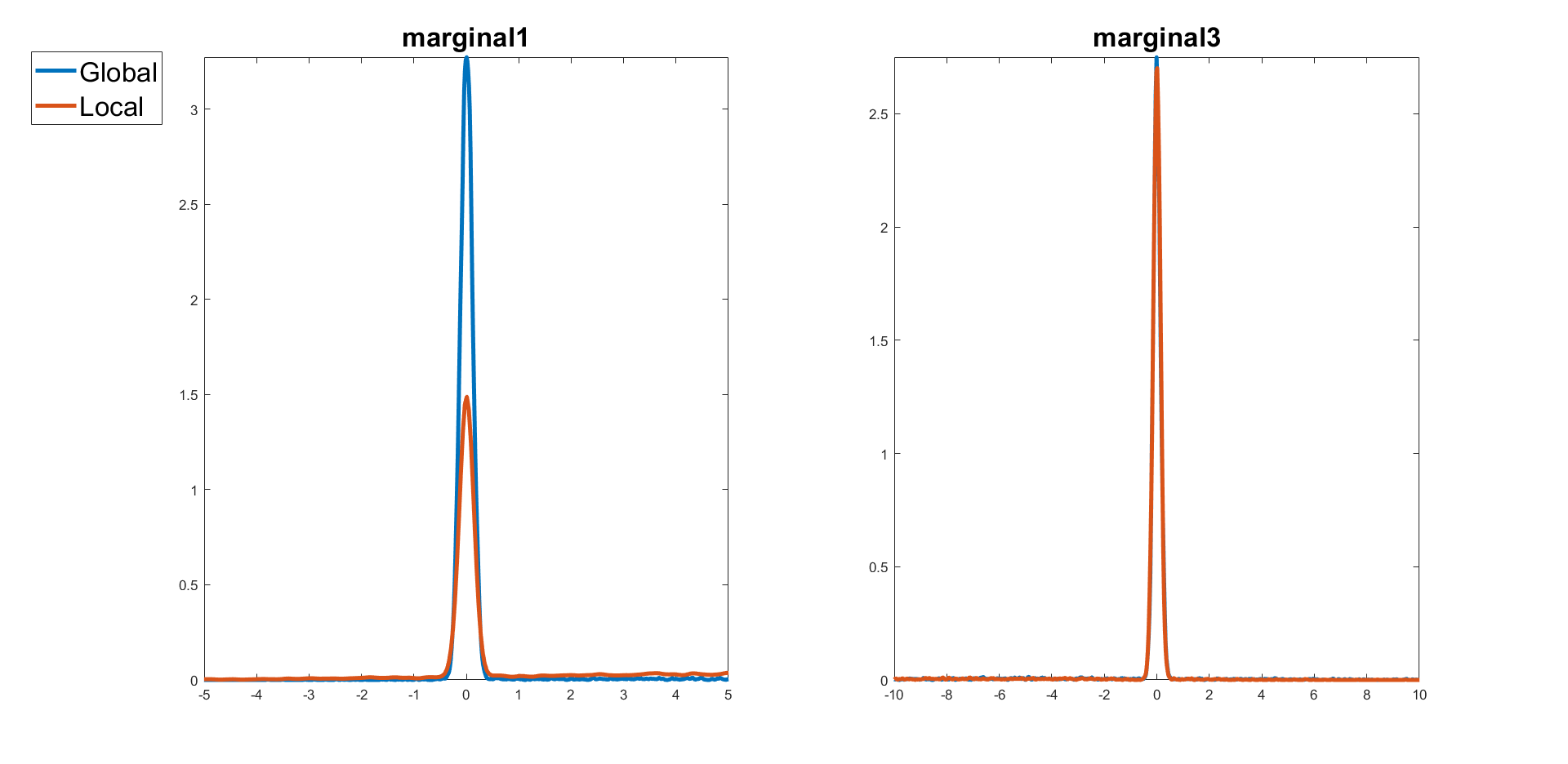}
\end{figure}

\begin{figure}[H]
\caption{Log of kernel density estimates of two of the marginals of the latent variables estimated using the mixture of normals variational approximation with the optimal number of components chosen using $\widetilde{s}[K]$ values for both the global and local boosting steps for the logistic regression example with a mixture of normals (left panel) and t (right panel) priors for the first $20$ latent variables. \label{fig:logdensityestimatesmixnom_t_global_local_some_statespace}}
\centering{}\includegraphics[width=15cm,height=7cm]{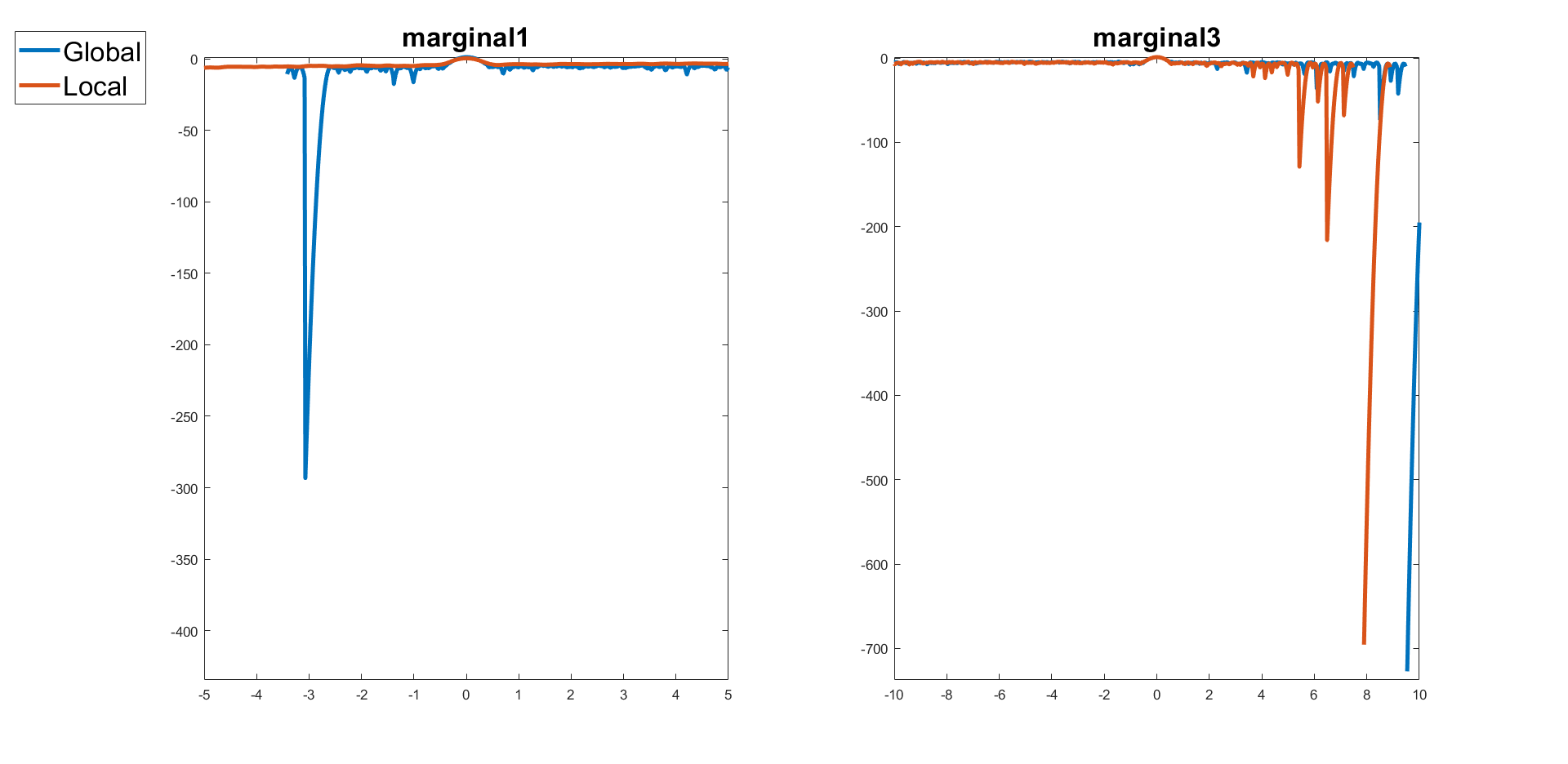}
\end{figure}

Figures \ref{fig:densityestimatesmixnom_t_global_local_some_statespace} and \ref{fig:logdensityestimatesmixnom_t_global_local_some_statespace} show the kernel density estimates and the log kernel density estimates, respectively, of the first and third marginal latent variables
estimated using the mixture of normals variational approximation for  global and local boosting steps for the time-varying logistic
regression example with a mixture of normals and $t$ priors for the first $20$ latent variables, respectively. The figure shows that the mixture of normals variational
approximation can capture the heavy-tailed features of the posterior distributions of the latent variables. The densities obtained using global and local boosting steps are quite similar for the $t$ prior and slightly different for the mixture of normals prior. 

%, with the mixture of normals variational approximation with local boosting steps having slightly thicker tails

\section{Initialising a new mixture component \label{sec:Initialising-a-New}}

Introducing a new component requires setting the initial values
for the new component parameters $\left(\muvec_{K+1}[K+1], \ellvec_{K+1}[K+1]\right)$
and the mixing weight vector $\pivec[K+1]$. A good initial value for the
new mixture component should be located in the region of the target
posterior distribution $p(\thetavec|\yvec)$ that is not well represented
by the existing mixture approximation $q_{\lambdavec[K+1]}\left(\thetavec\right)$.
There are many ways to set the initial value. We now discuss some 
that work well in our examples. For global and local boosting moves, we set $\pi$ equal to $0.5$. 
Algorithm \ref{alg:initial values for mu_K+1 and L_K+1}
gives initial values for  $\muvec_{K+1b}[K+1]$ and the diagonal elements in $L^{'}_{K+1b}[K+1]$. Algorithm \ref{alg:initial values for mu_K+1G and L_K+1G}  gives initial values for $\muvec_{K+1G}[K+1]$ and the diagonal elements of $L^{'}_{K+1G}[K+1]$. 

% The diagonal elements in $L_{K+1}[K+1]$ are
%initialized by $10$, and the non-zero non-diagonal elements
%in $L_{K+1}[K+1]$ are initialized by 0.001. These values ensure that the optimisation %algorithm is stable. 
%is initialized by $1/(K+1)$.

%$\varphi_{s}\sim q_{\lambda}^{\left(K\right)}\left(\theta\right)$

\begin{algorithm} [H]
\caption{Initial values for $\muvec_{K+1b}[K+1]$ and the diagonal elements of $L^{'}_{K+1b}[K+1]$. \label{alg:initial values for mu_K+1 and L_K+1}}

Input: $\left\{ \pi_{k}[K],\muvec_{k}[K],L^{'}_{k}[K]\right\} _{k=1}^{K}$

Output: initial values for $\muvec_{K+1b}[K+1]$ and the diagonal elements of $L^{'}_{K+1b}[K+1]$.

\begin{itemize}
\item We now discuss how to obtain the initial values for $\muvec_{K+1i}[K+1]$ and the diagonal elements of $L^{'}_{K+1i}[K+1]$ for $i=1,...,n$.  For $i=1$ to $n$,
\begin{itemize}
%Draw $c\in \{1,\dots, K\}$ with $P(c=j)=\pi_j[K]$
\item Select the mixture component $c$ with the highest weight.  
For notational simplicity, we assume the components are relabelled 
so that $c=K$. Then, after relabelling the draw,
$$\thetavec_G\sim N(\muvec_{KG}[K],\Omega_{KG}[K]^{-1}).$$  

\item Construct $R_1$ grid values of $\widetilde{\muvec}_{K+1,i,r_1}[K+1]$ and $R_2$ grid values of $\textrm{diag}\widetilde{(L^{'})}_{K+1,i,r_2}[K+1]$.  
    
\item For $r_1=1,...,R_1$ and for $r_2=1,...,R_2$, construct $S$ grid points for $\boldsymbol{b}_{i,j}$ for $ j=1,...,S$,
and compute the log difference between the current marginal approximation and the conditional posterior density of $\bvec_i|\thetavec_G$,

\begin{align}
 \widetilde{r}_{(r_1,r_2)}(\bvec_{i,j}) & =\log \left\{p(\bvec_{i,j}|\thetavec_G)p(\yvec_i|\bvec_{i,j},\thetavec_G)\right\} - \log \left\{q_{\lambdavec[K]}(\bvec_{i,j}|\thetavec_G)\right\},  \label{logratio}
\end{align}
and calculate $\textrm{Var}(\widetilde{r}_{(r_1,r_2)}(\bvec_{i,j}))$

\item Set $\muvec_{K+1,i}[K+1]=\muvec_{K+1,i,r_1}$ and $\textrm{diag}(L^{'})_{K+1,i}[K+1] = \textrm{diag}\widetilde{L^{'}}_{K+1,i,r_2}[K+1]$   with the minimum  $\textrm{Var} (\widetilde{r}_{(r_1,r_2)}(\bvec_{i,j}))$ for $r_1=1,...,R_1$ and for $r_2=1,...,R_2$.
\end{itemize}
\end{itemize}

\end{algorithm}

\begin{algorithm} [H]
\caption{Initial values for $\muvec_{K+1G}[K+1]$ and the diagonal elements of $L^{'}_{K+1G}[K+1]$. \label{alg:initial values for mu_K+1G and L_K+1G}}

Input: $\left\{ \pi_{k}[K],\muvec_{k}[K],L^{'}_{k}[K]\right\} _{k=1}^{K}$

Output: initial values for $\muvec_{K+1G}[K+1]$ and the diagonal elements of ${L^{'}_{K+1G}}[K+1]$.

%Draw $S$ samples from the proposal distribution $h(\thetavec_{G})$

\begin{itemize}
\item We now discuss how to obtain the initial values for $\muvec_{K+1Gi}$ and the diagonal elements of $L^{'}_{K+1Gi}[K+1]$ for $i=n+1,...,n+m_G$, where $m_G$ is the number of global parameters.  For $i=n+1$ to $n+m_G$,
\begin{itemize}
\item Create grid points, $\thetavec_{G,i,j}$, for $ j=1,...,S$. Let $\thetavec_{j}^{*}$ be a vector containing $\thetavec_{G,i,j}$ and other parameters and latent variables fixed at their means or some other reasonable values. Next, compute the log difference, 

\begin{align}
 \widetilde{r}(\thetavec_{G,i,j}) & = \log \left\{p(\yvec|\thetavec^{*}_{j})p(\thetavec^{*}_{j})\right\} - \log \left\{q_{\lambdavec[K]}(\thetavec^{*}_{j})\right\},  \label{logratio}
\end{align}

\item Set $\mu_{K+1G,n+i}=\thetavec_{G,i,j}$ with the largest $\widetilde{r}(\thetavec_{G,i,j})$.
\end{itemize}
\item Each diagonal element of $L^{'}_{K+1Gi}[K+1]$ is set to $100$ for $i=n+1,...,n+m_G$.
\end{itemize}
\end{algorithm}

\begin{singlespace}

\bibliographystyle{apalike}
\addcontentsline{toc}{section}{\refname}
\bibliography{VB-Boosting}

\end{singlespace}
\end{document}